%% file: chapter_main.tex
\documentclass[a4paper,11pt]{article}
\usepackage{aaskaiid}
\usepackage{amsmath}
\usepackage{xcolor}

\DeclareMathOperator{\sech}{sech}

\newcommand{\hicap } {{\rm H}\,{\scriptsize\rm I}}
\newcommand{\hi } {{\rm H}\,{\small\rm I}}
\newcommand{\ci } {[{\rm C}\,{\small\rm I}]}
\newcommand{\oii } {[{\rm O}\,{\small\rm II}]}
\newcommand{\oiii } {[{\rm O}\,{\small\rm III}]}

\title{Testing dark matter models and modified gravity theories with spatially resolved H\,{\Large I} observations}
\ShortTitle{Testing dark matter models and modified gravity theories}
\author[1]{Federico~Lelli}
\ShortName{F. Lelli et al.} 
\emailAdd{federico.lelli@inaf.it}
\author[2]{Antonino~Marasco}
\emailAdd{antonino.marasco@inaf.it}
\author[3]{Timothy~A.~Davis}
\emailAdd{davist@cardiff.ac.uk}
\author[4]{Nathan~Deg}
\emailAdd{nathan.deg@queensu.ca}
\author[5]{Jonathan~Freundlich}
\emailAdd{jonathan.freundlich@astro.unistra.fr}
\author[6,7]{Gyula~I.~G.~J\`ozsa}
\emailAdd{gjozsa@mpifr-bonn.mpg.de}
\author[8]{Peter~Kamphuis}
\emailAdd{peterkamphuisastronomy@gmail.com}
\author[9]{Pavel~E.~Mancera~Pi\~na}
\emailAdd{pavel@strw.leidenuniv.nl}
\author[10]{Stacy~S.~McGaugh}
\emailAdd{stacy.mcgaugh@case.edu}
\author[11]{Amidou~Sorgho}
\emailAdd{asorgho@iaa.es}
\author[4]{Kristine~Spekkens}
\emailAdd{kristine.spekkens@queensu.ca}

\affiliation[1]{INAF, Osservatorio Astrofisico di Arcetri, Largo Enrico Fermi 5, 50125, Firenze, Italy}
\affiliation[2]{INAF, Osservatorio Astronomico di Padova, Vicolo dell'Osservatorio 5, 35122, Padova}
\affiliation[3]{Cardiff Hub for Astrophysics Research \&\ Technology, School of Physics \&\ Astronomy, Cardiff University, Queens Buildings, Cardiff, CF24 3AA, UK}
\affiliation[4]{Department of Physics, Engineering Physics, and Astronomy,Queen’s University,
Kingston ON K7L 3N6, Canada}
\affiliation[5]{Observatoire Astronomique de Strasbourg, Université de Strasbourg, CNRS UMR 7550, F-67000 Strasbourg, France}
\affiliation[6]{Max-Planck-Institut f\"ur Radioastronomie, Auf dem H\"ugel 69, D-53121 Bonn, Germany}
\affiliation[7]{Centre for Radio Astronomy Techniques \& Technologies (RATT), Department of Physics and Electronics, Rhodes University, Makhanda, EC, South Africa}
\affiliation[8]{Ruhr University Bochum, Faculty of Physics and Astronomy, Astronomical Institute (AIRUB), 44780, Bochum, Germany}
\affiliation[9]{Leiden Observatory, Leiden University, P.O. Box 9513, 2300 RA, Leiden, The Netherlands}
\affiliation[10]{Department of Astronomy, Case Western Reserve University, 10900 Euclid Avenue, Cleveland, OH 44106}
\affiliation[11]{Instituto Astrof\'isica Andaluc\'ia,
    Glorieta de la Astronomía s/n, 18008, Granada, Spain
}
\abstract{
Understanding the nature of dark matter (DM) is one of the most pressing questions of modern physics. The flat rotation curves of galaxies, especially the extended \hi\ rotation curves from radio observations at 21 cm, played a key historical role to establish the DM problem and continue to be a major tool to test different galaxy formation models, particle DM models, and modified gravity theories. 

The SKA observatory will revolutionize the study of \hi\ rotation curves thanks to its unprecedented angular resolution, sensitivity, and survey speed. In combination with optical and near-infrared ancillary data, \hi\ surveys with SKA-Mid AA4 will allow us to (1) obtain \hi\ data at spatial resolutions of $1''-2''$, probing the inner galaxy dynamics with unprecedented details (e.g., the cusp-core problem); (2) move from representative samples of a few hundreds of objects to statistical samples of tens of thousands of objects, allowing us to study the dynamical scaling laws of galaxies and their DM properties with unprecedented statistical power across different cosmic environments; (3) move from galaxies at $z\simeq0$ to galaxies at $z\simeq1$, probing DM halos at earlier cosmic epochs and exploring the redshift evolution of the dynamical laws across about 8 Gyrs.

The massive SKA data, however, will pose important technical challenges for the derivation of \hi\, rotation curves with automated kinematic software. Developing such kinematic software in the SKA era will be crucial to reach major scientific goals, such as discriminating between different solutions to the DM problem.}

\begin{document}
\maketitle

\section{Introduction}\label{sec:intro}

The nature of dark matter (DM) is one of the biggest mysteries of the Universe. DM is invoked to explain a variety of astrophysical phenomena, such as the internal dynamics of galaxies and galaxy systems (groups and clusters), the strength of gravitational lensing, and the power spectrum of the cosmic microwave background \citep[CMB; see][for historical perspectives]{Sanders2010, Bertone2018}. Nonbaryonic DM is a major ingredient of the $\Lambda$ Cold Dark Matter ($\Lambda$CDM) cosmological model and is essential to promote structure formation, starting from the tiny density perturbations imprinted on the CMB up to present-day galaxies. DM is generally thought to be a new elementary particle that has no electromagnetic interactions, beyond the standard model of particle physics. For many years, the leading DM candidates have been weakly interacting massive particles (WIMPs), but a plethora of alternative DM models exist, such as self-interacting DM made of particles interacting with each other, and fuzzy DM made of ultra-light Axion-like particles \citep[see][for a review]{2022PTEP.2022h3C01W}. Despite four decades of dedicated searches and experiments, particle DM has eluded clear detections so far \citep[e.g.,][]{Billard2022}. Alternatively, the DM effect may indicate the need to revise Einstein's theory of general relativity and its Newtonian limit. In this context, Modified Newtonian Dynamics \citep[MOND,][]{Milgrom1983a} and its relativistic extensions \citep{Skordis2021, Blanchet2024} appear to be most promising. Without a doubt, understanding the nature of DM is one of the major challenges of modern physics.

\begin{figure}[h]
    \centering
	\includegraphics[width=0.495\textwidth]{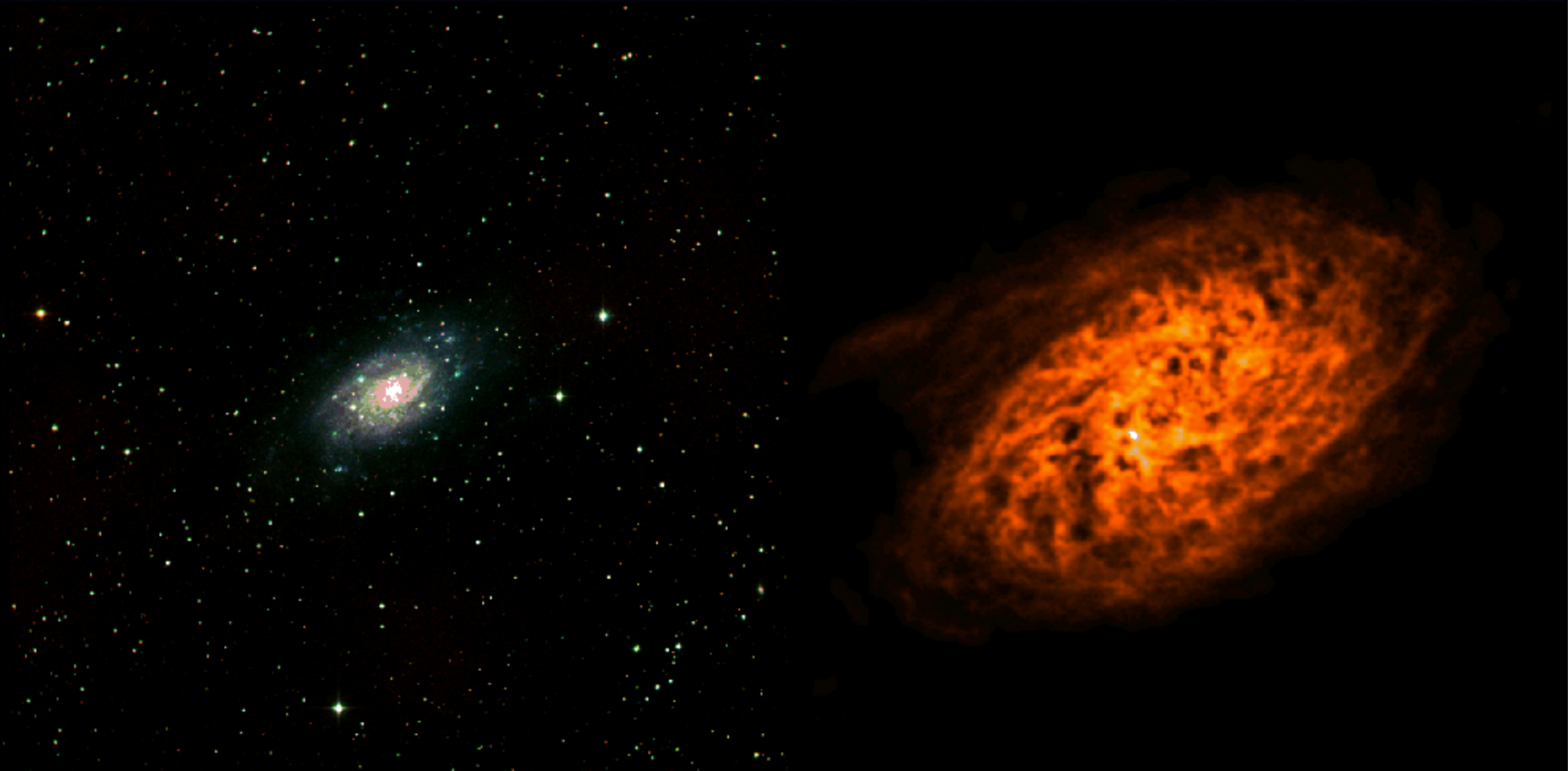}
    \includegraphics[width=0.495\textwidth]{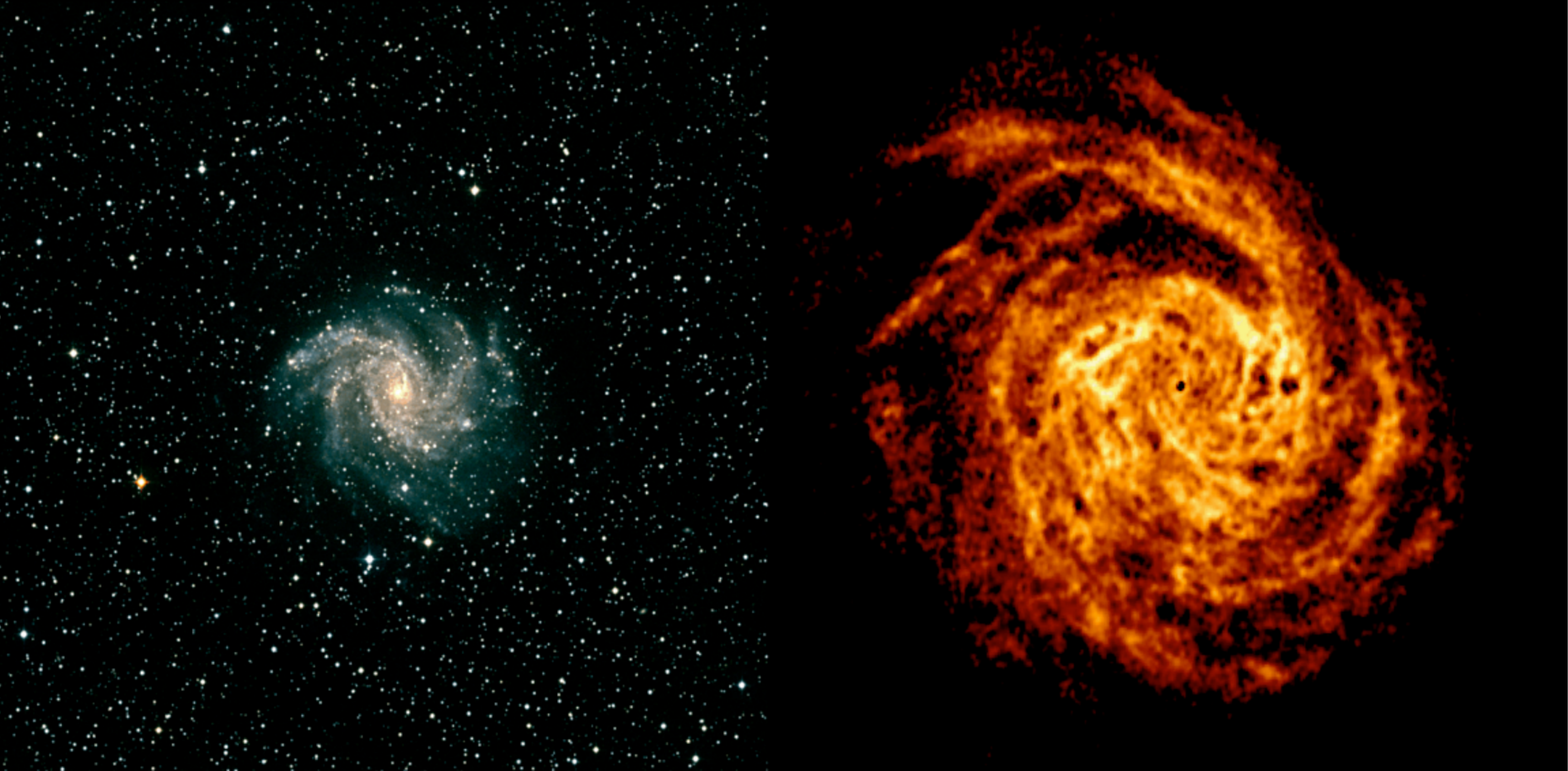}\\
    \includegraphics[width=0.495\textwidth]{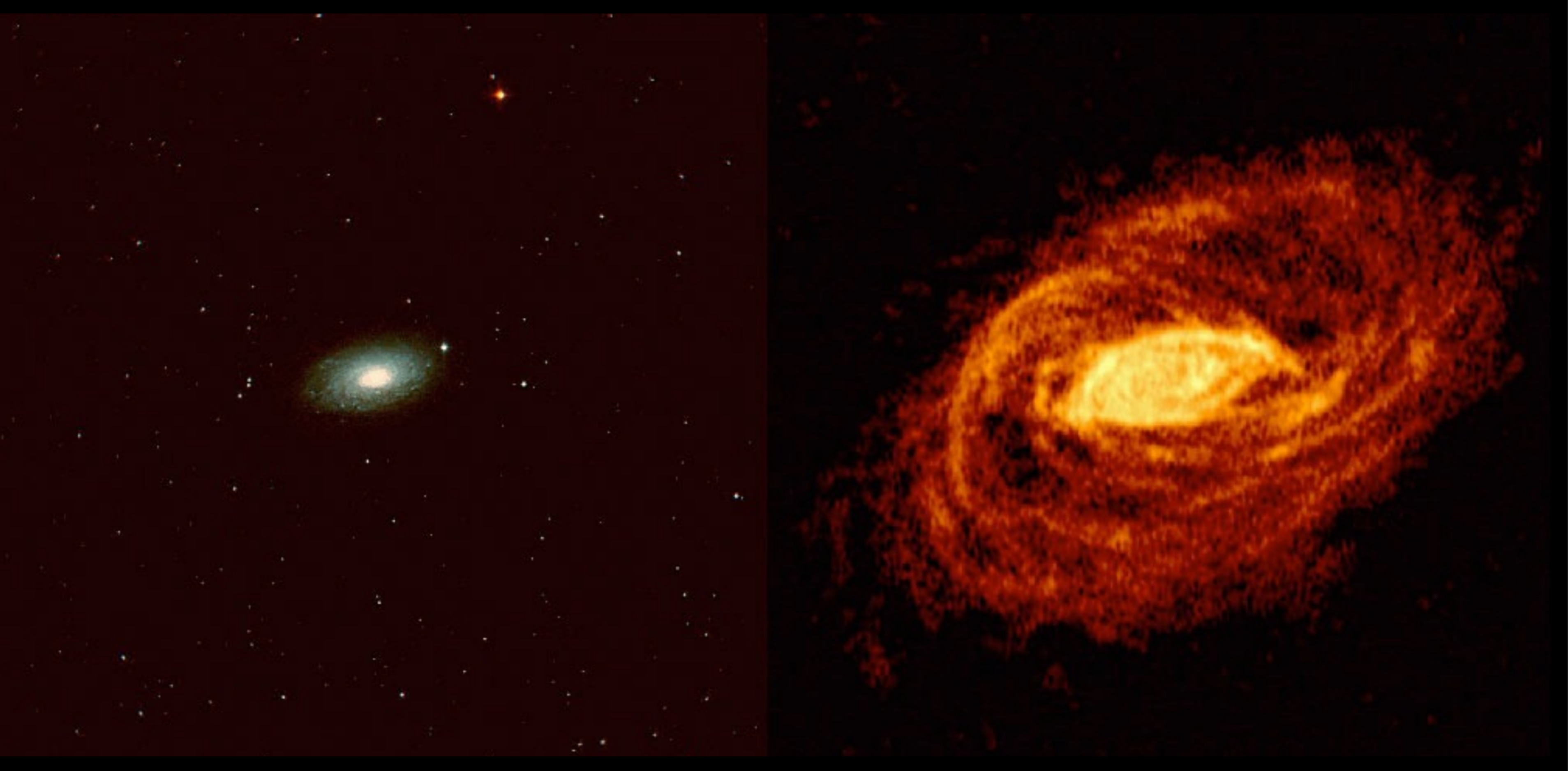}
    \includegraphics[width=0.495\textwidth]{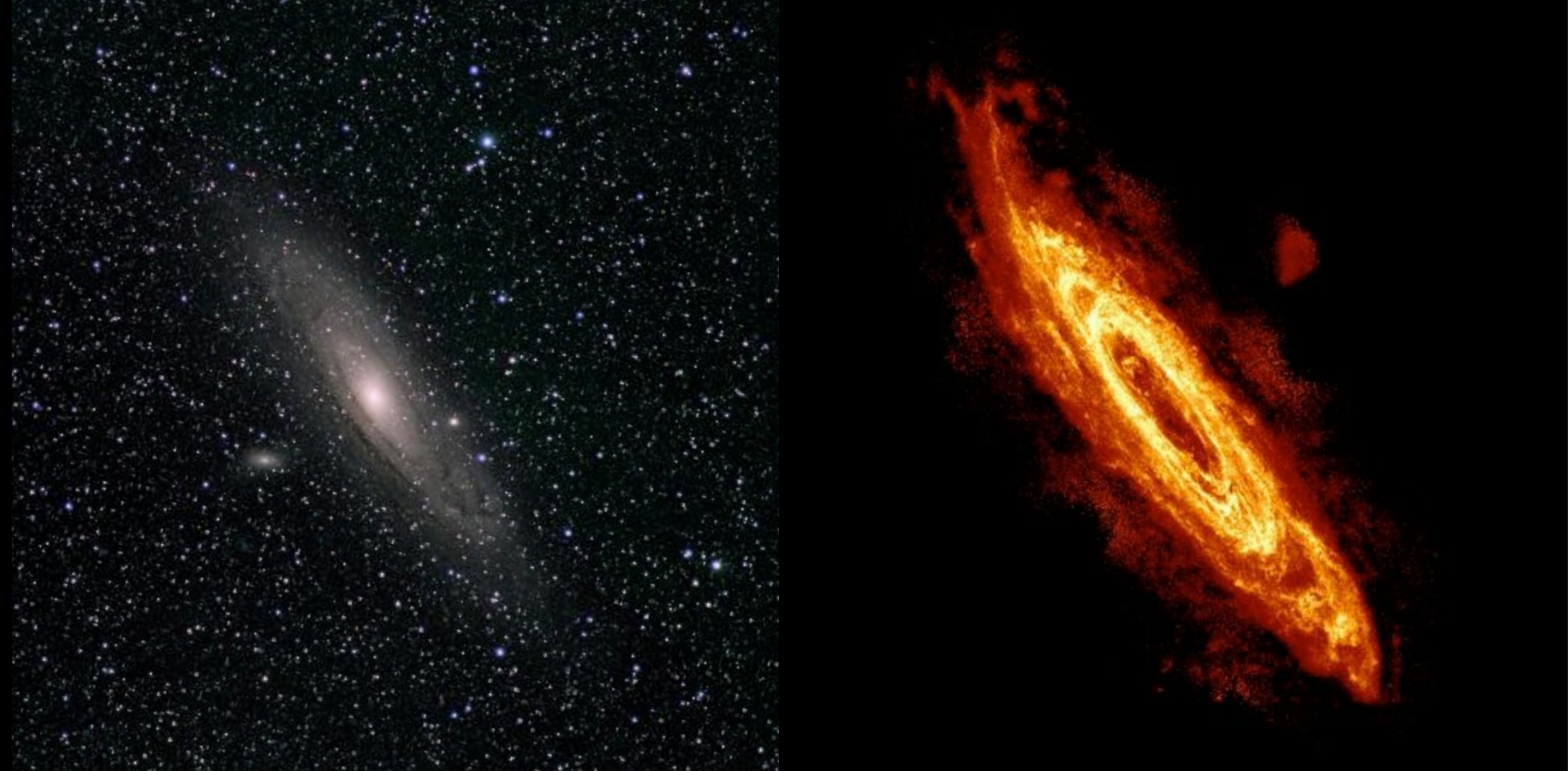}
    \caption{Optical images (left panels) and \hicap\ maps (right panels) on the same physical scale for four typical spirals: NGC\,2403 \citep[top left,][]{Fraternali02}, NGC\,6946 \citep[top right,][]{Boomsma2008}, NGC\,5055 \citep[bottom left,][]{Battaglia2006}, and M31 \citep[bottom right,][]{Braun2004}. Considering standard observations, the \hicap\, disk is generally more extended than the stellar component. Mass models of galaxies are usually built using optical and/or NIR images to compute the stellar gravitational contribution, and \hicap\, maps to compute the gas gravitational contribution (see Fig.\,\ref{fig:massmodel}). Images kindly provided by M. Verheijen.}
    \label{fig:himaps}
\end{figure}

The flat rotation curves of disk galaxies played a key historical role in the establishment of the DM problem \citep{Bosma1978, Rubin1978}. A key complementary role was also played by optical and near-infrared (NIR) surface photometry, providing the stellar gravitational contribution to the rotation curve \citep{Casertano1983}. In fact, the proper computation of the baryonic gravitational field of a disk-like system showed that H$\alpha$ rotation curves measured from optical spectroscopy were \emph{not} extended enough to unambiguously probe the DM effect \citep{Kalnajs1983, Kent1986}. The decisive evidence came from radio observations of the 21-cm line of atomic hydrogen (\hi). The \hi\ disks of galaxies are generally twice as extended as their stellar components \citep[e.g.,][see Fig.\,\ref{fig:himaps}]{Swaters2002a, Lelli2016b}, so they are uniquely suited to probe the outermost regions where the DM effect is dominant \citep{vanAlbada1985, vanAlbada1986, Kent1987}. 

\begin{figure}[t]
    \centering
	\includegraphics[width=\textwidth]{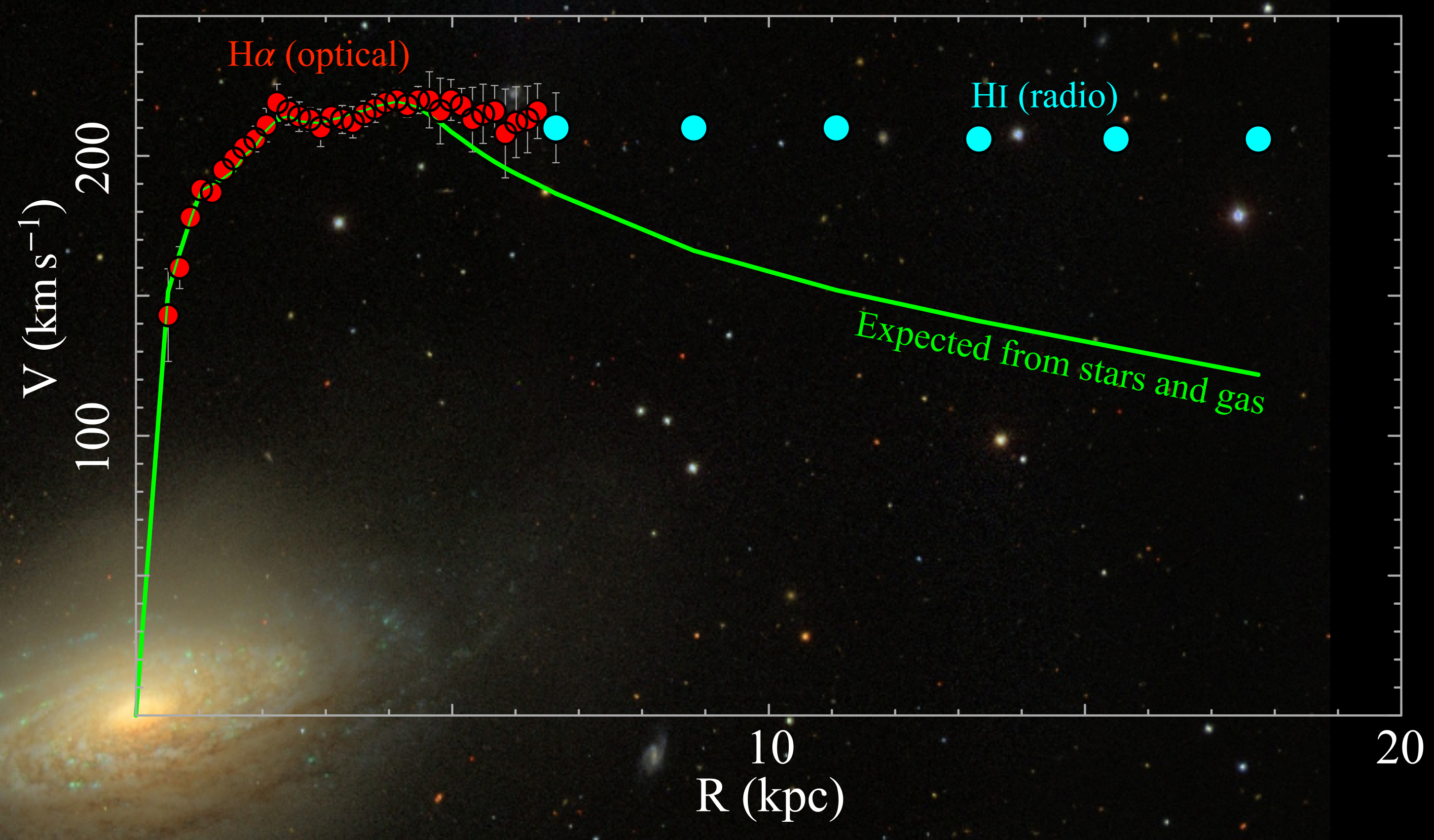}
    \caption{A classic evidence for the DM effect: the observed rotation curve of the spiral galaxy NGC\,3521 differs from that expected from the baryonic distribution. The baryonic contribution \citep[green line;][]{Lelli2016b} is computed solving the Poisson's equation for the observed distribution of stars (trace by NIR photometry) and gas (traced by \hicap\ interferometry). The H$\alpha$ rotation curve \citep[red points;][]{Daigle2006} implies almost no mass discrepancy; the DM effect is evident only at the larger radii probed by the \hicap\ rotation curve \citep[cyan points;][]{Begeman1987}. The background image is from the SDSS-DR9 \citep{Ahn2012}.}
    \label{fig:massmodel}
\end{figure}

Over the past 40 years, rotation curves have been derived for hundreds of nearby galaxies \citep[e.g.,][]{Begeman1987, Persic1996, Verheijen2001, deBlok2008, Swaters2009}. Ideally, one combines H$\alpha$ and/or CO data at high angular resolutions ($\sim$1$''-$2$''$) in the inner parts with \hi\ data at lower angular resolutions ($\sim$10$''-$30$''$) in the outer regions \citep[e.g.,][]{Bosma1978, deBlok2002, Noordermeer2005, Lelli2016b, Shelest2020}. In addition, NIR surface photometry provides the best possible tracer of the stellar mass distribution \citep[e.g.,][]{Verheijen2001b, McGaugh2014, Schombert2019}, so it can be used in concert with \hi\ and CO surface density maps to construct full mass models, separating the gravitational contributions of stars, gas, and DM (see Fig. \ref{fig:massmodel} for an example). 

The Spitzer Photometry and Accurate Rotation Curves (SPARC) database has been a major advance in this effort, collecting \hi+H$\alpha$ rotation curves from the literature \citep[see][for references]{Lelli2016b} and complementing them with NIR surface photometry \citep{Schombert2014}. The SPARC sample consists of 175 galaxies at $z\simeq0$ and has been used to test different $\Lambda$CDM models of galaxy formation \citep{Katz2017, Li2020} as well as alternative particle DM models, such as self-interacting DM \citep{Ren2019}, fuzzy DM \citep{Bar2022}, and superfluid DM \citep{Mistele2022}. The SPARC data have also been used to test modified gravity theories, such as MOND \citep{Li2018}, massive gravity \citep{Panpanich2018}, f($R$) gravity \citep{Naik2019}, conformal gravity \citep{LiModesto2020}, and others. However, the SPARC sample is too small and heterogeneous to firmly distinguish between these different theories. In addition, it lacks very low-mass galaxies (with baryonic masses smaller than $\sim$10$^8$ M$_\odot$), which are probed by smaller complementary samples, such as LITTLE-THINGS \citep{Oh2015, Iorio2017}, SHIELD \citep{McNichols2016}, and the compilation of \cite{ManceraPina2025}. 

To distinguish among different DM models and/or modified gravity theories, we need much larger samples (tens of thousands of galaxies) with both \hi\ and NIR data, probing galaxies in different cosmic environments. The first steps in this direction are represented by the WALLABY survey with the Australian SKA pathfinder \citep[ASKAP;][]{Westmeier2022, Deg2022, Murugeshan2024}, the AWES survey with APERTIF \citep{Adams2022}, and the upcoming BIG-SPARC database \citep{Haubner2024}. These projects will provide spatially resolved \hi\ data for a few thousand galaxies at $z\lesssim0.05$, increasing current samples by about one order of magnitude. Another avenue to test different DM models and modified gravity theories is probing the cosmic evolution of the baryon-DM connection by measuring \hi\ kinematics at different $z$, possibly up to $z\simeq1$. Pioneering \hi\ observations at $z\gtrsim0.05$ are provided by the CHILES survey with the Very Large Array \citep[VLA;][]{Hess2019}, the DINGO survey with ASKAP \citep{Rhee2023}, and the MIGHTEE-HI survey \citep{Jarvis2025, Varasteanu2025} and LADUMA survey \citep{KazemiMoridani2025} with the SKA-Mid precursor telescope MeerKAT.

The Square Kilometre Array (SKA) observatory, especially SKA-Mid in South Africa, will revolutionize the study of \hi\ rotation curves due to its unprecedented angular resolution, sensitivity, and survey speed. This chapter describes the main prospects expected from the SKA-Mid Array Assembly 4 (AA4). We start with a description of the main open questions in galaxy dynamics, namely the origin of the dynamical scaling laws (Sect. \ref{sec:scalinglaws}) and the long-standing $\Lambda$CDM problems on small scales (Sect. \ref{sec:cuspcore}). Next, we discuss the SKA prospects to obtain \hi\ observations at unprecedented angular resolutions of $\sim$1$''-$2$''$ (Sect. \ref{sec:highres}), for statistical samples of tens of thousands of spatially resolved galaxies (Sect. \ref{sec:largesamples}), and for galaxies at $z\simeq1$ that have been inaccessible in \hi\ emission so far (Sect.\,\ref{sec:highz}). Then, we discuss next-generation fitting tools and techniques to model the \hi\ kinematics in the SKA era (Sect. \ref{sec:tools}). We conclude with a brief summary (Sect.\,\ref{sec:summary}).

\section{Open questions}\label{sec:questions}

\subsection{The baryon-DM coupling and the dynamical laws of galaxies}\label{sec:scalinglaws}

A fundamental but somewhat underappreciated piece of evidence in the DM puzzle is that the dark mass and the luminous mass of galaxies appear tightly coupled to each other. Empirically, the observed dynamics, which is generally dominated by DM, is closely related to the distribution of baryons \citep[e.g.,][]{Swaters2012}. One of the pioneers of \hi\ kinematics in galaxies, Renzo Sancisi, summarized the evidence with a concise statement, which is often referred to as Renzo's Rule or Sancisi's Law: ``For any feature in the luminosity profile there is a corresponding feature in the rotation curve and vice versa'' \citep{Sancisi2004}. An example of this phenomenon is the correspondence in Fig.\,\ref{fig:massmodel} between the ``bump`` at 2 kpc in the observed rotation curve and that in the baryonic contribution, pointing to a ``maximum disk'' scenario in which baryons entirely dominate the inner galaxy regions \citep{vanAlbada1986}. Another important quote from \citet{Sancisi2004} is the following one: ``The unavoidable conclusion from the observed correspondence between the shapes of the rotation curves and those of the luminosity profiles is that the gravitational potential is strongly correlated with the distribution of luminous matter''.

\begin{figure}[t]
    \centering
	\includegraphics[width=\textwidth]{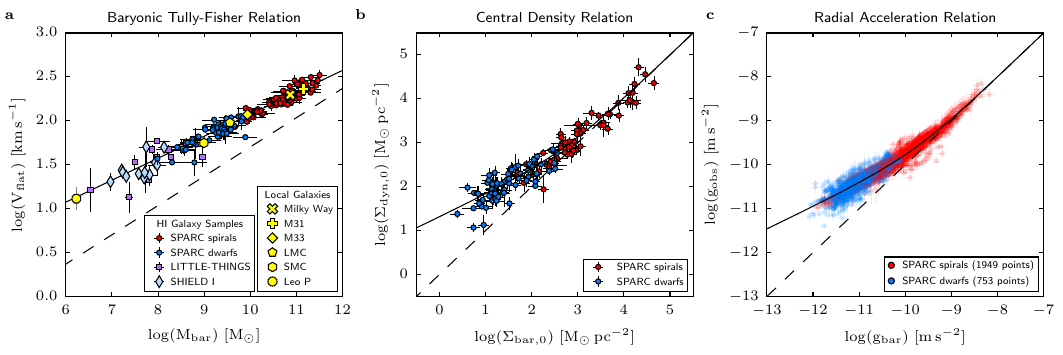}
    \caption{The dynamical scaling laws of galaxies \citep[adapted from][]{Lelli2022}. \textbf{Panel a}: The baryonic Tully-Fisher relation (BTFR). The dashed line shows the simplest $\Lambda$CDM prediction, with slope of 3 and intercept set by the cosmological ratio $\Omega_{\rm bar}/\Omega_{\rm CDM}$ \citep[see, e.g.,][]{McGaugh2012}. The solid line shows a one-parameter fit fixing the slope to 4, as predicted by MOND. \textbf{Panel b}: The central density relation (CDR). The dashed line corresponds to the 1:1 line. The solid line shows a one-parameter fit using the MOND interpolation function. \textbf{Panel c}: The radial acceleration relation (RAR). The dashed line corresponds to the 1:1 line. The solid line shows a one-parameter fit using the MOND interpolation function. In the BTFR and CDR, each point represents a single galaxy. In the RAR, each point represents a spatially-resolved measurement at different radii in 153 different galaxies. In all cases, the observed scatter is comparable to that expected from observational uncertainties, leaving little room for intrinsic scatter. \hicap\ surveys with SKA-Mid will allow studying these dynamical laws with large statistical samples as a function of environment and cosmic time.}
    \label{fig:scalinglaws}
\end{figure}
The empirical baryon-DM coupling can be quantified by a set of ``dynamical laws'' that connect a baryonic property (inferred from the observed distribution of stars and gas) and a dynamical one (inferred from the shape of the rotation curve), making no assumption about the nature of DM. Figure \ref{fig:scalinglaws} shows three of the main dynamical scaling laws of galaxies \citep[see][for details]{Lelli2022}: (1) the baryonic Tully-Fisher relation (BTFR) that links the total baryonic mass to the ``flat'' rotation velocity along the outer part of the rotation curve \citep{McGaugh2000, Verheijen2001b}; (2) the central density relation (CDR) that links the central baryonic surface density to the central dynamical-mass surface density inferred from the inner rising part of the rotation curve \citep{Lelli2013, Lelli2016c, Swaters2014}; and (3) the radial acceleration relation (RAR) that links the baryonic gravitational field measured from the baryonic distribution to the observed centripetal acceleration at each radius \citep{McGaugh2016, Lelli2017}. These dynamical laws are extremely tight by the usual standards of extragalactic astronomy: the observed scatter around the mean relations is consistent with the observational uncertainties, implying that any intrinsic scatter must be tiny \citep[e.g.,][]{Lelli2016a, Lelli2019, Li2018, Ponomareva2018, Ponomareva2021, Desmond2023}. The properties of these dynamical laws represent one of the major ``small-scale challenges'' for $\Lambda$CDM models of galaxy formation \citep{Bullock2017}.

The dynamical laws of galaxies raised a lively debate in the community because they can be interpreted in drastically different ways. In a particle DM scenario (irrespectively of the nature of the DM particle), the physical processes driving galaxy evolution inside DM halos must conspire to produce these scaling relations at $z=0$, which would then represent an ``attractor solution'' for baryonic physics \citep[e.g.,][]{DiCintio2016, Ludlow2017}. In a modified gravity scenario, instead, these scaling relations would represent new ``Laws of Nature'' akin to Kepler’s laws for planetary systems. Intriguingly, the existence and properties of these dynamical laws were predicted in advance by MOND \citep{Milgrom1983b}, a paradigm in which the nonrelativistic laws of gravity and/or inertia are modified below a characteristic acceleration scale $a_0$ \citep[see][for reviews]{Famaey2012, Milgrom2014, Banik2022}. 

Interestingly, particle DM models and MOND theories make diametrically different predictions about how the properties of the dynamical laws (slope, normalization, and scatter) should depend on large-scale environment and cosmic time. In general, particle DM models predict that the dynamical laws should (1) have measurable intrinsic scatter because they emerge from the chaotic process of galaxy formation \citep{DiCintio2016, Garaldi2018, Dutton2019}, (2) evolve with cosmic time \citep{Keller2017, Pillepich2019, Mayer2023}, and (3) show ``positive'' residual correlations with large-scale environment \citep{Paranjape2022}. Here, the term ``environment'' acquires the specific physical meaning of average gravitational acceleration due to the large-scale structure in a given region of the Universe \citep{Desmond2018}. In contrast, MOND predicts that the dynamical laws should (1) have zero or tiny intrinsic scatter \citep{Milgrom1983b, Milgrom1994}, (2) hold at any cosmic epoch \citep{Hossenfelder2018}, and (3) show ``negative'' residual correlations with the environment \citep{Chae2020, Chae2021, Chae2022a}. To test these predictions, one needs high-quality rotation curves and mass models for statistical galaxy samples in different environments and at different redshifts. In Sect.\,\ref{sec:prospects}, we describe how the synergy between \hi\ data from SKA-Mid surveys and optical/NIR images from wide-field photometric surveys (e.g., Euclid, Rubin) will provide the necessary data to test these predictions.

\subsection{$\Lambda$CDM small-scale problems: cusp-core, rotation-curve diversity, and too-big-to-fail}\label{sec:cuspcore}

According to DM-only cosmological simulations, CDM halos are singular (``cuspy'') at their centers with volume densities scaling as $\rho\propto r^{-1}$ \citep{Dubinski1991, Warren1992}, and can be described by a universal density profile, the NFW profile \citep{Navarro1996}. In contrast, the rotation curves of galaxies -- especially those of DM-dominated low-mass and low-surface-brightness galaxies -- often reveal much shallower density profiles or even constant density ``cores'' \citep[see][for a review]{deBlok2010}. For a long time, this ``cusp-core'' discrepancy has been considered a major challenge to the standard $\Lambda$CDM paradigm \citep[see][for a review]{Bullock2017}. Proposed solutions either advocate for fundamental changes in the physics of the dark sector, or focus on the role of baryonic processes during galaxy formation and evolution. The first type of solutions comprises alternatives to CDM such as warm DM (e.g., \citealt{Colin2000} but see also \citealt{Maccio2012}), self-interacting DM \citep[e.g.,][]{Spergel2000}, and fuzzy DM \citep[e.g.,][]{Goodman2000, Hu2000}. The second type of solutions generally invokes strong baryonic feedback from supernova explosions and/or active galactic nuclei (AGN) to produce massive gas outflows, which lead to non-adiabatic oscillations in the gravitational potential, eventually transforming DM cusps into DM cores \citep[e.g.,][]{Governato2010, Governato2012, Pontzen2012, El-Zant2016, Freundlich2020a}.

Some authors consider the cusp-core discrepancy ``solved'' in the $\Lambda$CDM context due to these ad-hoc adjustments in the modeling of baryonic physics. However, different hydrodynamical simulations disagree on both the intensity of baryonic feedback and its effect on the distribution of DM, with some simulations predicting cores in a given mass range, while others not forming cores at all \citep[e.g.,][]{Bose2019}. A general prediction of feedback-driven core-formation models is that the presence of a DM core should correlate with the star formation history of a galaxy \citep[e.g.][]{Muni2025}. This prediction could be tested by combining DM density profiles inferred from high-resolution \hi\ rotation curves with, e.g., star formation histories inferred from color-magnitude diagrams of resolved stellar populations \citep[e.g.,][]{McQuinn2010a} and/or multi-wavelength integrated photometry \citep[e.g.,][]{Schombert2019}. On the other hand, self-interacting DM and fuzzy DM appear rather insensitive to baryonic processes \citep{Robles2017} and predict different core-baryon relationships with respect to $\Lambda$CDM feedback models \citep{Jiang2023}, so there is hope to distinguish between these two proposed solutions to the cusp-core problem.

A related issue in the $\Lambda$CDM context is the so-called ``rotation-curve diversity'' problem \citep{Oman2015, Ghari2019}. At a fixed galaxy mass, observations show a diversity in the inner rotation curve shapes $-$ so in the DM density profiles $-$ contrary to $\Lambda$CDM expectations if feedback processes were the same for galaxies of similar mass. More generally, the observed diversity of dwarf galaxies is not reproduced in current $\Lambda$CDM simulations, which may again be related to the implementation of baryonic feedback and the resulting core-formation process. For example, the NIHAO simulations produce both DM cores and a population of ``ultra-diffuse'' galaxies, but no compact dwarfs \citep{Jiang2019}. Importantly, the observed diversity in the inner rotation-curve shapes of dwarf galaxies corresponds to a diversity in their inner baryonic surface densities \citep[see][]{Lelli2014b}, as quantified by the CDR (Fig.\ref{fig:scalinglaws}, middle panel). The true underlying problem, therefore, is understanding the origin of the empirical baryon-DM coupling (Sect.\,\ref{sec:scalinglaws}). This issue was pointed out by \citet{Sancisi2004} more than 20 years ago: ``why, in systems of comparable luminosity, are some DM halos cuspy (like the light) and others (also like the light) are not?''.

Another related issue in the $\Lambda$CDM context is the ``too-big-to-fail'' problem, whereby there is a mismatch between the predicted and observed number of satellite galaxies that have DM masses (measured within the stellar effective radius) above a given threshold. This would lead to the paradox of the most massive DM sub-halos not hosting a satellite galaxy and being completely dark, unless their DM mass within the stellar effective radius is substantially reduced by baryonic effects \citep[e.g.,][]{Sales2022}. To some extent, this problem is a combination of the ``cusp-core'' and the ``missing satellites'' problems, extrapolated to the lowest mass galaxies \citep{Bullock2017}. Obtaining \hi\ rotation curves for large samples of gas-rich satellite galaxies may clarify this problem and its relation to the ``cusp-core'' and ``rotation-curve diversity'' problems.

Finally, we point out that these various problems (``cusp-core'', ``rotation curve diversity'', ``too-big-to-fail'') are specific to $\Lambda$CDM and do not apply in a modified gravity context. In general, modified gravity theories are required to reproduce the observed rotation curves starting from the observed baryonic distribution, so the diversity in the rotation curve shapes (e.g., steeply or slowing rising) must naturally correspond to a diversity in the baryonic mass distribution (e.g., cuspy or cored). The inner parts of galaxies remain very interesting in a MOND context because it is where the predictions of different theories display the largest differences, albeit these are generally limited to a small 10$-$20\% effect \citep[e.g.,][]{Brada1995, Lelli2017b, Petersen2020, Chae2022b}. Constructing a large sample of galaxies with high-resolution \hi\ data (providing both the rotation curve and the gas distribution) and ancillary optical/NIR photometry (providing the stellar distribution) is crucial to distinguish between different modified-gravity theories.

\section{Observational prospects with SKA-Mid}\label{sec:prospects}

\hi\ surveys with the SKA-Mid AA4 will allow us to address the open questions described in the previous section. However, different questions require different survey strategies. For example, to study the inner shapes of rotation curves in high detail, we need targeted \hi\ observations of nearby galaxies at resolutions of about 1$''$-2$''$ and 3 km/s, reaching \hi\ column densities of about 5-10 M$_\odot$ pc$^{-2}$ (typical of the innermost galaxy regions). Instead, to investigate the possible dependency of the dynamical laws and DM properties on cosmic environment, we need a large statistical sample (tens of thousands of galaxies) with ``standard'' medium-shallow \hi\ data, for example reaching a 3$\sigma$ column density of 1 M$_\odot$ pc$^{-2}$ at resolutions of $5''-10''$ and $5-10$ km/s. Finally, to investigate the possible evolution of the dynamical scaling laws with $z$, we need an ultra-deep pencil-beam \hi\ survey in a well-studied cosmological field. This section describes these potential \hi\ surveys and highlights the importance of SKA-Mid AA4 to achieve the requested resolutions and/or sensitivities.

\subsection{Reaching higher spatial resolutions}\label{sec:highres}

For many years, the main limitation of \hi\ rotation curves has been their low angular resolutions of $15''-30''$ from ``standard'' observations with historical radio interferometers, such as the Westerbork Synthesis Radio Telescope (WSRT), the Very Large Array (VLA), and the Australia Telescope Compact Array (ATCA). Such angular resolutions are significantly worse than those of $1''-2''$ available for ancillary optical/NIR images (used to compute the stellar gravitational contribution), optical spectroscopy (used to measure H$\alpha$ rotation curves), and mm/submm interferometry (used to measure CO rotation curves). High-resolution H$\alpha$ and CO kinematic data have been instrumental in establishing the presence of DM cores in low-mass galaxies \citep[e.g.,][]{deBlok2002, Simon2003, Simon2005, KuziodeNaray2006} because they allow minimizing beam-smearing effects (see Sect.\,\ref{sec:beamsmearing}) and detecting potential noncircular motions (see Sect.\,\ref{sec:noncircular}). However, at very low masses ($M_{\rm bar} \lesssim 10^8$ M$_\odot$),  H$\alpha$ and CO observations are \emph{not} ideal to trace rotation curves for several reasons: (1) both H$\alpha$ and CO emission are ``clumpy'' in dwarf galaxies, complicating the construction of kinematic models (see Sect.\,\ref{sec:clumpy}); (2) the relatively high velocity dispersions of H$\alpha$ gas ($\sim$15$-$30 km s$^{-1}$, e.g., \citealt{Barat2020}) require large and uncertain asymmetric-drift corrections to measure the circular velocity tracing the gravitational potential (see Sect.\,\ref{sec:pressure}); and (3) the low metallicites of dwarf galaxies make the CO emission very hard to detect, even with the Atacama Large Millimeter/submillimeter Array \citep[ALMA, e.g.,][]{Hunt2023}. Obtaining high-resolution \hi\ data is needed to study the inner dynamics of dwarf galaxies in detail.

\begin{figure}[t]
    \centering
	\includegraphics[width=\textwidth]{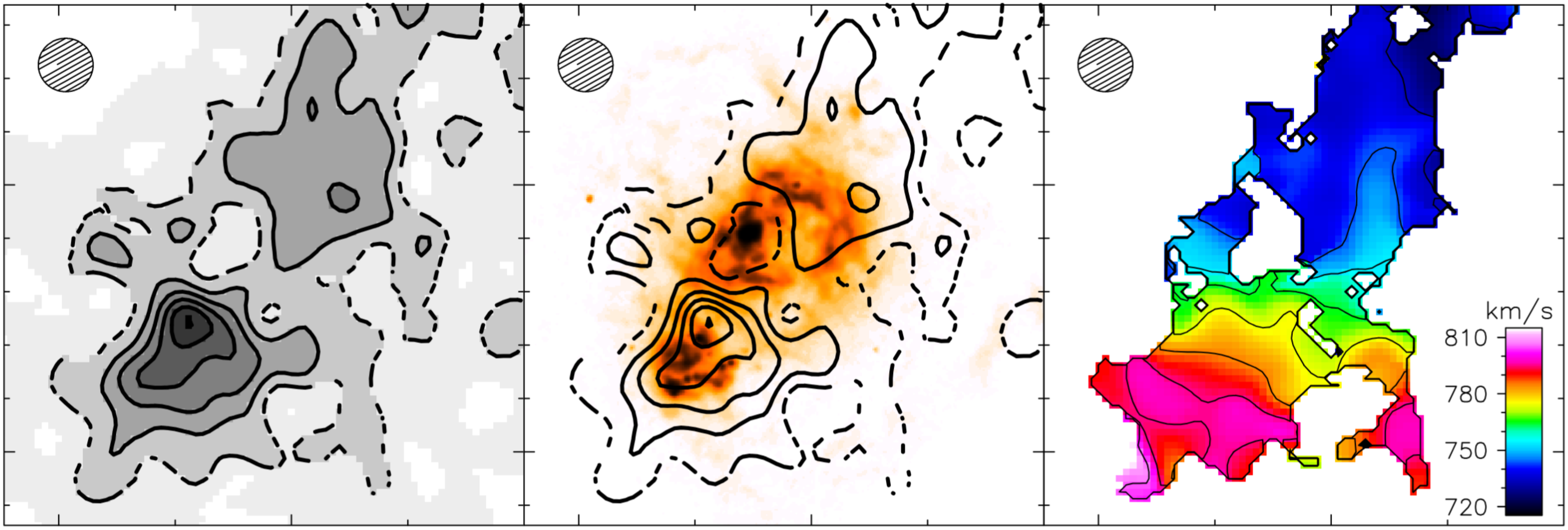}
    \caption{\hicap\ data at 2$''$ resolution from multi-configuration VLA data of the dwarf galaxy I Zw 18 \citep[from][]{Lelli2012a}. \textbf{Left Panel}: Total intensity \hicap\, map, showing clumps and holes that are not visible at lower resolutions of 5$''$ (see Fig. 2 in \citealt{Lelli2012a}).  Contours are at 3 (dashed), 6, 9, 12, 15 $\times$ 10$^{21}$ atoms cm$^{-2}$. \textbf{Middle panel}: Total intensity \hicap\ map superimposed on an H$\alpha$ image \citep[from][]{Cannon2004}, showing that the spatial relation between ionized and atomic gas is complex. \textbf{Right panel}: \hicap\ velocity field showing evidence for a rotating disk. SKA-Mid AA4 will allow us to obtain \hicap\ data with spatial resolutions of $1''-2''$ for significant samples of galaxies.}
    \label{fig:Izw18}
\end{figure}
In this context, a major advance was represented by the THINGS survey \citep{Walter2008}, which used the VLA in multiple configurations (B, C, D) to reach angular resolutions of $\sim$6$''$ in 34 nearby galaxies. Clearly, high-resolution and high-sensitivity \hi\ observations do not automatically allow for the derivation of accurate rotation curves because the \hi\ disk needs to be reasonably regular and properly inclined (see Sect.\,\ref{sec:tools} for modeling details). In the case of THINGS, rotation curves could be derived for 19 galaxies, after excluding low-inclination disks \citep{deBlok2008}. After THINGS, several other \hi\ surveys used the VLA with a similar strategy \citep[e.g.,][]{Hunter2012, Ott2012, Lelli2014b}. To our knowledge, the highest angular resolution ever achieved with \hi\ observations to date is 2$''$ using VLA observations in A, B, C, D configurations \citep[see Fig.\,\ref{fig:Izw18},][]{Lelli2012a}. These types of high-resolution \hi\ observations have been quite successful, but they are not very efficient because they require (1) long integration times due to the low sensitivity to diffuse \hi\ emission on long baselines, and (2) long observing campaigns spanning several years so that the VLA antennas can be moved into different array configurations. 

Due to its long baselines, SKA-Mid AA4 will completely change the game because it will allow us to routinely obtain \hi\ data at $1''-2''$ resolution. An \hi\ column density of ${\sim}$5 M$_\odot$ pc$^{-2}$ (typical of the inner galaxy regions) over 16 km s$^{-1}$ could be detected with spatial and spectral resolutions of ${\sim}1.3''$ and ${\sim}3.1\rm\,km\,s^{-1}$, respectively, with an integration of about 20 hours with SKA-Mid AA4 Band-2. For example, it would be conceivable to perform a legacy survey of 50 nearby galaxies in about 1,000 hours of SKA-Mid to obtain \hi\ data with similar angular resolutions of H$\alpha$ integral-field spectroscopy (e.g., from MaNGA or SAMI), CO interferometry (e.g., from ALMA), and NIR imaging (e.g., from Spitzer). This type of high-resolution \hi\ survey would have many scientific goals, such as relating the distribution and kinematics of atomic gas with those of molecular and ionized gas to understand the star-formation process and the related baryon cycling. In terms of the DM problem, there will be two major advances. Firstly, we could measure the properties of DM cores (sizes, volume densities) with unprecedented accuracy, even in the smallest \hi-bearing galaxies, providing major constraints to $\Lambda$CDM models of galaxy formation as well as to alternative DM models (e.g., warm DM, self-interacting DM, fuzzy DM). Secondly, we could test different modified-gravity theories in the inner galaxy regions, where they predict the largest differences \citep[e.g.,][]{Brada1995, Lelli2017b, Petersen2020, Chae2022b}.

\subsection{Building larger galaxy samples}\label{sec:largesamples}

A comprehensive understanding of the relationship between cosmic environment and galaxy dynamics (rotation-curve shapes, DM content, and so on) requires large, homogeneous, statistical samples of spatially resolved galaxies across a full range of environmental densities, from rich clusters, to loose groups, to truly isolated field galaxies. As we mentioned in Sect.\,\ref{sec:intro}, existing \hi\ samples of spatially resolved galaxies have limited statistics. In addition, they do not capture the full environmental diversity necessary to study the possible relations between the DM content of galaxies and their location in the large-scale structure of the Universe.

With SKA-Mid AA4, it will be possible to spatially resolve the \hi\ kinematics of large galaxy samples in a time-efficient way. To reach this goal, there are two possible survey strategies: 
\begin{enumerate}
    \item An untargeted medium-shallow \hi\ survey of a large fraction of the sky, similarly to APERTIF \citep{Adams2022} and WALLABY \citep{Westmeier2022}, but reaching higher angular resolutions ($\sim$5$''$ versus $\sim15''-30''$) at similar 3$\sigma$ column density sensitivities ($\sim$1 M$_\odot$ pc$^{-2}$ over 20 km s$^{-1}$), so that a much larger number of \hi\ disks will be spatially resolved.
    \item A ``snapshot'' targeted survey of $\sim$20,000 galaxies with known optical/NIR properties (e.g., galaxies with Euclid images), so that the integration time can be individually set according to the desired sensitivity and angular resolution to resolve the \hi\ disk. The target sample of 20,000 galaxies is about two orders of magnitudes larger than SPARC and about one order of magnitude larger than what we expect from WALLABY, APERTIF, and BIG-SPARC.
\end{enumerate}
Given that SKA-Mid will have standard single-pixel receivers, the field of view (FoV) at 21 cm remains rather limited, so it is not immediately obvious which of the two survey strategies is more effective, also taking into account commensal science goals.

In the case of an untargeted wide-field medium-shallow \hi\ survey with Band 2 of SKA-Mid AA4, we make a rough estimate considering the \hi\ mass function of \citet{Jones2018} and the \hi\ mass-size relation of \citet{Wang2016}. If we set a 3$\sigma$ column density sensitivity of 1 M$_\odot$ pc$^{-2}$ for a synthesized beam of 4.3$''$ (Briggs' Robust parameter of 1) and spectral channels of 3.1 km s$^{-1}$, we expect that we will spatially resolve about 20,000 galaxies by covering 150 square degrees in about 900 hours of SKA-Mid AA4. Importantly, a spatial resolution of $4.3''$ will allow us to spatially resolve galaxies with a \hi\ mass of $\sim$10$^7$ M$_\odot$ within a distance of about 20 Mpc, pushing down in terms of the BTFR and other dynamical scaling laws. However, the drawback is that the sample will be heavily dominated by giant galaxies with \hi\ masses of a few $10^{10}$ M$_\odot$ (near the ``knee'' of the \hi\ mass function), while low-mass galaxies with \hi\ masses below $10^9$ M$_\odot$ will be severely under-represented (less than 10$\%$ of the full sample). Thus, the dynamical scaling laws will not be sampled in a uniform way across their dynamic ranges, limiting our accuracy in determining their slope and/or intrinsic scatter as a function of mass, rotation velocity, surface density, and so on.

A pointed \hi\ survey may be more time-expensive than a wide-field \hi\ survey to spatially resolve 20,000 galaxies. For example, if we consider a mean integration of 15 minutes per galaxy, it will take 5000 hours to observe 20,000 galaxies. This time estimate is quite conservative: the exact numbers will depend on the exact definition of the galaxy sample and may be substantially smaller because a single SKA-Mid pointing may simultaneously probe multiple galaxies in clusters and/or groups, maximizing the survey efficiency. A lower survey efficiency, instead, will occur when targeting truly isolated galaxies in voids, albeit several background and foreground galaxies may be detected and spatially resolved along the line of sight. A pointed \hi\ survey will have the main advantage of obtaining $-$ by construction $-$ a galaxy sample that \emph{uniformly} covers the widest possible range in baryonic mass, surface brightness, star-formation rates, environmental properties, and so on. In addition, a pointed \hi\ survey may maximize the synergy with other existing multi-wavelength data, such as (1) imaging from Euclid, HST, JWST, Spitzer, Rubin, and so on, (2) precise distances from Cepheids, the tip of the red giant branch method, supernovae Ia, surface brightness fluctuations, and so on, and (3) complementary kinematic data from H$\alpha$ integral-field spectroscopy and/or interferometric CO observations.

Thinking farther ahead in the future, installing phased-array feeds on SKA-Mid antennas (similarly to ASKAP and APERTIF) will substantially enlarge the instantaneous FoV of the telescope. In this way, wide-field untargetted surveys will be the obvious operational mode to increase the number of galaxies with spatially resolved \hi\ kinematics by another 1-2 orders of magnitudes.

\subsection{Moving towards higher redshifts}\label{sec:highz}

The intrinsic faintness of the \hi\ emission line, combined with the limited spectral bandpasses and radio frequency interference (RFI) in the L-band, have historically limited \hi\ detections of individual galaxies to the local universe.  This paradigm is beginning to shift with a new generation of \hi\ surveys with SKA pathfinders \citep{Hess2019, Ponomareva2021, Xi2024, Jarvis2025}, but the existing facilities do not have the requested sensitivity and angular resolution to map galaxies beyond $z \gtrsim 0.1$, enabling detailed kinematic modeling. To our knowledge, the current record for the highest $z$, well-measured ($\gtrsim$4 independent points) \hi\ rotation curve is held by the giant galaxy Malin\,1 at $z\simeq0.0826$ \citep{Pickering1997, Lelli2010}.

\begin{figure}[t!]
\begin{center}
\includegraphics[width=0.485\textwidth]{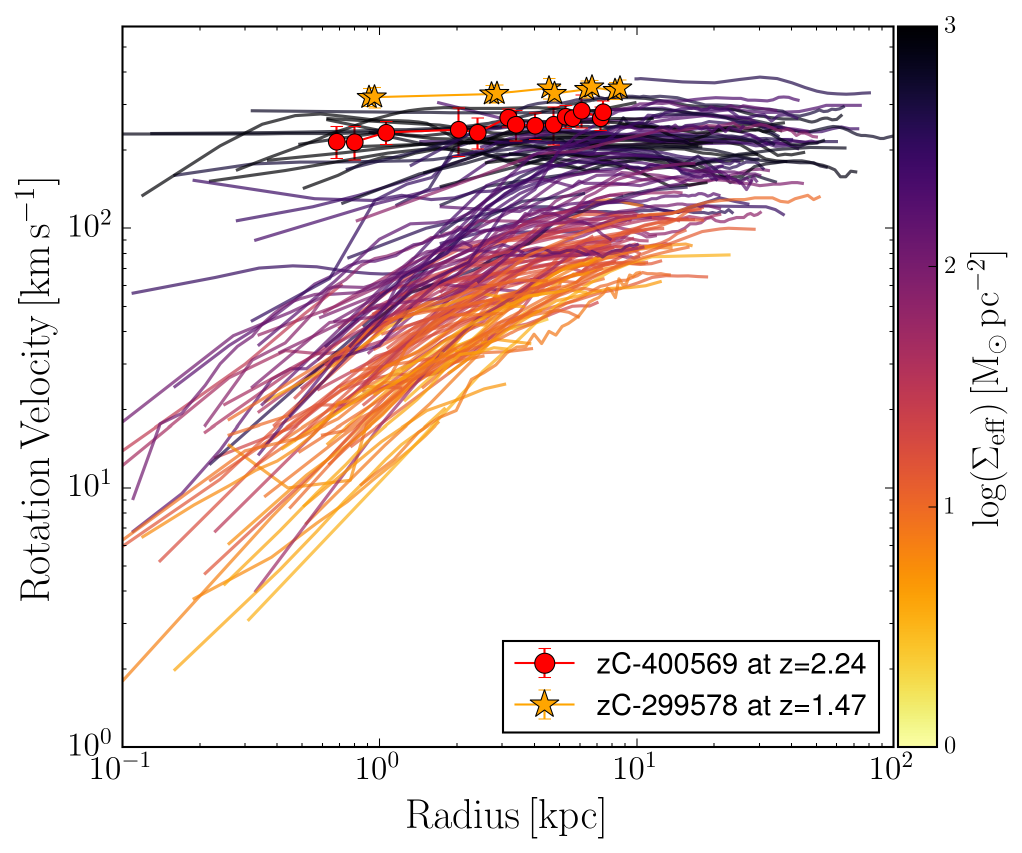}
\includegraphics[width=0.445\textwidth]{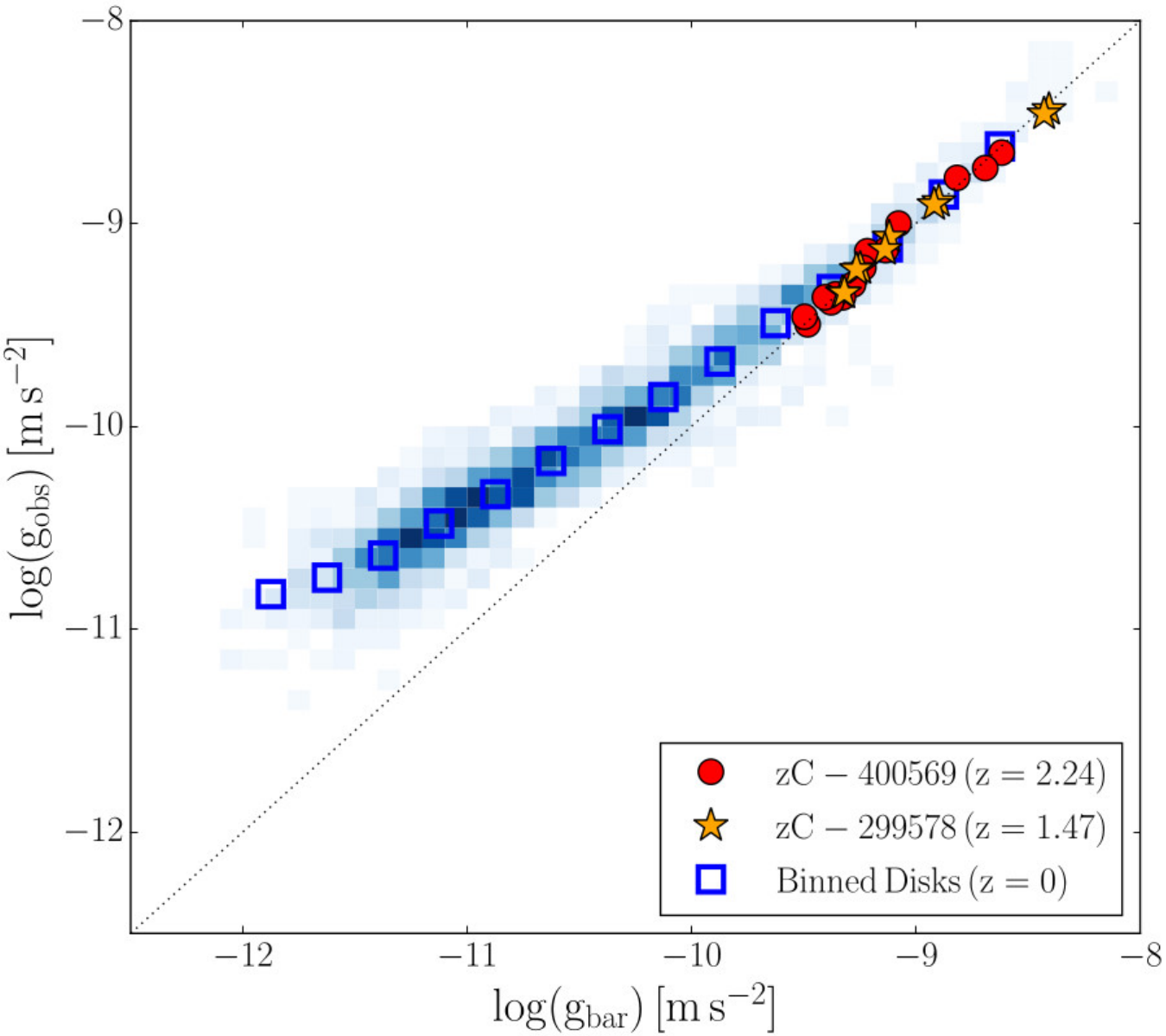}
    \caption{Comparison between the dynamical properties of SPARC galaxies at $z\simeq0$ with those of two best-studied galaxies at cosmic noon, having top-quality H$\alpha$ data from VLT/SINFONI and multi-line CO data from ALMA \citep[adapted from][]{Lelli2023}. \textbf{Left panel:} The rotation curves of cosmic-noon galaxies (red circles and orange stars) are much less extended than those of local galaxies of comparable mass and rotation velocity (solid lines color-coded by stellar surface density) due to the lack of \hicap\ data. \textbf{Right panel:} The location of cosmic-noon galaxies on the radial acceleration relation at $z\simeq0$ shows that their H$\alpha$+CO rotation curves are probing only the inner, high-acceleration, baryon-dominated regions. To firmly probe the DM effect, we need extended \hicap\ rotation curves probing the outer, low-acceleration, DM-dominated regions.}
    \label{fig:highzrotcurves}
\end{center}
  \end{figure}

At intermediate redshifts ($z\simeq0.3-1$) and cosmic noon ($z\simeq1-3$), rotation-curve studies currently rely on other kinematic tracers, such as molecular gas using mm/submm interferometric observations of CO and \ci\ lines \citep{Lelli2023, Rizzo2023, Lin2025}, or ionized gas using integral-field spectroscopy of the H$\alpha$, \oii, and \oiii\ lines \citep[e.g.,][]{DiTeodoro2016, DiTeodoro2018, Price2021, Puglisi2023}. Interferometric mm/submm observations with ALMA and/or NOEMA are relatively time-costly, so galaxy samples with resolved CO or \ci\ kinematics remain limited to a few tens of objects. Integral-field observations with the ESO Very Large Telescope (VLT), instead, are more time efficient, so they allowed probing ionized gas kinematics in several hundreds of galaxies \citep[e.g.,][]{Tiley2019, Nestor2023}. In physical terms, however, ionized gas has some intrinsic complications: (1) it probes a hotter ISM phase than molecular or atomic gas, so the pressure support to the gravitational potential becomes important to properly model (see Sect.\,\ref{sec:pressure}), and (2) non-circular motions (inflows, outflows) may affect the ionized gas kinematics in more complex ways than for the neutral gas \citep[e.g.,][]{Lelli2018}. Most importantly, both ionized and molecular gas tracers have an intrinsic physical limitation: the ionized and molecular gas disks of galaxies generally do not extend beyond their stellar components, so one can only trace rotation curves with a limited radial extent (see Fig.\,\ref{fig:highzrotcurves}). 

In relatively massive galaxies ($M_\star > 10^{10}$ M$_\odot$), the inner high-acceleration regions are generally baryon dominated. This fact prevents a firm detection of the DM effect in massive high-$z$ galaxies due to the disk-halo degeneracy: mass models with only baryons and mass models with a DM halo can fit the observed rotation curves with similar accuracy, given the significant uncertainties on stellar and gas masses \citep{Lelli2023}. The current observational situation at cosmic noon is similar to that at $z\simeq0$ in the 1980's: one needs extended \hi\ rotation curves to firmly probe the DM effect in the outer parts \citep{Kalnajs1983, vanAlbada1985, Kent1986, Kent1987}. 

In lower mass galaxies ($M_\star < 10^{10}$ M$_\odot$), the inner rotation curves may potentially probe the DM effect, similarly to DM-dominated low-mass galaxies at $z\simeq0$ \citep[see][for a review]{Lelli2022}. However, molecular lines (CO and \ci) are hard to detect in such low-mass galaxies, even with ALMA. Regarding ionized gas kinematics, the typical spatial resolution of integral-field spectrographs ($0.5-1\,\rm{arcsec}$) becomes a major limitation in low-mass galaxies, together with the limited spectral resolution of $R=\lambda/\Delta\lambda \simeq 3500-4000$ that corresponds to $\sigma\simeq40\,\rm{km/s}$ for the H$\alpha$ line at $z\approx1$ \citep{Stott2016,Wisnioski2019}. All these facts preclude the study of the dynamical scaling laws over cosmic time down to low masses, low surface densities, and low accelerations, which provide the strongest tests of $\Lambda$CDM models and alternative theories (see Sect.\,\ref{sec:scalinglaws}).

\begin{figure}[t!]
\begin{center}
    \includegraphics[width=\textwidth]{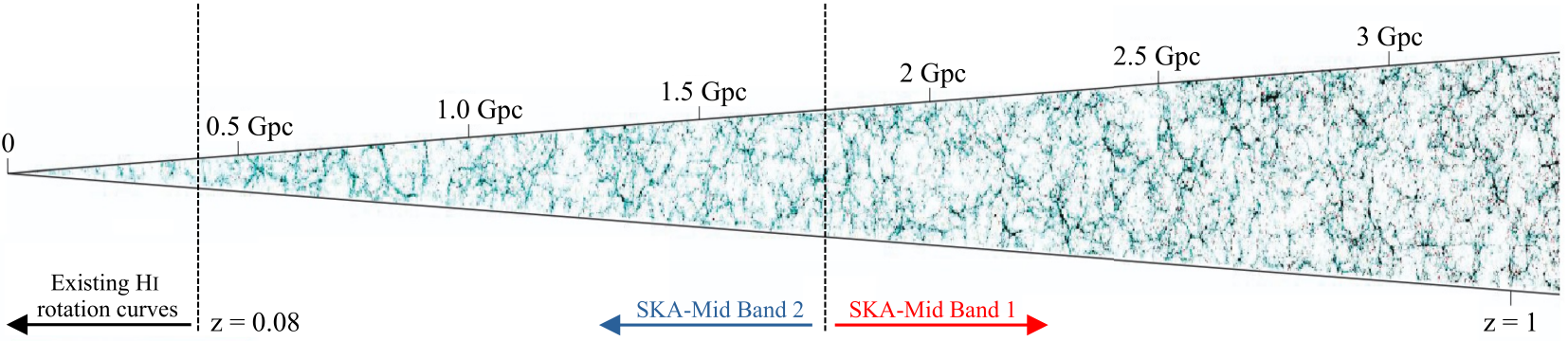}
    \caption{Prospects for measuring \hicap\ rotation curves at $z>0$ with SKA-Mid. A mock-observed simulated cone of the Universe is shown up to $z\simeq1$ \citep[adapted from][]{Obreschkow2009}. The dots represent simulated galaxies with detectable \hi\ emission. To date, \hicap\ rotation curves are limited to galaxies with $D\lesssim 350$ Mpc ($z \lesssim 0.08$). A pencil-beam survey of about 10,000-hours with Band 1 of SKA-Mid AA4 could spatially resolve galaxies with \hicap\ masses of about 10$^{10}$ M$_\odot$ out to $z \simeq 1$. }
    \label{fig:cosmictime}
\end{center}
\end{figure} 

SKA-Mid AA4 will revolutionize our view of the \hi\ kinematics in galaxies at $z>0.1$ thanks to its unique combination of angular resolution and sensitivity. To trace reliable \hi\ rotation curves out to the DM-dominated regime, a column density sensitivity of $\sim1\,\mathrm{M_\odot/pc}^2$ is required. In addition, to constrain the geometric parameters of the \hi\ disk (see Section~\ref{sec:tools}), the disk major axis should be sampled with at least 5 resolution elements. If we (1) adopt this combination of sensitivity and resolution as a metric for measuring \hi\ rotation curves, (2) assume the local \hi\ mass-size relation \citep[e.g.,][]{Wang2016} to predict the size of the \hi\ disk, and (3) account for cosmological redshift broadening and dimming \citep{Meyer2017}, we find that a deep SKA-Mid AA4 integration of about 10,000 hours is capable of mapping galaxies with $M_{\hi} \simeq 10^{10}$ M$_\odot$ out to $z\simeq1$. According to \hi\ stacking analyses \citep{Chowdhury2020, Chowdhury2022a, Chowdhury2022b, Bera2023, Bianchetti2025}, the $M_\star-M_{\rm HI}$ relation evolves with $z$, so an \hi\ mass of $10^{10}$ M$_\odot$ at $z\simeq1$ corresponds to galaxies with a similar stellar mass. SKA-Mid AA4, therefore, may probe the \hi\ content of galaxies with baryonic masses of a few 10$^{10}$ M$_\odot$, lower than the present-day mass of the Milky Way. In addition to providing extended rotation curves up to $z\simeq1$, a deep \hi\ survey with SKA-Mid AA4 will also allow us to estimate the gravitational contribution of atomic gas, which may dominate over that of stars and molecular gas at large radii \citep{Chowdhury2022b}.

Figure~\ref{fig:cosmictime} illustrates the dramatic leap in our ability to study \hi\ kinematics across more than half of the Universe's lifetime. A deep Band-2 survey with SKA-Mid AA$*$ could spatially resolve \hi\ disks in the redshift range $0 \lesssim z \lesssim 0.5$, considering that the maximum achievable angular resolution of $\sim$3$''$ corresponds to $\sim$20 kpc at $z\simeq0.5$. A very deep ($\sim$10,000 hours) Band-1 survey with SKA-Mid AA4 is needed to spatially resolve massive ($M_{\hi} \simeq 10^{10}$ M$_\odot$) disks at $z\simeq1$, given the better angular resolution of $\sim$1$''$ achievable with AA4. However, parts of the redshift range in Fig.\,\ref{fig:cosmictime} may be severely impacted by RFI, prohibiting the detection of the \hi\ line. For example, the range $0.1\lesssim z \lesssim 0.2$ is known to be strongly affected by RFI from satellites and ground-based radar \citep{Hess2019}. The overall RFI situation needs to be reassessed for the long baselines of SKA-Mid AA4. In general, the SKA observatory is the only facility under construction or being planned that will enable dynamical studies of \hi\ disks across cosmic time, for the first time.

\begin{figure}[t!]
\begin{center}
    \includegraphics[width=\textwidth]{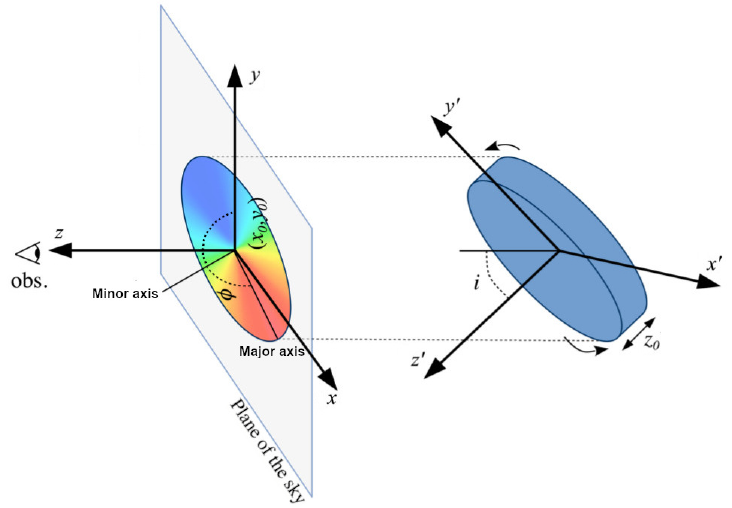}
    \caption{Geometrical parameters of a rotating disk model \citep[adapted from][]{DiTeodoro2015}. A thick disk in the physical coordinate system ($x', y', z'$) is projected into an ellipse in the plane of the sky ($x, y$). The inclination angle $i$ is taken with respect to the line of sight (the $z$ axis). The position angle $\phi$ identifies the position of the major axis on the receding half of the galaxy and it is taken counterclockwise from the North direction (along the $y$ axis). For a symmetric disk with purely circular orbits, the systemic velocity $V_{\rm sys}$ (the galaxy redshift) corresponds to the projected velocities along the minor axis.}
    \label{fig:diskmodel}
\end{center}
\end{figure} 

\section{Modeling prospects with next-generation software}
\label{sec:tools}

\subsection{State-of-the-art on kinematic fitting tools}

The \hi\ component of galaxies can be described by a rotating disk with nearly circular orbits. In this approximation, the projected line-of-sight velocity $V_{\rm l.o.s.}$ at a sky position $(x, y)$ is given by
\begin{equation}\label{eq:Vlos}
 V_{\rm l.o.s.}(x, y) = V_{\rm sys} + V_{\rm rot}(R)\sin(i)\cos(\theta),
\end{equation}
where $V_{\rm sys}$ is the systemic velocity (due to the Hubble expansion and peculiar motions), $V_{\rm rot}$ is the rotational velocity at radius $R$ in the disk plane ($x', y'$), $i$ is the inclination angle between the normal $z'$ to the disk plane and the line of sight $z$, and $\theta$ is the azimuthal angle in the disk plane (see Fig.\,\ref{fig:diskmodel}), given by the following equation:
\begin{equation}\label{eq:theta}
 \cos(\theta) = \dfrac{-(x - x_{0})\sin(\phi) + (y - y_{0})\cos(\phi)}{R},
\end{equation}
where ($x_0$, $y_0$) are the sky coordinates of the disk center and $\phi$ is the position angle taken in the anti-clockwise direction between the North direction and the receding half of the major axis of the projected disk. Thus, the rotation curve $V_{\rm rot}(R)$ can be measured knowing five geometric parameters: $V_{\rm sys}$, $x_0, y_0$, $\phi$, and $i$. In the following, we describe standard methods to measure $V_{\rm rot}(R)$ from emission-line data: tilted-ring and parametric modeling implemented via two dimensional (2D) or three dimensional (3D) approaches. In full 3D approaches, in addition to the parameters in Eq.\,\ref{eq:Vlos} and Eq.\,\ref{eq:theta}, it is possible to fit also for the intrinsic gas velocity dispersion $\sigma_{\rm V}(R)$, the intrinsic gas surface density $\Sigma(R)$, and the disk thickness $z_0$.

\subsubsection{Basic techniques: tilted-ring and parametric modeling}\label{sec:techniques}

The tilted-ring modeling approach \citep{Rogstad1974, Begeman1989} divides the gas disk into a number of rings (usually set by the spatial resolution of the observations). In principle, the parameters in Eq.\,\ref{eq:Vlos} and \ref{eq:theta} can be fitted independently in each ring. However, ring-by-ring fits are often under-determined and/or degenerate, so the best-fit parameters may unrealistically jump from one ring to another. One approach is that,  after a first preliminary fit, ($x_0$, $y_0$) and $V_{\rm sys}$ are measured as the mean (or median) values across the rings and kept fixed in subsequent iterations. The angles $\phi$ and $i$ can be estimated in the same way unless the gas disk is warped: in such cases, $\phi(R)$ and $i(R)$ vary from ring to ring and their radial trend can be appropriately modeled with a smooth function to obtain a reliable $V_{\rm rot}(R)$ without unrealistic jumps. Another approach is to perform a global fit to the disk by grouping various parameters (such as $x_0$, $y_0$, and $V_{\rm sys}$) to have a single value across all rings. The details of these implementations depend on the specific tilted-ring fitting software.

The parametric modeling approach uses parametric functions to describe $V_{\rm rot}(R)$, such as an $\arctan$ function \citep{Epinat2010}, and fits a global model to the whole disk. In several circumstances, parametric modeling can require fewer free parameters to describe $V_{\rm rot}(R)$ than typical tilted-ring models. This eases the use of Markov Chain Monte Carlo (MCMC) algorithms \citep[e.g.,][]{Bouche2015, TD17, Lee2025} and neural networks \citep{Dawson2021} in a Bayesian framework (see Sect.\,\ref{sss:bayes} and \ref{sss:neuralnetworks}). 
On the other hand, parametric modeling cannot account for disk warps in an automated way because specific functions for the radial variation in $\phi$ and $i$ need to be defined a-priori. Parametric modeling may also neglect possible real features in the rotation curve because a smooth shape is imposed on $V_{\rm rot}(R)$. While tilted-ring models provide the most empirical derivation of the rotation curve itself, some modeling tools allow for hybrid approaches, mixing tilted-ring-style approaches for some parameters and parametric functions for others.

Special types of parametric models are those in which $V_{\rm rot}(R)$ is described by constraining the gravitational potential of different components (stars, gas, and DM halo), so the fitting involves physical parameters (such as masses and characteristic radii) together with geometric ones. Ideally, this approach permits to build dynamically self-consistent models (for instance, by assuming hydrostatic equilibrium) because the properties of the gravitational potential and those of the gas tracer are constrained at the same time. In addition, this approach bypasses issues in the mass modeling of rotation curves due to correlations between $V_{\rm rot}(R)$ measurements \citep{Posti22}.
The main drawback of this approach is its reliance on pre-implemented (cosmologically-motivated) models for the DM distribution, which may not capture the observational reality and/or the full complexity of galaxy dynamics. In addition, these parametric models have relatively heavy computational costs when dynamical self-consistency is requested, and they require refitting the full data (in 2D or 3D) several times for testing different DM halos or modified-gravity theory. On the contrary, it is computationally fast and conceptually straightforward to fit different DM halos and/or modified gravity theories to the empirical 1D rotation curve from a titled-ring model. Although promising, \hi\ modeling tools in a dynamically self-consistent framework have not yet been implemented.

\subsubsection{From rotation velocity to circular velocity}\label{sec:pressure}

When modeling the mass distribution of galaxies, we are interested in the circular velocity of a test particle ($V_{\rm c}^2 = -R\,\nabla \Phi$) tracing the total gravitational potential $\Phi$. In general, at a given radius $R$, an ensemble of particles (stars or gas clouds) rotates with a velocity $V_{\rm rot}$ that is lower than $V_{\rm c}$ due to the additional outwards acceleration provided by the pressure ($\rho \sigma_{\rm V}^2$) gradient. Contrary to the case of stars and/or ionized gas, pressure support is often negligible in the \hi\ disks of spiral galaxies because the velocity dispersion is typically between $5-15$ km\,s$^{-1}$, so effectively $V_{\rm c} \simeq V_{\rm rot}$.

For galaxies in which $V_{\rm rot}/\sigma_{\rm V}\simeq1-4$ (e.g., dwarf galaxies), pressure support can be taken into account using the so-called asymmetric-drift correction. Assuming that (i) $\Phi$ is axisymmetric depending only on the cylindrical coordinates ($R$, $z$), and (ii) the velocity dispersion $\sigma_{\rm V}$ is isotropic as expected for collisional gas, the circular velocity is given by \citep{BT94}
\begin{equation}\label{eq:drift1}
V_{\rm c} = \sqrt{V_{\rm rot}^2 - \sigma_{\rm V}^2 \left( \dfrac{\partial \ln \rho}{\partial \ln R} + \dfrac{\partial \ln \sigma_{\rm V}^2}{\partial \ln R} \right)},
\end{equation}
where $\rho(R,z)$ is the gas volume density, which is not directly observable.

A common approach \citep{Meurer1996, Lelli2014a} is to assume that the vertical density distribution does not depend on radius, so $\partial \ln \rho / \partial \ln R = \partial \ln \Sigma / \partial \ln R$, where the gas surface density $\Sigma(R)$ can be traced from emission-line maps. Then, one may either fit the product $\Sigma \sigma_{\rm V}^2$ with a smooth functional form \citep{Iorio2017} or treat $\Sigma(R)$ and $\sigma_{\rm V}(R)$ independently. The main caveat with this approach is that the thickness of the gas disk is expected to increase with radius from theoretical considerations based on vertical hydrostatic equilibrium \citep{Romeo1992, Olling1995, Bacchini2020}, so $\partial \ln \rho / \partial \ln R > \partial \ln \Sigma / \partial \ln R$ and the correction may be underestimated.

Another approach \citep{Burkert2010} is to use the Spitzer's solution for a self-gravitating disk in vertical equilibrium \citep{Spitzer1942}, so $\rho(R, z) = \rho_{0}(R) \sech^2(z/z_{\rm d})$ where $\rho_{0}(R) = \pi G \Sigma(R)^2/2\sigma_{\rm V}(R)$ is the gas volume density at $z=0$. The gaseous disks of galaxies, however, are not self-gravitating in any realistic case. A self-consistent asymmetric-drift correction should use an iterative approach to (i) obtain $\Phi(R, z)$ fitting $V_{\rm c}(R)$ with a mass model for a given DM halo shape, and (ii) measure the variation of the disk thickness with radius by solving the vertical hydrostatic equation for the given $\Phi(R, z)$ \citep{Bacchini2020}. This approach may be implemented in future tools and initial efforts are underway \citep[e.g.,][]{ManceraPina2025}. 
It may also be possible to calibrate the asymmetric-drift correction using numerical simulations, for example to obtain prescriptions informed by galaxy morphology \citep{Kretschmer2021}. Parametric models built in the dynamically self-consistent framework can potentially account for the flaring of the disk, providing direct computation of pressure support via Eq.\,\ref{eq:drift1}.

At any rate, the dominant uncertainty in the asymmetric-drift correction is driven by $\sigma_{\rm V}(R)$, not by the vertical density distribution. In fact, one often has to assume that $\sigma_{\rm V}$ is constant with radius because the velocity dispersion profile cannot be accurately measured in the lowest mass galaxies. Then, for an exponential gaseous disk with scale length $R_{\rm d}$, Eq.\,\ref{eq:drift1} simplifies to 
\begin{equation}\label{eq:drift3}
V_c =\sqrt{V_{\rm rot}^2 + \sigma_{\rm V}^2(R/R_{\rm d})}.
\end{equation}
Eq.\,\ref{eq:drift3} can be used as a rough first-order approximation in poorly resolved galaxies, which are the focus of the next section.

\subsection{Modeling of poorly resolved disks}\label{sec:mod_lowres}

\hi\ surveys with SKA-Mid may provide tens or even hundreds of thousands of galaxies that are just barely resolved. Unsurprisingly, preliminary results from the WALLABY survey indicate that barely resolved systems will dominate any wide-field \hi\ survey: only $\sim$10$\%$ of the \hi\ detections in WALLABY are sufficiently resolved for detailed kinematic modeling \citep{Westmeier2022,Murugeshan2024}. \hi\ modeling in the barely resolved regime demands specific techniques, such as the full treatment of beam smearing in 3D (Sec \ref{sec:beamsmearing}) and algorithms for the automated acceptence/rejection of kinematic fits (Sec. \ref{sec:accept/reject}). At low spatial resolutions and signal-to-noise ratio, measuring the intrinsic gas velocity dispersion is particularly difficult and often impossible; uncertainties in $\sigma_{\rm V}(R)$ do not materially affect the measured rotation curve $V_{\rm rot}(R)$ \citep{Deg2022}, but can have an impact on the circular velocity $V_{\rm c}(R)$ via the asymmetric-drift correction, especially for low-mass galaxies (see Sec. \ref{sec:pressure}). In addition, in the regime of barely resolved galaxies, disk warps and other complexities (see Sect.\,\ref{sec:mod_highres}) cannot be easily identified, so they are generally neglected in the kinematic modeling. This decreases the number of free parameters, which is an advantage in modeling large samples of galaxies, but possible systematic effects need to be carefully investigated \citep[e.g.,][]{Deg2022, Gasser2023}.

\subsubsection{Beam-smearing effects: 2D and 3D approaches}\label{sec:beamsmearing}

Tilted-ring and parametric modeling can be applied either to 2D velocity maps ($V_{\rm l.o.s.}$ at each sky position) or directly to the 3D emission-line cubes. The 2D approach is robust for well-resolved galaxies with tens or hundreds of resolution elements \citep{Begeman1987, Begeman1989}, but becomes unreliable in less resolved galaxies as is often the case for dwarf galaxies \citep{Swaters2009}. In this situation, the derivation of velocity maps is ambiguous due to beam-smearing effects, altering the intrinsic shape of the line profiles. Gas emission with intrinsically different $V_{\rm l.o.s}$ is flux-averaged within the resolution element, so the line profiles tend to become broad and non-Gaussian with long tails of emission towards the galaxy systemic velocity \citep[e.g.,][]{deBlok2008}. In such cases, if one estimates $V_{\rm l.o.s}$ as the moment-one along the profile or as the central velocity of a Gaussian fit, the rotational velocity will be underestimated, especially in the central parts \citep[e.g.,][]{Swaters2009, Lelli2010}. Moreover, if one estimates the gas velocity dispersion $\sigma_{\rm V}$ as the moment-two along the profile or as the width of a Gaussian fit, the intrinsic velocity dispersion will be severely overestimated \citep{DiTeodoro2015}. For galaxies with high inclinations, the minor axis is always poorly resolved due to the disk projection along the line of sight. Therefore, similar effects occur when analyzing edge-on galaxies, limiting 2D methods to galaxies with inclinations $< 75^{\circ}$. There is no simple way to fully correct for beam smearing using a 2D approach because it depends on the interplay between intrinsic rotation-curve shape, gas distribution, disk thickness, and disk inclination \citep{Begeman1989}.

The 3D approach builds model cubes using Eq.\,\ref{eq:Vlos} together with the gas surface density profile, the gas vertical structure, and the intrinsic $\sigma_{\rm V}$ profile \citep{Corbelli1997, Sicking1997}. The model cubes are then smoothed to the same spectral and spatial resolution of the observations (reproducing beam smearing effects) and compared with the observed one until a best-fit is found. The 3D approach can then recover the intrinsic $V_{\rm rot}(R)$ and $\sigma_{\rm V}(R)$ profiles \citep{Jozsa2007, Swaters2009, DiTeodoro2015}. Over the past years, several 3D parametric fitting tools have been developed, such as KinMS \citep{Davis13}, GalPak-3D \citep{Bouche2015}, DysmalPy \citep{Price2021, Lee2025} and Geko \citep{Danhaive2025}. Importantly, 3D parameteric modeling may not always capture the correct beam-smearing effects because it imposes a specific rotation-curve shape. For example, an intrinsically flat rotation curve leads to more severe beam-smearing effects than an intrinsically solid-body one. A 3D tilted-ring modeling is often the best solution, but requires good sensitivity in each ring. Several titled-ring 3D fitting tools are now available, such as TiRiFiC \citep{Jozsa2007} and its automated version FAT \citep{Kamphuis2015}, $^{\rm 3D}$Barolo \citep{DiTeodoro2015}, and a new version of KinMS \citep{Davis20}. These 3D kinematic modeling tools allows for the derivation of rotation curves even for poorly resolved galaxies (with only $\sim3-4$ independent elements along the major axis), which cannot be reliably modeled in 2D. However, even these packages may still be affected by beam-smearing, making the innermost rings still somewhat uncertain (see, e.g., \citealt{Deg2022}). In principle, the best way to remove many of the uncertainties discussed above would be to model the emission of the kinematic tracer in the $uv$-plane, as we discuss in Section \ref{sss:uvmodeling}.

\subsubsection{Automated acceptance/rejection criteria}\label{sec:accept/reject}

A key use of large \hi\, surveys is the comparison with cosmological simulations. These comparisons require that the same modeling approach and the same software are applied to both mock and real observations. A significant challenge in this context is understanding the ``selection function'', that is, how to select simulated galaxies that \textit{could} have been observed and modeled in the same way as a given survey. The traditional approach of accepting models based on visual inspection \citep[e.g.,][]{Deg2022} introduces an hard-to-quantify effect to the selection function.  Moreover, even if this effect could be quantified and/or repeated, visual inspection may simply be unfeasible for the large galaxy samples expected from SKA-Mid surveys. However, visual inspection by expert users could still play an important role in the SKA era; for example, we may train machine-learning algorithms using the assessments from visual inspection as a robust training set.

An automated approach based on a suite of observations and/or test models must be adopted for large \hi\ surveys with SKA-Mid. This approach must take into account the limits of modeling software in terms of resolution, sensitivity, inclination, and even galaxy mass or rotation velocity. In fact, the reliability of kinematic models strongly depends on disk inclination and galaxy mass, being progressively worse for low-inclination, low-mass galaxies. For example, \citet{Deg2025} demonstrated that, at velocities below $\sim$50 km s$^{-1}$, kinematic models for galaxies resolved by $5$ beams were reliable only for inclinations larger than $40^{\circ}$ \citep[see also][]{Rhee2004, KuziodeNaray2006, Read2016b}. A critical assessment in terms of data quality and disk inclination is crucial to study the dynamical laws of galaxies (Sect.\,\ref{sec:scalinglaws}) and potential outliers \citep[e.g.,][]{Lelli2024}.  

A major challenge in any automated approach is the presence in the final catalog of ``contaminants'': rotation curves that do not represent the underlying gravitational potential of the galaxy because of technical issues (e.g., the choice of the wrong geometric parameters) and/or intrinsic physical issues (e.g., disks out of equilibrium due to interactions, mergers, asymmetries, etc.). These contaminants can be found in large catalogs generated by both visual inspection (see, e.g., some of the low-velocity WALLABY models of \citealt{Deg2022}) and automated approaches. Automated accept/reject rules can be tuned to increase the number of total rotation curves at the cost of an increased contaminant fraction, so one needs to find the best compromise to minimize the contaminant fraction while maximizing the total number of acceptable rotation curves. Despite the numerous challenges of modeling large numbers of marginally resolved galaxies with low signal-to-noise ratio (SNR), the scientific benefits are well worth the effort. They will provide statistically significant samples of kinematic models across a wide range of environments and, in the SKA era, a range of redshifts.

\subsection{Modeling of highly resolved disks}\label{sec:mod_highres}

SKA-Mid AA4 will allow for \hi\ observations at 1$''$-2$''$ resolution (see Sect.\,\ref{sec:highres}). In nearby galaxies, these high spatial resolutions will likely uncover details in the gas distribution and kinematics that require more complex modeling approaches than a simple, axisymmetric, rotating disk. These complexities may include highly asymmetric gas distributions (clumps, shells, holes, and so on), non-circular motions (bar-driven flows, inflows, outflows, and so on), and complex vertical structures (extraplanar gas and disk flares). We discuss potential ideas to tackle these challenges.

\subsubsection{Asymmetric gas distributions}\label{sec:clumpy}

Where gas distributions are clumpy and/or strongly asymmetric, traditional approaches (which impose axisymmetry) will fail to reproduce the observed gas distribution. The impact of this effect on the resulting rotation curve can vary wildly. In systems that are clumpy, but in which the gas emission fills much of the radial ring, axisymetric models generally perform well \citep[e.g.,][]{Davis2014}. However, in extreme cases, the best-fit model may create artificial spikes in the rotation curve. Various techniques can be used to avoid such issues, but they all have some drawbacks.

Three main techniques have been used to deal with clumpy distributions. The simplest one is to normalize a-posteriori the flux in a given spatial pixel of the 3D model to match the observed intensity (moment-zero) map \citep{Lelli2012a}. The main drawback of this method is that it does not fully account for beam-smearing effects from the intrinsic variations in the surface density distribution within the beam. The second technique mitigates this issue by using the ``clean components'' determined during the imaging of the $uv$ data to determine the flux distribution, and fitting the kinematics of these components \citep{Smith2019}. The ``trusted'' flux in the model is then reproduced by definition. The third technique adopts a more drastic solution: it  allows the kinematics and positions of a cloud of point sources to be fitted to the data. While very powerful, this technique could have millions of free parameters and is thus extremely computationally expensive. In addition, all these methods lose nearly all the constraining power on the disk inclination because they are forced to reproduce the observed gas distribution. The best strategy, therefore, is a two-step approach: first fitting a model that is axisymmetric in both the gas distribution and kinematics to constrain the disk geometry (especially the inclination), then fitting a model that is not axisymmetric in the gas distribution to constrain the rotation curve \citep[e.g.,][]{Lelli2012a}.

In the future, new machine learning-based techniques, such as diffusion models, may be used to determine the gas distribution, although such efforts are in their early stages at the present time. Another way to account for clumpy gas distribution may be to do kinematic modeling directly in the $uv$ plane, modeling all the recoverable scales at the same time, as we discuss in Sect.\,\ref{sss:uvmodeling}.

\subsubsection{Non-circular motions in the disk plane}\label{sec:noncircular}

The simplest form of non-circular motion is a radial flow ($V_{\rm rad}$) in the disk plane, giving an additional term $V_{\rm rad}(R)\sin(i)\sin(\theta)$ in the right part of Eq.\,\ref{eq:Vlos}. More generally, $V_{\rm rot}$ and $V_{\rm rad}$ are the first-order terms of a harmonic expansion of the line-of-sight velocity \citep{Schoen1997}:
\begin{equation}
  V_{\rm l.o.s.}(x, y) = V_{\rm sys} + \sum_{m=1} \left [ c_{m}(R)\cos(m\theta) + s_{m}(R)\sin (m\theta) \right ],
\end{equation}
where the harmonic order $m=1$ gives $c_{1} = V_{\rm rot}\sin(i)$ and $s_{1} = V_{\rm rad}\sin(i)$. A perturbation of the gravitational potential of order $m$ causes harmonics of order $m-1$ and $m+1$ in projection \citep{Schoen1997}. For example, bar-like and oval distortions with $m=2$ give $m=1$ and $m=3$ terms in $V_{\rm l.o.s}$ \citep{Spekkens2007}. The harmonic decomposition increases the number of free parameters, which are often degenerate, so it can be applied only to high-quality observations. This method has been implemented with 2D approaches \citep{Gentile2005, Gentile2007, Trachternach2008, Marasco2018}, but further progress can be made with a 3D approach \citep{Hogarth24}. In particular, it will be important to investigate strategies to tackle the various degeneracies between the free parameters of the harmonic expansion, for example by using a Bayesian approach with physically motivated priors, as those discussed in Sect.\,\ref{sss:bayes}.

\subsubsection{Vertical structure: extraplanar gas and flares}\label{sec:vertical}

A non-negligible fraction ($\simeq15\%$) of \hi\ in star-forming galaxies resides in a thick structure that surrounds the thin disk, extending typically a few kpc above and below the midplane. The presence of ``extraplanar'' gas clearly emerges in deep \hi\ observations, reaching column densities of 0.1 M$_\odot$ pc$^{-2}$ or less \citep{Fraternali02, Oosterloo07}, which will be routinely reached by SKA-Mid. The extraplanar \hi\ gas is often characterized by peculiar kinematics, showing a slower rotation than the \hi\ in the thin disk and a global inflow motion with typical speeds of a few tens km\,s$^{-1}$ \citep[e.g.,][]{Fraternali02, Oosterloo07, Marasco2019}. These peculiar properties are broadly consistent with a galactic fountain triggered by stellar feedback \citep{Fraternali17}. To date, the peculiar kinematics of the extraplanar \hi\ gas has been modeled in 3D with substantial human intervention, given the increased number of free parameters and overall complexity. In the SKA era, we may need to explore automated approaches to account for extraplanar \hi\ gas in large galaxy samples. Substantial work needs to be done in this context.

As discussed in Sect.\,(\ref{sec:pressure}), \hi\ disks are naturally flared, with a scale-height that is expected to reach up to $\sim$1 kpc in the outskirts \citep{OBrien2010, Bacchini2020, ManceraPina2022}.
High-resolution observations should be able to distinguish between extraplanar gas and flaring of the \hi\ disk because these features occur in distinct regions (inner versus outer disk, respectively) and have different kinematics. In fact, extraplanar gas is often visible as residual emission after the subtraction of the best-fit kinematic model, including potential disk flaring \citep[e.g.,][]{Oosterloo07}.  However, in marginally resolved galaxies, the contamination from extraplanar gas might be inevitable, leading to a systematic underestimate of $V_{\rm rot}(R)$. Although there is no obvious way to circumvent this issue, tests using mock degraded data of nearby galaxies will be valuable to quantify the magnitude of this effect and to elaborate effective correction strategies.

\subsection{Next-generation modeling techniques}

\subsubsection{Bayesian techniques: improvements and optimization}\label{sss:bayes}

Both parametric and non-parametric fitting methods make use of a substantial number of free parameters, which are often degenerate. The most important degeneracy is between $V_{\rm rot}$ and inclination, but additional degeneracies arise when one goes beyond a standard, thin, rotating disk model, for example by adding non-circular motions (Sect.\,\ref{sec:noncircular}) and/or extraplanar gas (Sect.\,\ref{sec:vertical}). Degeneracies can be a concern even for high-resolution data with high SNR, and becomes a very severe problem for barely resolved, low-SNR data. Ideally, next-generation modeling tools must be able not only to infer a particular galaxy property, but also to determine a realistic observational uncertainty. A robust assessment of the uncertainties, including potential covariances, is key to test DM models and modified-gravity theories.

In principle, the use of Bayesian inference to fully explore the parameter space should automatically account for all possible degeneracies, providing realistic uncertainties for all the parameters depending on their intrinsic correlation and on the quality of the data themselves. Therefore, it is natural to expect that Bayesian techniques will be adopted in the modeling of SKA-Mid \hi\ data (but see below for a possible alternative). However, these techniques have two important drawbacks that we must overcome before considering them as the ``golden standard''. 

The first drawback is that, by construction, Bayesian inference is based on the likelihood function, which fully shapes the inferred posterior distributions, thus our measurements and associated uncertainties. The definition of this function is somewhat arbitrary and complicated by intrinsic difference between the model, which is smooth and typically axisymmetric, and the data, which show a higher degree of complexity especially at high resolutions, as well as by the presence of correlated noise in the datacube \citep[e.g.,][]{Davis+17}. 

The second drawback is the high computational cost needed for Bayesian analysis, which generally uses Monte Carlo methods and requires building $10^5-10^6$ models to robustly sample the parameter space. One way to optimize the existing algorithms is mapping the underlying probability distribution function and strategically sample it to improve both efficiency and accuracy. Possible avenues to do so include importance sampling, adaptive sampling, variance reduction techniques, optimal weighting, and multiphase Monte Carlo. Another, complementary way to optimize the process is to resort to GPU-acceleration, both for the creation of individual models and for the exploration of the parameter space. Libraries that simplify GPU-accelerated programming are now available in several programming languages \citep[e.g., the JAX library in python,][]{jax2018github}, and can substantially boost the performances. Preliminary tests using KinMS \citep{Davis13,Davis20} indicate a factor $\sim20$ speed-up in the model creation.

\subsubsection{Machine learning: artificial and Bayesian neural networks}\label{sss:neuralnetworks}

An alternative approach to Bayesian fitting is to directly infer the desired quantities (e.g., flat rotation velocity, or DM density profile) directly from the 2D or 3D data using machine learning techniques, such artificial neural networks (ANN). Neural networks have been shown to be able to detect and characterize sources from 2D or 3D \hi\ data, notably as part of the first two SKA data challenges \citep{Bonaldi2021, Hartley2023, Cornu2024, Cornu2025}. This approach would require generating mock data sampling different galaxy and DM parameters to create a large and representative training set. Once the ANN is trained, it should in principle provide fast predictions and hence be applicable to large datasets. However, the high dimensionality of the parameter space may induce degeneracies such that not all information may be recovered. In addition, ANNs generally lack explicit uncertainty quantification. By incorporating principles of Bayesian inference in a neural network, Bayesian neural networks (BNN) may constitute a promising way to device the new generation of dynamical modeling tools. In fact, BNNs provide full posterior distributions over the model parameters, offering a way to quantify uncertainty. Furthermore, they may provide a more physically-tractable way to emulate current forward-modeling tools. Self-supervised, physics-aware, BNNs have been developed to fit gas kinematics in 2D \citep{Dawson2021} and show significant promise when extended to 3D. 

It may also be possible to adopt hybrid approaches, such as using an ANN for a fast initial parameter estimation, then a more computationally intensive Bayesian method to refine the results and quantify uncertainties. Both classical and Bayesian machine learning methods have been developed to yield dynamical masses \citep[e.g.][]{Ntampaka2015, Calderon2019, Villanueva-Domingo2022, Chu2024, Hahn2024}, although not specifically from \hi\ data so far. Finally, since evaluating a rotation curve or a DM density profile from 2D or 3D data involves deconvolution steps, attention-based neural networks are likely to be well-suited to the task. Indeed, such neural network dynamically weigh the relevance of the different input elements, directly correlate all parts of the input, and provide increased interpretability through the attention weights. Furthermore, architectures such as transformers enable parallelization and hence faster training on modern hardware.

\subsubsection{Kinematic modeling in the \emph{uv} plane}\label{sss:uvmodeling}

Interferometric observations consist in a sparse sampling of the Fourier angular space (or \emph{uv}-plane), providing the so-called visibilities.
Conversion to the real space (the sky plane) requires a manipulation of the visibilities (e.g., gridding, weighting, deconvolution, and cleaning) that is not unique, and that inevitably alters the intrinsic properties of the original dataset. The ideal approach to circumvent these issues is to model the data directly in the \emph{uv}-plane. This allows using the data in their original space, which provides two main advantages for modeling purposes. First, the uncertainties in the visibilities are well defined.
Second, convolution with the synthesized beam (which is often the computational bottleneck in the model generation) is bypassed: the model and the data will probe exactly the same angular scales, by construction. This makes this method especially promising when dealing with barely resolved observations.

Several kinematic studies have explored this method \citep[e.g.,][]{Dye2018, Rizzo20, Powell21, Amvrosiadis25}, but the publicly available modeling packages have yet to implement it. Although promising, this method needs to be robustly implemented and carefully tested in order to assess its applicability to the unprecedented data volume that SKA-Mid will provide. Importantly, while the mathematical approach of a direct modeling of the visibilities is the most accurate one, it has two practical drawbacks: (1) the computational costs, and (2) the storage capacity costs. Regarding the first drawback, the computational cost can be large because, in a typical application, the disk model is first conceived in the real-space and only later transported to the $uv$ plane with a Fast Fourier Transform (FFT) to be compared with the observed visibilities \citep[e.g.,][]{Amvrosiadis25}.
In this case, the original model must be created with sufficient resolution to account for the full range of possible scales probed by the data, which might be unpractical. This problem has been mitigated by applying a non-uniform FFT \citep{Rizzo20, Powell21}. The second drawback is a logistic one: the SKA observatory should store and provide the (raw or gridded) visibilities rather than more practical sub-cubes with the individual \hi\ detections. An approach that would not require a storage of the visibilities, but only the dirty data cubes and the dirty beams, would be to work on the gridded visibilities instead of the measured ones. The fit quality would then be calculated in the Fourier domain by comparing a FFT of the dirty cube (done once in the complete process) with the FFT of the model multiplied with the gridded sampling function, i.e., the FFT of the dirty beam. This method has not been tested yet, but is theoretically even less computationally expensive than the 3D fitting method in the image domain, which requires two FFTs (back and forth) for the convolution with the clean beam. However, it would require the storage of the dirty cube and the dirty beam.

\section{Conclusions}
\label{sec:summary}

Understanding the nature of DM is one of the most outstanding challenges of modern physics. A prime way to test $-$ and possibly rule out $-$ different DM models and/or modified gravity theories is to study the gravitational potential of galaxies using extended \hi\ rotation curves (Sect.\,\ref{sec:intro}). During the past four decades, the systematic construction of mass models of galaxies revealed a surprising coupling between baryons and dynamics (Sect.\,\ref{sec:scalinglaws}), and raised several outstanding problems for the $\Lambda$CDM cosmological model (Sect.\,\ref{sec:cuspcore}). Many questions remain unanswered and could be tackled with a synergy between the SKA observatory and state-of-the-art optical/NIR facilities.

SKA-Mid AA4 will revolutionize the study of \hi\ rotation curves in several ways because it will give us (1) a sharper view of nearby galaxies, probing the \hi\ kinematics at exceptional resolutions of $1''-2''$ (Sect.\,\ref{sec:highres}), (2) a broader view of the nearby Universe, providing statistical samples of tens of thousands of galaxies in different cosmic environments (Sect.\,\ref{sec:largesamples}), and (3) a deeper view of the distant Universe, probing the \hi\ kinematics of galaxies across cosmic time, possibly up to $z\simeq1$ for the first time (Sect.\,\ref{sec:highz}). These observational advances will allow us to test DM models and modified gravity theories in new regimes, in which different theories make different predictions.

The unprecedented volume and quality of the SKA data will pose important technical challenges in the modeling of the \hi\ kinematics of galaxies. Substantial work has already been done, but many technical aspects remain to be investigated in the coming years (Sect.\,\ref{sec:tools}). 

The extended \hi\ rotation curves of galaxies provided one of the first unambiguous observations of the DM effect \citep[see][for such history]{Sanders2010, Bertone2018}. Future \hi\ legacy surveys with SKA-Mid may ``close the loop'', providing key evidence on what DM is and is not.

\bibliographystyle{abbrvnat-maxbibnames4}
\bibliography{GasDynamics} 

\end{document}

%% file: chapter_main.bbl
\begin{thebibliography}{213}
\providecommand{\natexlab}[1]{#1}
\providecommand{\url}[1]{\texttt{#1}}
\expandafter\ifx\csname urlstyle\endcsname\relax
  \providecommand{\doi}[1]{doi: #1}\else
  \providecommand{\doi}{doi: \begingroup \urlstyle{rm}\Url}\fi

\bibitem[{Adams} et~al.(2022){Adams}, {Adebahr}, {de Blok}, {D{\'e}nes},
  {Hess}, {van der Hulst}, {Kutkin}, {Lucero}, {Morganti}, {Moss}, {Oosterloo},
  {Orr{\'u}}, {Schulz}, {van Amesfoort}, {Berger}, {Boersma}, {Bouwhuis}, {van
  den Brink}, {van Cappellen}, {Connor}, {Coolen}, {Damstra}, {van Diepen},
  {Dijkema}, {Ebbendorf}, {Grange}, {de Goei}, {Gunst}, {Holties}, {Hut},
  {Ivashina}, {J{\'o}zsa}, {van Leeuwen}, {Loose}, {Maan}, {Mancini}, {Mika},
  {Mulder}, {Norden}, {Offringa}, {Oostrum}, {Pastor-Marazuela}, {Pisano},
  {Ponomareva}, {Romein}, {Ruiter}, {Schoenmakers}, {van der Schuur}, {Sluman},
  {Smits}, {Stuurwold}, {Verstappen}, {Vilchez}, {Vohl}, {Wierenga},
  {Wijnholds}, {Woestenburg}, {Zanting}, and {Ziemke}]{Adams2022}
E.~A.~K. {Adams} et al.
\newblock \emph{\aap}, 667:\penalty0 A38, Nov. 2022.
\newblock \doi{10.1051/0004-6361/202244007}.

\bibitem[{Ahn} et~al.(2012){Ahn}, {Alexandroff}, {Allende Prieto}, {Anderson},
  {Anderton}, {Andrews}, {Aubourg}, {Bailey}, {Balbinot}, {Barnes}, {Bautista},
  {Beers}, {Beifiori}, {Berlind}, {Bhardwaj}, {Bizyaev}, {Blake}, {Blanton},
  {Blomqvist}, {Bochanski}, {Bolton}, {Borde}, {Bovy}, {Brandt}, {Brinkmann},
  {Brown}, {Brownstein}, {Bundy}, {Busca}, {Carithers}, {Carnero}, {Carr},
  {Casetti-Dinescu}, {Chen}, {Chiappini}, {Comparat}, {Connolly}, {Crepp},
  {Cristiani}, {Croft}, {Cuesta}, {da Costa}, {Davenport}, {Dawson}, {de
  Putter}, {De Lee}, {Delubac}, {Dhital}, {Ealet}, {Ebelke}, {Edmondson},
  {Eisenstein}, {Escoffier}, {Esposito}, {Evans}, {Fan}, {Femen{\'\i}a
  Castell{\'a}}, {Fern{\'a}ndez Alvar}, {Ferreira}, {Filiz Ak}, {Finley},
  {Fleming}, {Font-Ribera}, {Frinchaboy}, {Garc{\'\i}a-Hern{\'a}ndez},
  {Garc{\'\i}a P{\'e}rez}, {Ge}, {G{\'e}nova-Santos}, {Gillespie}, {Girardi},
  {Gonz{\'a}lez Hern{\'a}ndez}, {Grebel}, {Gunn}, {Guo}, {Haggard}, {Hamilton},
  {Harris}, {Hawley}, {Hearty}, {Ho}, {Hogg}, {Holtzman}, {Honscheid},
  {Huehnerhoff}, {Ivans}, {Ivezi{\'c}}, {Jacobson}, {Jiang}, {Johansson},
  {Johnson}, {Kauffmann}, {Kirkby}, {Kirkpatrick}, {Klaene}, {Knapp}, {Kneib},
  {Le Goff}, {Leauthaud}, {Lee}, {Lee}, {Long}, {Loomis}, {Lucatello},
  {Lundgren}, {Lupton}, {Ma}, {Ma}, {MacDonald}, {Mack}, {Mahadevan}, {Maia},
  {Majewski}, {Makler}, {Malanushenko}, {Malanushenko}, {Manchado},
  {Mandelbaum}, {Manera}, {Maraston}, {Margala}, {Martell}, {McBride},
  {McGreer}, {McMahon}, {M{\'e}nard}, {Meszaros}, {Miralda-Escud{\'e}},
  {Montero-Dorta}, {Montesano}, {Morrison}, {Muna}, {Munn}, {Murayama},
  {Myers}, {Neto}, {Nguyen}, {Nichol}, {Nidever}, {Noterdaeme}, {Nuza},
  {Ogando}, {Olmstead}, {Oravetz}, {Owen}, {Padmanabhan},
  {Palanque-Delabrouille}, {Pan}, {Parejko}, {Parihar}, {P{\^a}ris},
  {Pattarakijwanich}, {Pepper}, {Percival}, {P{\'e}rez-Fournon},
  {P{\'e}rez-R{\`a}fols}, {Petitjean}, {Pforr}, {Pieri}, {Pinsonneault}, {Porto
  de Mello}, {Prada}, {Price-Whelan}, {Raddick}, {Rebolo}, {Rich}, {Richards},
  {Robin}, {Rocha-Pinto}, {Rockosi}, {Roe}, {Ross}, {Ross}, {Rossi},
  {Rubi{\~n}o-Martin}, {Samushia}, {Sanchez Almeida}, {S{\'a}nchez},
  {Santiago}, {Sayres}, {Schlegel}, {Schlesinger}, {Schmidt}, {Schneider},
  {Schultheis}, {Schwope}, {Sc{\'o}ccola}, {Seljak}, {Sheldon}, {Shen}, {Shu},
  {Simmerer}, {Simmons}, {Skibba}, {Skrutskie}, {Slosar}, {Sobreira}, {Sobeck},
  {Stassun}, {Steele}, and {Steinmetz}]{Ahn2012}
C.~P. {Ahn} et al.
\newblock \emph{\apjs}, 203\penalty0 (2):\penalty0 21, Dec. 2012.
\newblock \doi{10.1088/0067-0049/203/2/21}.

\bibitem[{Amvrosiadis} et~al.(2025){Amvrosiadis}, {Wardlow}, {Birkin}, {Smail},
  {Swinbank}, {Nightingale}, {Bertoldi}, {Brandt}, {Casey}, {Chapman}, {Chen},
  {Cox}, {da Cunha}, {Dannerbauer}, {Dudzevi{\v{c}}i{\={u}}t{\.{e}}},
  {Gullberg}, {Hodge}, {Knudsen}, {Menten}, {Walter}, and {van der
  Werf}]{Amvrosiadis25}
A.~{Amvrosiadis} et al.
\newblock \emph{\mnras}, 536\penalty0 (4):\penalty0 3757--3783, Feb. 2025.
\newblock \doi{10.1093/mnras/stae2760}.

\bibitem[{Bacchini} et~al.(2020){Bacchini}, {Fraternali}, {Pezzulli}, and
  {Marasco}]{Bacchini2020}
C.~{Bacchini}, F.~{Fraternali}, G.~{Pezzulli}, and A.~{Marasco}.
\newblock \emph{Astron. Astrophys.}, 644:\penalty0 A125, Dec. 2020.
\newblock \doi{10.1051/0004-6361/202038962}.

\bibitem[{Banik} and {Zhao}(2022)]{Banik2022}
I.~{Banik} and H.~{Zhao}.
\newblock \emph{Symmetry}, 14\penalty0 (7):\penalty0 1331, June 2022.
\newblock \doi{10.3390/sym14071331}.

\bibitem[{Bar} et~al.(2022){Bar}, {Blum}, and {Sun}]{Bar2022}
N.~{Bar}, K.~{Blum}, and C.~{Sun}.
\newblock \emph{\prd}, 105\penalty0 (8):\penalty0 083015, Apr. 2022.
\newblock \doi{10.1103/PhysRevD.105.083015}.

\bibitem[{Barat} et~al.(2020){Barat}, {D'Eugenio}, {Colless}, {Sweet},
  {Groves}, and {Cortese}]{Barat2020}
D.~{Barat} et al.
\newblock \emph{Mon. Not. R. Astron. Soc.}, 498\penalty0 (4):\penalty0
  5885--5903, Nov. 2020.
\newblock \doi{10.1093/Mon. Not. R. Astron. Soc./staa2716}.

\bibitem[{Battaglia} et~al.(2006){Battaglia}, {Fraternali}, {Oosterloo}, and
  {Sancisi}]{Battaglia2006}
G.~{Battaglia}, F.~{Fraternali}, T.~{Oosterloo}, and R.~{Sancisi}.
\newblock \emph{Astron. Astrophys.}, 447:\penalty0 49--62, Feb. 2006.
\newblock \doi{10.1051/0004-6361:20053210}.

\bibitem[Begeman(1987)]{Begeman1987}
K.~Begeman.
\newblock \emph{HI Rotation Curves of Spiral Galaxies}.
\newblock PhD thesis, University of Groningen, NL, 1987.

\bibitem[{Begeman}(1989)]{Begeman1989}
K.~G. {Begeman}.
\newblock \emph{Astron. Astrophys.}, 223:\penalty0 47--60, Oct. 1989.

\bibitem[{Bera} et~al.(2023){Bera}, {Kanekar}, {Chengalur}, and
  {Bagla}]{Bera2023}
A.~{Bera}, N.~{Kanekar}, J.~N. {Chengalur}, and J.~S. {Bagla}.
\newblock \emph{\apjl}, 950\penalty0 (2):\penalty0 L18, June 2023.
\newblock \doi{10.3847/2041-8213/acd0b3}.

\bibitem[{Bertone} and {Hooper}(2018)]{Bertone2018}
G.~{Bertone} and D.~{Hooper}.
\newblock \emph{Reviews of Modern Physics}, 90\penalty0 (4):\penalty0 045002,
  Oct. 2018.
\newblock \doi{10.1103/RevModPhys.90.045002}.

\bibitem[{Bianchetti} et~al.(2025){Bianchetti}, {Sinigaglia}, {Rodighiero},
  {Elson}, {Vaccari}, {Pisano}, {Luber}, {Prandoni}, {Hess}, {Baes}, {Adams},
  {Maccagni}, {Renzini}, {Bisigello}, {Yun}, {Momjian}, {Gim}, {Pan},
  {Oosterloo}, {Dodson}, {Lucero}, {Frank}, {Ilbert}, {Davies}, {Khostovan},
  and {Salvato}]{Bianchetti2025}
A.~{Bianchetti} et al.
\newblock \emph{\apj}, 982\penalty0 (2):\penalty0 82, Apr. 2025.
\newblock \doi{10.3847/1538-4357/adb1b8}.

\bibitem[{Billard} et~al.(2022){Billard}, {Boulay}, {Cebri{\'a}n}, {Covi},
  {Fiorillo}, {Green}, {Kopp}, {Majorovits}, {Palladino}, {Petricca},
  {Roszkowski (chair)}, and {Schumann}]{Billard2022}
J.~{Billard} et al.
\newblock \emph{Reports on Progress in Physics}, 85\penalty0 (5):\penalty0
  056201, May 2022.
\newblock \doi{10.1088/1361-6633/ac5754}.

\bibitem[Binney and Tremaine(1994)]{BT94}
J.~Binney and S.~Tremaine.
\newblock \emph{Galactic Dynamics}.
\newblock Princeton University Press, Princeton, USA, first edition, 1994.

\bibitem[{Blanchet} and {Skordis}(2024)]{Blanchet2024}
L.~{Blanchet} and C.~{Skordis}.
\newblock \emph{\jcap}, 2024\penalty0 (11):\penalty0 040, Nov. 2024.
\newblock \doi{10.1088/1475-7516/2024/11/040}.

\bibitem[{Bonaldi} et~al.(2021){Bonaldi}, {An}, {Br{\"u}ggen}, {Burkutean},
  {Coelho}, {Goodarzi}, {Hartley}, {Sandhu}, {Wu}, {Yu}, {Zhoolideh Haghighi},
  {Ant{\'o}n}, {Bagheri}, {Barbosa}, {Barraca}, {Bartashevich}, {Bergano},
  {Bonato}, {Brand}, {de Gasperin}, {Giannetti}, {Dodson}, {Jain}, {Jaiswal},
  {Lao}, {Liu}, {Liuzzo}, {Lu}, {Lukic}, {Maia}, {Marchili}, {Massardi},
  {Mohan}, {Morgado}, {Panwar}, {Prabhakar}, {Ribeiro}, {Rygl}, {Sabz Ali},
  {Saremi}, {Schisano}, {Sheikhnezami}, {Vafaei Sadr}, {Wong}, and
  {Wong}]{Bonaldi2021}
A.~{Bonaldi} et al.
\newblock \emph{\mnras}, 500\penalty0 (3):\penalty0 3821--3837, Jan. 2021.
\newblock \doi{10.1093/mnras/staa3023}.

\bibitem[{Boomsma} et~al.(2008){Boomsma}, {Oosterloo}, {Fraternali}, {van der
  Hulst}, and {Sancisi}]{Boomsma2008}
R.~{Boomsma} et al.
\newblock \emph{\aap}, 490\penalty0 (2):\penalty0 555--570, Nov. 2008.
\newblock \doi{10.1051/0004-6361:200810120}.

\bibitem[{Bose} et~al.(2019){Bose}, {Frenk}, {Jenkins}, {Fattahi}, {G{\'o}mez},
  {Grand}, {Marinacci}, {Navarro}, {Oman}, {Pakmor}, {Schaye}, {Simpson}, and
  {Springel}]{Bose2019}
S.~{Bose} et al.
\newblock \emph{\mnras}, 486\penalty0 (4):\penalty0 4790--4804, July 2019.
\newblock \doi{10.1093/mnras/stz1168}.

\bibitem[{Bosma}(1978)]{Bosma1978}
A.~{Bosma}.
\newblock \emph{{The distribution and kinematics of neutral hydrogen in spiral
  galaxies of various morphological types}}.
\newblock PhD thesis, University of Groningen, 1978.

\bibitem[{Bouch{\'e}} et~al.(2015){Bouch{\'e}}, {Carfantan}, {Schroetter},
  {Michel-Dansac}, and {Contini}]{Bouche2015}
N.~{Bouch{\'e}} et al.
\newblock \emph{Astron. J.}, 150\penalty0 (3):\penalty0 92, Sept. 2015.
\newblock \doi{10.1088/0004-6256/150/3/92}.

\bibitem[{Brada} and {Milgrom}(1995)]{Brada1995}
R.~{Brada} and M.~{Milgrom}.
\newblock \emph{\mnras}, 276\penalty0 (2):\penalty0 453--459, Sept. 1995.
\newblock \doi{10.1093/mnras/276.2.453}.

\bibitem[Bradbury et~al.(2018)Bradbury, Frostig, Hawkins, Johnson, Leary,
  Maclaurin, Necula, Paszke, VanderPlas, Wanderman-Milne, and
  Zhang]{jax2018github}
J.~Bradbury et al.
\newblock {JAX: composable transformations of Python+NumPy programs}, 2018.
\newblock URL \url{http://github.com/google/jax}.

\bibitem[{Braun} and {Thilker}(2004)]{Braun2004}
R.~{Braun} and D.~A. {Thilker}.
\newblock \emph{\aap}, 417:\penalty0 421--435, Apr. 2004.
\newblock \doi{10.1051/0004-6361:20034423}.

\bibitem[{Bullock} and {Boylan-Kolchin}(2017)]{Bullock2017}
J.~S. {Bullock} and M.~{Boylan-Kolchin}.
\newblock \emph{Ann. Rev. Astron. Astrophys.}, 55\penalty0 (1):\penalty0
  343--387, Aug. 2017.
\newblock \doi{10.1146/annurev-astro-091916-055313}.

\bibitem[{Burkert} et~al.(2010){Burkert}, {Genzel}, {Bouch{\'e}}, {Cresci},
  {Khochfar}, {Sommer-Larsen}, {Sternberg}, {Naab}, {F{\"o}rster Schreiber},
  {Tacconi}, {Shapiro}, {Hicks}, {Lutz}, {Davies}, {Buschkamp}, and
  {Genel}]{Burkert2010}
A.~{Burkert} et al.
\newblock \emph{Astrophys. J.}, 725\penalty0 (2):\penalty0 2324--2332, Dec.
  2010.
\newblock \doi{10.1088/0004-637X/725/2/2324}.

\bibitem[{Calderon} and {Berlind}(2019)]{Calderon2019}
V.~F. {Calderon} and A.~A. {Berlind}.
\newblock \emph{\mnras}, 490\penalty0 (2):\penalty0 2367--2379, Dec. 2019.
\newblock \doi{10.1093/mnras/stz2775}.

\bibitem[{Cannon} et~al.(2004){Cannon}, {McClure-Griffiths}, {Skillman}, and
  {C{\^o}t{\'e}}]{Cannon2004}
J.~M. {Cannon}, N.~M. {McClure-Griffiths}, E.~D. {Skillman}, and
  S.~{C{\^o}t{\'e}}.
\newblock \emph{Astrophys. J.}, 607:\penalty0 274--284, May 2004.
\newblock \doi{10.1086/383408}.

\bibitem[{Casertano}(1983)]{Casertano1983}
S.~{Casertano}.
\newblock \emph{Mon. Not. R. Astron. Soc.}, 203:\penalty0 735--747, May 1983.

\bibitem[{Chae}(2022)]{Chae2022b}
K.-H. {Chae}.
\newblock \emph{\apj}, 941\penalty0 (1):\penalty0 55, Dec. 2022.
\newblock \doi{10.3847/1538-4357/ac93fc}.

\bibitem[{Chae} et~al.(2020){Chae}, {Lelli}, {Desmond}, {McGaugh}, {Li}, and
  {Schombert}]{Chae2020}
K.-H. {Chae} et al.
\newblock \emph{Astrophys. J.}, 904\penalty0 (1):\penalty0 51, Nov. 2020.
\newblock \doi{10.3847/1538-4357/abbb96}.

\bibitem[{Chae} et~al.(2021){Chae}, {Desmond}, {Lelli}, {McGaugh}, and
  {Schombert}]{Chae2021}
K.-H. {Chae} et al.
\newblock \emph{\apj}, 921\penalty0 (2):\penalty0 104, Nov. 2021.
\newblock \doi{10.3847/1538-4357/ac1bba}.

\bibitem[{Chae} et~al.(2022){Chae}, {Lelli}, {Desmond}, {McGaugh}, and
  {Schombert}]{Chae2022a}
K.-H. {Chae} et al.
\newblock \emph{\prd}, 106\penalty0 (10):\penalty0 103025, Nov. 2022.
\newblock \doi{10.1103/PhysRevD.106.103025}.

\bibitem[{Chowdhury} et~al.(2020){Chowdhury}, {Kanekar}, {Chengalur}, {Sethi},
  and {Dwarakanath}]{Chowdhury2020}
A.~{Chowdhury} et al.
\newblock \emph{\nat}, 586\penalty0 (7829):\penalty0 369--372, Oct. 2020.
\newblock \doi{10.1038/s41586-020-2794-7}.

\bibitem[{Chowdhury} et~al.(2022{\natexlab{a}}){Chowdhury}, {Kanekar}, and
  {Chengalur}]{Chowdhury2022a}
A.~{Chowdhury}, N.~{Kanekar}, and J.~N. {Chengalur}.
\newblock \emph{\apj}, 937\penalty0 (2):\penalty0 103, Oct. 2022{\natexlab{a}}.
\newblock \doi{10.3847/1538-4357/ac7d52}.

\bibitem[{Chowdhury} et~al.(2022{\natexlab{b}}){Chowdhury}, {Kanekar}, and
  {Chengalur}]{Chowdhury2022b}
A.~{Chowdhury}, N.~{Kanekar}, and J.~N. {Chengalur}.
\newblock \emph{\apjl}, 935\penalty0 (1):\penalty0 L5, Aug. 2022{\natexlab{b}}.
\newblock \doi{10.3847/2041-8213/ac8150}.

\bibitem[{Chu} et~al.(2024){Chu}, {Tang}, {Xu}, {Lu}, and {Long}]{Chu2024}
J.~{Chu} et al.
\newblock \emph{\mnras}, 528\penalty0 (4):\penalty0 6354--6369, Mar. 2024.
\newblock \doi{10.1093/mnras/stae406}.

\bibitem[{Col{\'\i}n} et~al.(2000){Col{\'\i}n}, {Avila-Reese}, and
  {Valenzuela}]{Colin2000}
P.~{Col{\'\i}n}, V.~{Avila-Reese}, and O.~{Valenzuela}.
\newblock \emph{\apj}, 542\penalty0 (2):\penalty0 622--630, Oct. 2000.
\newblock \doi{10.1086/317057}.

\bibitem[{Corbelli} and {Schneider}(1997)]{Corbelli1997}
E.~{Corbelli} and S.~E. {Schneider}.
\newblock \emph{Astrophys. J.}, 479\penalty0 (1):\penalty0 244--257, Apr. 1997.
\newblock \doi{10.1086/303849}.

\bibitem[{Cornu} et~al.(2024){Cornu}, {Salom{\'e}}, {Semelin}, {Marchal},
  {Freundlich}, {Aicardi}, {Lu}, {Sainton}, {Mertens}, {Combes}, and
  {Tasse}]{Cornu2024}
D.~{Cornu} et al.
\newblock \emph{\aap}, 690:\penalty0 A211, Oct. 2024.
\newblock \doi{10.1051/0004-6361/202449548}.

\bibitem[{Cornu} et~al.(2026){Cornu}, {Semelin}, {Salom{\'e}}, {Lu}, {Aicardi},
  {Freundlich}, {Mertens}, {Marchal}, {Sainton}, {Combes}, and
  {Tasse}]{Cornu2025}
D.~{Cornu} et al.
\newblock \emph{\aap}, 707:\penalty0 A203, Mar. 2026.
\newblock \doi{10.1051/0004-6361/202557257}.

\bibitem[{Daigle} et~al.(2006){Daigle}, {Carignan}, {Amram}, {Hernandez},
  {Chemin}, {Balkowski}, and {Kennicutt}]{Daigle2006}
O.~{Daigle} et al.
\newblock \emph{\mnras}, 367\penalty0 (2):\penalty0 469--512, Apr. 2006.
\newblock \doi{10.1111/j.1365-2966.2006.10002.x}.

\bibitem[{Danhaive} and {Tacchella}(2025)]{Danhaive2025}
A.~L. {Danhaive} and S.~{Tacchella}.
\newblock \emph{arXiv e-prints}, art. arXiv:2510.07369, Oct. 2025.
\newblock \doi{10.48550/arXiv.2510.07369}.

\bibitem[{Davis}(2014)]{Davis2014}
T.~A. {Davis}.
\newblock \emph{\mnras}, 443\penalty0 (1):\penalty0 911--918, Sept. 2014.
\newblock \doi{10.1093/mnras/stu1163}.

\bibitem[{Davis} et~al.(2013){Davis}, {Alatalo}, {Bureau}, {Cappellari},
  {Scott}, {Young}, {Blitz}, {Crocker}, {Bayet}, {Bois}, {Bournaud}, {Davies},
  {de Zeeuw}, {Duc}, {Emsellem}, {Khochfar}, {Krajnovi{\'c}}, {Kuntschner},
  {Lablanche}, {McDermid}, {Morganti}, {Naab}, {Oosterloo}, {Sarzi}, {Serra},
  and {Weijmans}]{Davis13}
T.~A. {Davis} et al.
\newblock \emph{\mnras}, 429\penalty0 (1):\penalty0 534--555, Feb. 2013.
\newblock \doi{10.1093/mnras/sts353}.

\bibitem[{Davis} et~al.(2017{\natexlab{a}}){Davis}, {Bureau}, {Onishi},
  {Cappellari}, {Iguchi}, and {Sarzi}]{Davis+17}
T.~A. {Davis} et al.
\newblock \emph{\mnras}, 468\penalty0 (4):\penalty0 4675--4690, July
  2017{\natexlab{a}}.
\newblock \doi{10.1093/mnras/stw3217}.

\bibitem[{Davis} et~al.(2017{\natexlab{b}}){Davis}, {Bureau}, {Onishi},
  {Cappellari}, {Iguchi}, and {Sarzi}]{TD17}
T.~A. {Davis} et al.
\newblock \emph{\mnras}, 468\penalty0 (4):\penalty0 4675--4690, July
  2017{\natexlab{b}}.
\newblock \doi{10.1093/mnras/stw3217}.

\bibitem[{Davis} et~al.(2020){Davis}, {Zabel}, and {Dawson}]{Davis20}
T.~A. {Davis}, N.~{Zabel}, and J.~M. {Dawson}.
\newblock {KinMS: Three-dimensional kinematic modeling of arbitrary gas
  distributions}.
\newblock Astrophysics Source Code Library, record ascl:2006.003, June 2020.

\bibitem[{Dawson} et~al.(2021){Dawson}, {Davis}, {Gomez}, and
  {Schock}]{Dawson2021}
J.~M. {Dawson}, T.~A. {Davis}, E.~L. {Gomez}, and J.~{Schock}.
\newblock \emph{Mon. Not. R. Astron. Soc.}, 503\penalty0 (1):\penalty0
  574--585, May 2021.
\newblock \doi{10.1093/Mon. Not. R. Astron. Soc./stab427}.

\bibitem[{de Blok}(2010)]{deBlok2010}
W.~J.~G. {de Blok}.
\newblock \emph{Advances in Astronomy}, 2010:\penalty0 789293, Jan. 2010.
\newblock \doi{10.1155/2010/789293}.

\bibitem[{de Blok} and {Bosma}(2002)]{deBlok2002}
W.~J.~G. {de Blok} and A.~{Bosma}.
\newblock \emph{\aap}, 385:\penalty0 816--846, Apr. 2002.
\newblock \doi{10.1051/0004-6361:20020080}.

\bibitem[{de Blok} et~al.(2008){de Blok}, {Walter}, {Brinks}, {Trachternach},
  {Oh}, and {Kennicutt}]{deBlok2008}
W.~J.~G. {de Blok} et al.
\newblock \emph{Astron. J.}, 136\penalty0 (6):\penalty0 2648--2719, Dec. 2008.
\newblock \doi{10.1088/0004-6256/136/6/2648}.

\bibitem[{Deg} et~al.(2022){Deg}, {Spekkens}, {Westmeier}, {Reynolds},
  {Venkataraman}, {Goliath}, {Shen}, {Halloran}, {Bosma}, {Catinella}, {de
  Blok}, {D{\'e}nes}, {Di Teodoro}, {Elagali}, {For}, {Howlett}, {J{\'o}zsa},
  {Kamphuis}, {Kleiner}, {Koribalski}, {Lee-Waddell}, {Lelli}, {Lin},
  {Murugeshan}, {Oh}, {Rhee}, {Scott}, {Staveley-Smith}, {van der Hulst},
  {Verdes-Montenegro}, {Wang}, and {Wong}]{Deg2022}
N.~{Deg} et al.
\newblock \emph{\pasa}, 39:\penalty0 e059, Nov. 2022.
\newblock \doi{10.1017/pasa.2022.43}.

\bibitem[{Deg} et~al.(2025){Deg}, {Spekkens}, {Arora}, {Dudley}, {White},
  {Helias}, {English}, {O'Beirne}, {Kilborn}, {Ferrand}, {Richardson},
  {Catinella}, {Cortese}, {D{\'e}nes}, {Elagali}, {For}, {Lee-Waddell}, {Rhee},
  {Shao}, {Shen}, {Staveley-Smith}, {Westmeier}, and {Wong}]{Deg2025}
N.~{Deg} et al.
\newblock \emph{\apj}, 994\penalty0 (2):\penalty0 232, Dec. 2025.
\newblock \doi{10.3847/1538-4357/ae1003}.

\bibitem[{Desmond}(2023)]{Desmond2023}
H.~{Desmond}.
\newblock \emph{\mnras}, 526\penalty0 (3):\penalty0 3342--3351, Dec. 2023.
\newblock \doi{10.1093/mnras/stad2762}.

\bibitem[{Desmond} et~al.(2018){Desmond}, {Ferreira}, {Lavaux}, and
  {Jasche}]{Desmond2018}
H.~{Desmond}, P.~G. {Ferreira}, G.~{Lavaux}, and J.~{Jasche}.
\newblock \emph{\mnras}, 474\penalty0 (3):\penalty0 3152--3161, Mar. 2018.
\newblock \doi{10.1093/mnras/stx3062}.

\bibitem[{Di Cintio} and {Lelli}(2016)]{DiCintio2016}
A.~{Di Cintio} and F.~{Lelli}.
\newblock \emph{Mon. Not. R. Astron. Soc.}, 456\penalty0 (1):\penalty0
  L127--L131, Feb. 2016.
\newblock \doi{10.1093/Mon. Not. R. Astron. Soc.l/slv185}.

\bibitem[{Di Teodoro} and {Fraternali}(2015)]{DiTeodoro2015}
E.~M. {Di Teodoro} and F.~{Fraternali}.
\newblock \emph{Mon. Not. R. Astron. Soc.}, 451\penalty0 (3):\penalty0
  3021--3033, Aug. 2015.
\newblock \doi{10.1093/Mon. Not. R. Astron. Soc./stv1213}.

\bibitem[{Di Teodoro} et~al.(2016){Di Teodoro}, {Fraternali}, and
  {Miller}]{DiTeodoro2016}
E.~M. {Di Teodoro}, F.~{Fraternali}, and S.~H. {Miller}.
\newblock \emph{\aap}, 594:\penalty0 A77, Oct. 2016.
\newblock \doi{10.1051/0004-6361/201628315}.

\bibitem[{Di Teodoro} et~al.(2018){Di Teodoro}, {Grillo}, {Fraternali},
  {Gobat}, {Karman}, {Mercurio}, {Rosati}, {Balestra}, {Caminha}, {Caputi},
  {Lombardi}, {Suyu}, {Treu}, and {Vanzella}]{DiTeodoro2018}
E.~M. {Di Teodoro} et al.
\newblock \emph{\mnras}, 476\penalty0 (1):\penalty0 804--813, May 2018.
\newblock \doi{10.1093/mnras/sty175}.

\bibitem[{Dubinski} and {Carlberg}(1991)]{Dubinski1991}
J.~{Dubinski} and R.~G. {Carlberg}.
\newblock \emph{Astrophys. J.}, 378:\penalty0 496, Sept. 1991.
\newblock \doi{10.1086/170451}.

\bibitem[{Dutton} et~al.(2019){Dutton}, {Macci{\`o}}, {Obreja}, and
  {Buck}]{Dutton2019}
A.~A. {Dutton}, A.~V. {Macci{\`o}}, A.~{Obreja}, and T.~{Buck}.
\newblock \emph{\mnras}, 485\penalty0 (2):\penalty0 1886--1899, May 2019.
\newblock \doi{10.1093/mnras/stz531}.

\bibitem[{Dye} et~al.(2018){Dye}, {Furlanetto}, {Dunne}, {Eales}, {Negrello},
  {Nayyeri}, {van der Werf}, {Serjeant}, {Farrah}, {Micha{\l}owski}, {Baes},
  {Marchetti}, {Cooray}, {Riechers}, and {Amvrosiadis}]{Dye2018}
S.~{Dye} et al.
\newblock \emph{\mnras}, 476\penalty0 (4):\penalty0 4383--4394, June 2018.
\newblock \doi{10.1093/mnras/sty513}.

\bibitem[{El-Zant} et~al.(2016){El-Zant}, {Freundlich}, and
  {Combes}]{El-Zant2016}
A.~A. {El-Zant}, J.~{Freundlich}, and F.~{Combes}.
\newblock \emph{\mnras}, 461\penalty0 (2):\penalty0 1745--1759, Sept. 2016.
\newblock \doi{10.1093/mnras/stw1398}.

\bibitem[{Epinat} et~al.(2010){Epinat}, {Amram}, {Balkowski}, and
  {Marcelin}]{Epinat2010}
B.~{Epinat}, P.~{Amram}, C.~{Balkowski}, and M.~{Marcelin}.
\newblock \emph{Mon. Not. R. Astron. Soc.}, 401\penalty0 (4):\penalty0
  2113--2147, Feb. 2010.
\newblock \doi{10.1111/j.1365-2966.2009.15688.x}.

\bibitem[{Famaey} and {McGaugh}(2012)]{Famaey2012}
B.~{Famaey} and S.~S. {McGaugh}.
\newblock \emph{Living Reviews in Relativity}, 15:\penalty0 10, Sept. 2012.
\newblock \doi{10.12942/lrr-2012-10}.

\bibitem[{Fraternali}(2017)]{Fraternali17}
F.~{Fraternali}.
\newblock In A.~{Fox} and R.~{Dav{\'e}}, editors, \emph{Gas Accretion onto
  Galaxies}, volume 430 of \emph{Astrophysics and Space Science Library}, page
  323, Jan. 2017.
\newblock \doi{10.1007/978-3-319-52512-9_14}.

\bibitem[{Fraternali} et~al.(2002){Fraternali}, {van Moorsel}, {Sancisi}, and
  {Oosterloo}]{Fraternali02}
F.~{Fraternali}, G.~{van Moorsel}, R.~{Sancisi}, and T.~{Oosterloo}.
\newblock \emph{\aj}, 123\penalty0 (6):\penalty0 3124--3140, June 2002.
\newblock \doi{10.1086/340358}.

\bibitem[{Freundlich} et~al.(2020){Freundlich}, {Dekel}, {Jiang}, {Ishai},
  {Cornuault}, {Lapiner}, {Dutton}, and {Macci{\`o}}]{Freundlich2020a}
J.~{Freundlich} et al.
\newblock \emph{\mnras}, 491\penalty0 (3):\penalty0 4523--4542, Jan. 2020.
\newblock \doi{10.1093/mnras/stz3306}.

\bibitem[{Garaldi} et~al.(2018){Garaldi}, {Romano-D{\'\i}az}, {Porciani}, and
  {Pawlowski}]{Garaldi2018}
E.~{Garaldi}, E.~{Romano-D{\'\i}az}, C.~{Porciani}, and M.~S. {Pawlowski}.
\newblock \emph{Phys. Rev. Lett.}, 120\penalty0 (26):\penalty0 261301, June
  2018.
\newblock \doi{10.1103/PhysRevLett.120.261301}.

\bibitem[{Gasser}(2023)]{Gasser2023}
A.~{Gasser}.
\newblock {Modelling Rotation Curves in Marginally Resolved Warped Galaxies}.
\newblock Master's thesis, Royal Military College of Canada, May 2023.
\newblock URL
  \url{https://espace.rmc-cmr.ca/jspui/handle/11264/1287?mode=full}.

\bibitem[{Gentile} et~al.(2005){Gentile}, {Burkert}, {Salucci}, {Klein}, and
  {Walter}]{Gentile2005}
G.~{Gentile} et al.
\newblock \emph{Astrophys. J.}, 634\penalty0 (2):\penalty0 L145--L148, Dec.
  2005.
\newblock \doi{10.1086/498939}.

\bibitem[{Gentile} et~al.(2007){Gentile}, {Salucci}, {Klein}, and
  {Granato}]{Gentile2007}
G.~{Gentile}, P.~{Salucci}, U.~{Klein}, and G.~L. {Granato}.
\newblock \emph{Mon. Not. R. Astron. Soc.}, 375\penalty0 (1):\penalty0
  199--212, Feb. 2007.
\newblock \doi{10.1111/j.1365-2966.2006.11283.x}.

\bibitem[{Ghari} et~al.(2019){Ghari}, {Famaey}, {Laporte}, and
  {Haghi}]{Ghari2019}
A.~{Ghari}, B.~{Famaey}, C.~{Laporte}, and H.~{Haghi}.
\newblock \emph{\aap}, 623:\penalty0 A123, Mar. 2019.
\newblock \doi{10.1051/0004-6361/201834661}.

\bibitem[{Goodman}(2000)]{Goodman2000}
J.~{Goodman}.
\newblock \emph{\na}, 5\penalty0 (2):\penalty0 103--107, Apr. 2000.
\newblock \doi{10.1016/S1384-1076(00)00015-4}.

\bibitem[{Governato} et~al.(2010){Governato}, {Brook}, {Mayer}, {Brooks},
  {Rhee}, {Wadsley}, {Jonsson}, {Willman}, {Stinson}, {Quinn}, and
  {Madau}]{Governato2010}
F.~{Governato} et al.
\newblock \emph{Nature}, 463:\penalty0 203--206, Jan. 2010.
\newblock \doi{10.1038/nature08640}.

\bibitem[{Governato} et~al.(2012){Governato}, {Zolotov}, {Pontzen},
  {Christensen}, {Oh}, {Brooks}, {Quinn}, {Shen}, and {Wadsley}]{Governato2012}
F.~{Governato} et al.
\newblock \emph{Mon. Not. R. Astron. Soc.}, 422:\penalty0 1231--1240, May 2012.
\newblock \doi{10.1111/j.1365-2966.2012.20696.x}.

\bibitem[{Hahn} et~al.(2024){Hahn}, {Bottrell}, and {Lee}]{Hahn2024}
C.~{Hahn}, C.~{Bottrell}, and K.-G. {Lee}.
\newblock \emph{\apj}, 968\penalty0 (2):\penalty0 90, June 2024.
\newblock \doi{10.3847/1538-4357/ad4344}.

\bibitem[{Hartley} et~al.(2023){Hartley}, {Bonaldi}, {Braun}, {Aditya},
  {Aicardi}, {Alegre}, {Chakraborty}, {Chen}, {Choudhuri}, {Clarke}, {Coles},
  {Collinson}, {Cornu}, {Darriba}, {Veneri}, {Forbrich}, {Fraga}, {Galan},
  {Garrido}, {Gubanov}, {H{\r{a}}kansson}, {Hardcastle}, {Heneka}, {Herranz},
  {Hess}, {Jagannath}, {Jaiswal}, {Jurek}, {Korber}, {Kitaeff}, {Kleiner},
  {Lao}, {Lu}, {Mazumder}, {Mold{\'o}n}, {Mondal}, {Ni}, {{\"O}nnheim},
  {Parra}, {Patra}, {Peel}, {Salom{\'e}}, {S{\'a}nchez-Exp{\'o}sito},
  {Sargent}, {Semelin}, {Serra}, {Shaw}, {Shen}, {Sj{\"o}berg}, {Smith},
  {Soroka}, {Stolyarov}, {Tolley}, {Toribio}, {van der Hulst}, {Sadr},
  {Verdes-Montenegro}, {Westmeier}, {Yu}, {Yu}, {Zhang}, {Zhang}, {Zhang},
  {Alberdi}, {Ashdown}, {Bom}, {Br{\"u}ggen}, {Cannon}, {Chen}, {Combes},
  {Conway}, {Courbin}, {Ding}, {Fourestey}, {Freundlich}, {Gao}, {Gheller},
  {Guo}, {Gustavsson}, {Jirstrand}, {Jones}, {J{\'o}zsa}, {Kamphuis}, {Kneib},
  {Lindqvist}, {Liu}, {Liu}, {Mao}, {Marchal}, {M{\'a}rquez}, {Meshcheryakov},
  {Olberg}, {Oozeer}, {Pandey-Pommier}, {Pei}, {Peng}, {Sabater}, {Sorgho},
  {Starck}, {Tasse}, {Wang}, {Wang}, {Xi}, {Yang}, {Zhang}, {Zhang}, {Zhao},
  and {Zuo}]{Hartley2023}
P.~{Hartley} et al.
\newblock \emph{\mnras}, 523\penalty0 (2):\penalty0 1967--1993, Aug. 2023.
\newblock \doi{10.1093/mnras/stad1375}.

\bibitem[{Haubner} et~al.(2026){Haubner}, {Lelli}, {Di Teodoro}, {Duey},
  {McGaugh}, {Schombert}, {Hess}, and {Apertif Team}]{Haubner2024}
K.~{Haubner} et al.
\newblock In D.~J. {Pisano}, K.~M. {Mogotsi}, J.~{Healy}, and S.~{Blyth},
  editors, \emph{IAU Symposium}, volume~20 of \emph{IAU Symposium}, pages
  119--125, July 2026.
\newblock \doi{10.1017/S1743921324002886}.

\bibitem[{Hess} et~al.(2019){Hess}, {Luber}, {Fern{\'a}ndez}, {Gim}, {van
  Gorkom}, {Momjian}, {Gross}, {Meyer}, {Popping}, {Davies}, {Hunt}, {Kreckel},
  {Lucero}, {Pisano}, {Sanchez-Barrantes}, {Yun}, {Dodson}, {Vinsen},
  {Wicenec}, {Wu}, {Bershady}, {Chung}, {Davis}, {Donovan Meyer}, {Henning},
  {Maddox}, {Smith}, {van der Hulst}, {Verheijen}, and {Wilcots}]{Hess2019}
K.~M. {Hess} et al.
\newblock \emph{\mnras}, 484\penalty0 (2):\penalty0 2234--2256, Apr. 2019.
\newblock \doi{10.1093/mnras/sty3421}.

\bibitem[{Hogarth} et~al.(2024){Hogarth}, {Saintonge}, {Davis}, {Ellison},
  {Lin}, {L{\'o}pez-Cob{\'a}}, {Pan}, and {Thorp}]{Hogarth24}
L.~M. {Hogarth} et al.
\newblock \emph{\mnras}, 528\penalty0 (4):\penalty0 6768--6785, Mar. 2024.
\newblock \doi{10.1093/mnras/stae377}.

\bibitem[{Hossenfelder} and {Mistele}(2018)]{Hossenfelder2018}
S.~{Hossenfelder} and T.~{Mistele}.
\newblock \emph{International Journal of Modern Physics D}, 27\penalty0
  (14):\penalty0 1847010, Jan. 2018.
\newblock \doi{10.1142/S0218271818470107}.

\bibitem[{Hu} et~al.(2000){Hu}, {Barkana}, and {Gruzinov}]{Hu2000}
W.~{Hu}, R.~{Barkana}, and A.~{Gruzinov}.
\newblock \emph{\prl}, 85\penalty0 (6):\penalty0 1158--1161, Aug. 2000.
\newblock \doi{10.1103/PhysRevLett.85.1158}.

\bibitem[{Hunt} et~al.(2023){Hunt}, {Belfiore}, {Lelli}, {Draine}, {Marasco},
  {Garc{\'\i}a-Burillo}, {Venturi}, {Combes}, {Wei{\ss}}, {Henkel}, {Menten},
  {Annibali}, {Casasola}, {Cignoni}, {McLeod}, {Tosi}, {Beltr{\'a}n}, {Concas},
  {Cresci}, {Ginolfi}, {Kumari}, and {Mannucci}]{Hunt2023}
L.~K. {Hunt} et al.
\newblock \emph{\aap}, 675:\penalty0 A64, July 2023.
\newblock \doi{10.1051/0004-6361/202245805}.

\bibitem[{Hunter} et~al.(2012){Hunter}, {Ficut-Vicas}, {Ashley}, {Brinks},
  {Cigan}, {Elmegreen}, {Heesen}, {Herrmann}, {Johnson}, {Oh}, {Rupen},
  {Schruba}, {Simpson}, {Walter}, {Westpfahl}, {Young}, and
  {Zhang}]{Hunter2012}
D.~A. {Hunter} et al.
\newblock \emph{Astron. J.}, 144:\penalty0 134, Nov. 2012.
\newblock \doi{10.1088/0004-6256/144/5/134}.

\bibitem[{Iorio} et~al.(2017){Iorio}, {Fraternali}, {Nipoti}, {Di Teodoro},
  {Read}, and {Battaglia}]{Iorio2017}
G.~{Iorio} et al.
\newblock \emph{Mon. Not. R. Astron. Soc.}, 466\penalty0 (4):\penalty0
  4159--4192, Apr. 2017.
\newblock \doi{10.1093/Mon. Not. R. Astron. Soc./stw3285}.

\bibitem[{Jarvis} et~al.(2025){Jarvis}, {Tudorache}, {Heywood}, {Ponomareva},
  {Baes}, {Maddox}, {Spekkens}, {V{\u{a}}r{\u{a}}{\c{s}}teanu}, {Hale},
  {Santos}, {Varadaraj}, {Adams}, {Bianchetti}, {Catinella}, {Delhaize},
  {Maksymowicz-Maciata}, {Pi{\~n}a}, {Pan}, {Saintonge}, {Sharma}, and
  {Wong}]{Jarvis2025}
M.~J. {Jarvis} et al.
\newblock \emph{\mnras}, Oct. 2025.
\newblock \doi{10.1093/mnras/staf1702}.

\bibitem[{Jiang} et~al.(2019){Jiang}, {Dekel}, {Freundlich}, {Romanowsky},
  {Dutton}, {Macci{\`o}}, and {Di Cintio}]{Jiang2019}
F.~{Jiang} et al.
\newblock \emph{\mnras}, 487\penalty0 (4):\penalty0 5272--5290, Aug. 2019.
\newblock \doi{10.1093/mnras/stz1499}.

\bibitem[{Jiang} et~al.(2023){Jiang}, {Benson}, {Hopkins}, {Slone}, {Lisanti},
  {Kaplinghat}, {Peter}, {Zeng}, {Du}, {Yang}, and {Shen}]{Jiang2023}
F.~{Jiang} et al.
\newblock \emph{\mnras}, 521\penalty0 (3):\penalty0 4630--4644, May 2023.
\newblock \doi{10.1093/mnras/stad705}.

\bibitem[Jones et~al.(2018)Jones, Espada, Verdes-Montenegro, Huchtmeier,
  Lisenfeld, Leon, Sulentic, Sabater, Jones, Sanchez, and Garrido]{Jones2018}
M.~G. Jones et al.
\newblock \emph{A\&A}, 609:\penalty0 17, 1 2018.
\newblock ISSN 14320746.
\newblock \doi{10.1051/0004-6361/201731448}.
\newblock URL
  \url{https://ui.adsabs.harvard.edu/abs/2018A%26A...609A..17J/abstract
  http://cdsarc.u-strasbg.fr/viz-bin/qcat?J/A+A/609/A17}.

\bibitem[{J{\'o}zsa} et~al.(2007){J{\'o}zsa}, {Kenn}, {Klein}, and
  {Oosterloo}]{Jozsa2007}
G.~I.~G. {J{\'o}zsa}, F.~{Kenn}, U.~{Klein}, and T.~A. {Oosterloo}.
\newblock \emph{Astron. Astrophys.}, 468\penalty0 (2):\penalty0 731--774, June
  2007.
\newblock \doi{10.1051/0004-6361:20066164}.

\bibitem[{Kalnajs}(1983)]{Kalnajs1983}
A.~{Kalnajs}.
\newblock In \emph{Internal Kinematics and Dynamics of Galaxies}, volume 100 of
  \emph{{IAU Symposia}}, pages 87--88, Jan. 1983.

\bibitem[{Kamphuis} et~al.(2015){Kamphuis}, {J{\'o}zsa}, {Oh}, {Spekkens},
  {Urbancic}, {Serra}, {Koribalski}, and {Dettmar}]{Kamphuis2015}
P.~{Kamphuis} et al.
\newblock \emph{Mon. Not. R. Astron. Soc.}, 452\penalty0 (3):\penalty0
  3139--3158, Sept. 2015.
\newblock \doi{10.1093/Mon. Not. R. Astron. Soc./stv1480}.

\bibitem[{Katz} et~al.(2017){Katz}, {Lelli}, {McGaugh}, {Di Cintio}, {Brook},
  and {Schombert}]{Katz2017}
H.~{Katz} et al.
\newblock \emph{Mon. Not. R. Astron. Soc.}, 466\penalty0 (2):\penalty0
  1648--1668, Apr. 2017.
\newblock \doi{10.1093/Mon. Not. R. Astron. Soc./stw3101}.

\bibitem[{Kazemi-Moridani} et~al.(2025){Kazemi-Moridani}, {Baker}, {Verheijen},
  {Gawiser}, {Blyth}, {Obreschkow}, {Chemin}, {Collier}, {Cook}, {Delhaize},
  {Elson}, {Frank}, {Glowacki}, {Hess}, {Holwerda}, {Hutchens}, {Jarvis},
  {Kaasinen}, {Makhathini}, {Mohapatra}, {Pan}, {Schr{\"o}der},
  {Stockenstroom}, {Vaccari}, {Westmeier}, {Wu}, and
  {Zwaan}]{KazemiMoridani2025}
A.~{Kazemi-Moridani} et al.
\newblock \emph{\apj}, 981\penalty0 (2):\penalty0 208, Mar. 2025.
\newblock \doi{10.3847/1538-4357/ad9f3f}.

\bibitem[{Keller} and {Wadsley}(2017)]{Keller2017}
B.~W. {Keller} and J.~W. {Wadsley}.
\newblock \emph{Astrophys. J.}, 835\penalty0 (1):\penalty0 L17, Jan. 2017.
\newblock \doi{10.3847/2041-8213/835/1/L17}.

\bibitem[{Kent}(1986)]{Kent1986}
S.~M. {Kent}.
\newblock \emph{Astron. J.}, 91:\penalty0 1301--1327, June 1986.
\newblock \doi{10.1086/114106}.

\bibitem[{Kent}(1987)]{Kent1987}
S.~M. {Kent}.
\newblock \emph{Astron. J.}, 93:\penalty0 816--832, Apr. 1987.
\newblock \doi{10.1086/114366}.

\bibitem[{Kretschmer} et~al.(2021){Kretschmer}, {Dekel}, {Freundlich},
  {Lapiner}, {Ceverino}, and {Primack}]{Kretschmer2021}
M.~{Kretschmer} et al.
\newblock \emph{\mnras}, 503\penalty0 (4):\penalty0 5238--5253, June 2021.
\newblock \doi{10.1093/mnras/stab833}.

\bibitem[{Kuzio de Naray} et~al.(2006){Kuzio de Naray}, {McGaugh}, {de Blok},
  and {Bosma}]{KuziodeNaray2006}
R.~{Kuzio de Naray}, S.~S. {McGaugh}, W.~J.~G. {de Blok}, and A.~{Bosma}.
\newblock \emph{\apjs}, 165\penalty0 (2):\penalty0 461--479, Aug. 2006.
\newblock \doi{10.1086/505345}.

\bibitem[{Lee} et~al.(2025){Lee}, {F{\"o}rster Schreiber}, {Price}, {Liu},
  {Genzel}, {Davies}, {Tacconi}, {Shimizu}, {Nestor Shachar}, {Espejo Salcedo},
  {Pastras}, {Wuyts}, {Lutz}, {Renzini}, {{\"U}bler}, {Herrera-Camus}, and
  {Sternberg}]{Lee2025}
L.~L. {Lee} et al.
\newblock \emph{\apj}, 978\penalty0 (1):\penalty0 14, Jan. 2025.
\newblock \doi{10.3847/1538-4357/ad90b5}.

\bibitem[{Lelli}(2022)]{Lelli2022}
F.~{Lelli}.
\newblock \emph{Nature Astronomy}, 6:\penalty0 35--47, Jan. 2022.
\newblock \doi{10.1038/s41550-021-01562-2}.

\bibitem[{Lelli}(2024)]{Lelli2024}
F.~{Lelli}.
\newblock \emph{\aap}, 689:\penalty0 L3, Sept. 2024.
\newblock \doi{10.1051/0004-6361/202451289}.

\bibitem[{Lelli} et~al.(2010){Lelli}, {Fraternali}, and {Sancisi}]{Lelli2010}
F.~{Lelli}, F.~{Fraternali}, and R.~{Sancisi}.
\newblock \emph{Astron. Astrophys.}, 516:\penalty0 A11, June 2010.
\newblock \doi{10.1051/0004-6361/200913808}.

\bibitem[{Lelli} et~al.(2012){Lelli}, {Verheijen}, {Fraternali}, and
  {Sancisi}]{Lelli2012a}
F.~{Lelli}, M.~{Verheijen}, F.~{Fraternali}, and R.~{Sancisi}.
\newblock \emph{Astron. Astrophys.}, 537:\penalty0 A72, Jan. 2012.
\newblock \doi{10.1051/0004-6361/201117867}.

\bibitem[{Lelli} et~al.(2013){Lelli}, {Fraternali}, and {Verheijen}]{Lelli2013}
F.~{Lelli}, F.~{Fraternali}, and M.~{Verheijen}.
\newblock \emph{Mon. Not. R. Astron. Soc.}, 433:\penalty0 L30--L34, June 2013.
\newblock \doi{10.1093/Mon. Not. R. Astron. Soc.l/slt053}.

\bibitem[{Lelli} et~al.(2014{\natexlab{a}}){Lelli}, {Fraternali}, and
  {Verheijen}]{Lelli2014b}
F.~{Lelli}, F.~{Fraternali}, and M.~{Verheijen}.
\newblock \emph{Astron. Astrophys.}, 563:\penalty0 A27, Mar.
  2014{\natexlab{a}}.
\newblock \doi{10.1051/0004-6361/201322658}.

\bibitem[{Lelli} et~al.(2014{\natexlab{b}}){Lelli}, {Verheijen}, and
  {Fraternali}]{Lelli2014a}
F.~{Lelli}, M.~{Verheijen}, and F.~{Fraternali}.
\newblock \emph{Astron. Astrophys.}, 566:\penalty0 A71, June
  2014{\natexlab{b}}.
\newblock \doi{10.1051/0004-6361/201322657}.

\bibitem[{Lelli} et~al.(2016{\natexlab{a}}){Lelli}, {McGaugh}, and
  {Schombert}]{Lelli2016a}
F.~{Lelli}, S.~S. {McGaugh}, and J.~M. {Schombert}.
\newblock \emph{Astron. J.}, 152\penalty0 (6):\penalty0 157, Dec.
  2016{\natexlab{a}}.
\newblock \doi{10.3847/0004-6256/152/6/157}.

\bibitem[{Lelli} et~al.(2016{\natexlab{b}}){Lelli}, {McGaugh}, and
  {Schombert}]{Lelli2016b}
F.~{Lelli}, S.~S. {McGaugh}, and J.~M. {Schombert}.
\newblock \emph{Astrophys. J.}, 816\penalty0 (1):\penalty0 L14, Jan.
  2016{\natexlab{b}}.
\newblock \doi{10.3847/2041-8205/816/1/L14}.

\bibitem[{Lelli} et~al.(2016{\natexlab{c}}){Lelli}, {McGaugh}, {Schombert}, and
  {Pawlowski}]{Lelli2016c}
F.~{Lelli}, S.~S. {McGaugh}, J.~M. {Schombert}, and M.~S. {Pawlowski}.
\newblock \emph{Astrophys. J.}, 827\penalty0 (1):\penalty0 L19, Aug.
  2016{\natexlab{c}}.
\newblock \doi{10.3847/2041-8205/827/1/L19}.

\bibitem[{Lelli} et~al.(2017{\natexlab{a}}){Lelli}, {McGaugh}, and
  {Schombert}]{Lelli2017b}
F.~{Lelli}, S.~S. {McGaugh}, and J.~M. {Schombert}.
\newblock \emph{\mnras}, 468\penalty0 (1):\penalty0 L68--L71, June
  2017{\natexlab{a}}.
\newblock \doi{10.1093/mnrasl/slx031}.

\bibitem[{Lelli} et~al.(2017{\natexlab{b}}){Lelli}, {McGaugh}, {Schombert}, and
  {Pawlowski}]{Lelli2017}
F.~{Lelli}, S.~S. {McGaugh}, J.~M. {Schombert}, and M.~S. {Pawlowski}.
\newblock \emph{Astrophys. J.}, 836\penalty0 (2):\penalty0 152, Feb.
  2017{\natexlab{b}}.
\newblock \doi{10.3847/1538-4357/836/2/152}.

\bibitem[{Lelli} et~al.(2018){Lelli}, {De Breuck}, {Falkendal}, {Fraternali},
  {Man}, {Nesvadba}, and {Lehnert}]{Lelli2018}
F.~{Lelli} et al.
\newblock \emph{\mnras}, 479\penalty0 (4):\penalty0 5440--5447, Oct. 2018.
\newblock \doi{10.1093/mnras/sty1795}.

\bibitem[{Lelli} et~al.(2019){Lelli}, {McGaugh}, {Schombert}, {Desmond}, and
  {Katz}]{Lelli2019}
F.~{Lelli} et al.
\newblock \emph{Mon. Not. R. Astron. Soc.}, 484\penalty0 (3):\penalty0
  3267--3278, Apr. 2019.
\newblock \doi{10.1093/Mon. Not. R. Astron. Soc./stz205}.

\bibitem[{Lelli} et~al.(2023){Lelli}, {Zhang}, {Bisbas}, {Lin}, {Papadopoulos},
  {Schombert}, {Di Teodoro}, {Marasco}, and {McGaugh}]{Lelli2023}
F.~{Lelli} et al.
\newblock \emph{\aap}, 672:\penalty0 A106, Apr. 2023.
\newblock \doi{10.1051/0004-6361/202245105}.

\bibitem[{Li} et~al.(2018){Li}, {Lelli}, {McGaugh}, and {Schombert}]{Li2018}
P.~{Li}, F.~{Lelli}, S.~{McGaugh}, and J.~{Schombert}.
\newblock \emph{Astron. Astrophys.}, 615:\penalty0 A3, July 2018.
\newblock \doi{10.1051/0004-6361/201732547}.

\bibitem[{Li} et~al.(2020){Li}, {Lelli}, {McGaugh}, and {Schombert}]{Li2020}
P.~{Li}, F.~{Lelli}, S.~{McGaugh}, and J.~{Schombert}.
\newblock \emph{Astrophys. J.S}, 247\penalty0 (1):\penalty0 31, Mar. 2020.
\newblock \doi{10.3847/1538-4365/ab700e}.

\bibitem[{Li} and {Modesto}(2020)]{LiModesto2020}
Q.~{Li} and L.~{Modesto}.
\newblock \emph{Gravitation and Cosmology}, 26\penalty0 (2):\penalty0 99--117,
  Aug. 2020.
\newblock \doi{10.1134/S0202289320020085}.

\bibitem[{Lin} et~al.(2025){Lin}, {Lelli}, {De Breuck}, {Man}, {Zhang},
  {Santini}, {Marasco}, {Castellano}, {Nesvadba}, {Bisbas}, {Huang}, and
  {Lehnert}]{Lin2025}
L.~{Lin} et al.
\newblock \emph{\aap}, 693:\penalty0 A91, Jan. 2025.
\newblock \doi{10.1051/0004-6361/202450814}.

\bibitem[{Ludlow} et~al.(2017){Ludlow}, {Ben{\'\i}tez-Llambay}, {Schaller},
  {Theuns}, {Frenk}, {Bower}, {Schaye}, {Crain}, {Navarro}, {Fattahi}, and
  {Oman}]{Ludlow2017}
A.~D. {Ludlow} et al.
\newblock \emph{Phys. Rev. Lett.}, 118\penalty0 (16):\penalty0 161103, Apr.
  2017.
\newblock \doi{10.1103/PhysRevLett.118.161103}.

\bibitem[{Macci{\`o}} et~al.(2012){Macci{\`o}}, {Paduroiu}, {Anderhalden},
  {Schneider}, and {Moore}]{Maccio2012}
A.~V. {Macci{\`o}} et al.
\newblock \emph{\mnras}, 424\penalty0 (2):\penalty0 1105--1112, Aug. 2012.
\newblock \doi{10.1111/j.1365-2966.2012.21284.x}.

\bibitem[{Mancera Pi{\~n}a} et~al.(2022){Mancera Pi{\~n}a}, {Fraternali},
  {Oosterloo}, {Adams}, {di Teodoro}, {Bacchini}, and {Iorio}]{ManceraPina2022}
P.~E. {Mancera Pi{\~n}a} et al.
\newblock \emph{\mnras}, 514\penalty0 (3):\penalty0 3329--3348, Aug. 2022.
\newblock \doi{10.1093/mnras/stac1508}.

\bibitem[{Mancera Pi{\~n}a} et~al.(2025){Mancera Pi{\~n}a}, {Read}, {Kim},
  {Marasco}, {Benavides}, {Glowacki}, {Pezzulli}, and {Lagos}]{ManceraPina2025}
P.~E. {Mancera Pi{\~n}a} et al.
\newblock \emph{\aap}, 699:\penalty0 A311, July 2025.
\newblock \doi{10.1051/0004-6361/202554381}.

\bibitem[{Marasco} et~al.(2018){Marasco}, {Oman}, {Navarro}, {Frenk}, and
  {Oosterloo}]{Marasco2018}
A.~{Marasco} et al.
\newblock \emph{Mon. Not. R. Astron. Soc.}, 476\penalty0 (2):\penalty0
  2168--2176, May 2018.
\newblock \doi{10.1093/Mon. Not. R. Astron. Soc./sty354}.

\bibitem[{Marasco} et~al.(2019){Marasco}, {Fraternali}, {Heald}, {de Blok},
  {Oosterloo}, {Kamphuis}, {J{\'o}zsa}, {Vargas}, {Winkel}, {Walterbos},
  {Dettmar}, and {Juẗte}]{Marasco2019}
A.~{Marasco} et al.
\newblock \emph{Astron. Astrophys.}, 631:\penalty0 A50, Nov. 2019.
\newblock \doi{10.1051/0004-6361/201936338}.

\bibitem[{Mayer} et~al.(2023){Mayer}, {Teklu}, {Dolag}, and {Remus}]{Mayer2023}
A.~C. {Mayer}, A.~F. {Teklu}, K.~{Dolag}, and R.-S. {Remus}.
\newblock \emph{\mnras}, 518\penalty0 (1):\penalty0 257--269, Jan. 2023.
\newblock \doi{10.1093/mnras/stac3017}.

\bibitem[{McGaugh}(2012)]{McGaugh2012}
S.~S. {McGaugh}.
\newblock \emph{Astron. J.}, 143:\penalty0 40, Feb. 2012.
\newblock \doi{10.1088/0004-6256/143/2/40}.

\bibitem[{McGaugh} and {Schombert}(2014)]{McGaugh2014}
S.~S. {McGaugh} and J.~M. {Schombert}.
\newblock \emph{\aj}, 148\penalty0 (5):\penalty0 77, Nov. 2014.
\newblock \doi{10.1088/0004-6256/148/5/77}.

\bibitem[{McGaugh} et~al.(2000){McGaugh}, {Schombert}, {Bothun}, and {de
  Blok}]{McGaugh2000}
S.~S. {McGaugh}, J.~M. {Schombert}, G.~D. {Bothun}, and W.~J.~G. {de Blok}.
\newblock \emph{Astrophys. J.}, 533:\penalty0 L99--L102, Apr. 2000.
\newblock \doi{10.1086/312628}.

\bibitem[{McGaugh} et~al.(2016){McGaugh}, {Lelli}, and
  {Schombert}]{McGaugh2016}
S.~S. {McGaugh}, F.~{Lelli}, and J.~M. {Schombert}.
\newblock \emph{Phys. Rev. Lett.}, 117\penalty0 (20):\penalty0 201101, Nov.
  2016.
\newblock \doi{10.1103/PhysRevLett.117.201101}.

\bibitem[{McNichols} et~al.(2016){McNichols}, {Teich}, {Nims}, {Cannon},
  {Adams}, {Bernstein-Cooper}, {Giovanelli}, {Haynes}, {J{\'o}zsa}, {McQuinn},
  {Salzer}, {Skillman}, {Warren}, {Dolphin}, {Elson}, {Haurberg}, {Ott},
  {Saintonge}, {Cave}, {Hagen}, {Huang}, {Janowiecki}, {Marshall}, {Thomann},
  and {Van Sistine}]{McNichols2016}
A.~T. {McNichols} et al.
\newblock \emph{Astrophys. J.}, 832\penalty0 (1):\penalty0 89, Nov. 2016.
\newblock \doi{10.3847/0004-637X/832/1/89}.

\bibitem[{McQuinn} et~al.(2010){McQuinn}, {Skillman}, {Cannon}, {Dalcanton},
  {Dolphin}, {Hidalgo-Rodr{\'{\i}}guez}, {Holtzman}, {Stark}, {Weisz}, and
  {Williams}]{McQuinn2010a}
K.~B.~W. {McQuinn} et al.
\newblock \emph{Astrophys. J.}, 721:\penalty0 297--317, Sept. 2010.
\newblock \doi{10.1088/0004-637X/721/1/297}.

\bibitem[{Meurer} et~al.(1996){Meurer}, {Carignan}, {Beaulieu}, and
  {Freeman}]{Meurer1996}
G.~R. {Meurer}, C.~{Carignan}, S.~F. {Beaulieu}, and K.~C. {Freeman}.
\newblock \emph{Astron. J.}, 111:\penalty0 1551--+, Apr. 1996.
\newblock \doi{10.1086/117895}.

\bibitem[{Meyer} et~al.(2017){Meyer}, {Robotham}, {Obreschkow}, {Westmeier},
  {Duffy}, and {Staveley-Smith}]{Meyer2017}
M.~{Meyer} et al.
\newblock \emph{\pasa}, 34:\penalty0 52, Nov. 2017.
\newblock \doi{10.1017/pasa.2017.31}.

\bibitem[{Milgrom}(1983{\natexlab{a}})]{Milgrom1983a}
M.~{Milgrom}.
\newblock \emph{Astrophys. J.}, 270:\penalty0 365--370, July
  1983{\natexlab{a}}.
\newblock \doi{10.1086/161130}.

\bibitem[{Milgrom}(1983{\natexlab{b}})]{Milgrom1983b}
M.~{Milgrom}.
\newblock \emph{Astrophys. J.}, 270:\penalty0 371--389, July
  1983{\natexlab{b}}.
\newblock \doi{10.1086/161131}.

\bibitem[{Milgrom}(1994)]{Milgrom1994}
M.~{Milgrom}.
\newblock \emph{Annals of Physics}, 229\penalty0 (2):\penalty0 384--415, Feb.
  1994.
\newblock \doi{10.1006/aphy.1994.1012}.

\bibitem[Milgrom(2014)]{Milgrom2014}
M.~Milgrom.
\newblock \emph{Scholarpedia}, 9\penalty0 (6):\penalty0 31410, 2014.
\newblock \doi{10.4249/scholarpedia.31410}.
\newblock revision \#196839.

\bibitem[{Mistele} et~al.(2022){Mistele}, {McGaugh}, and
  {Hossenfelder}]{Mistele2022}
T.~{Mistele}, S.~{McGaugh}, and S.~{Hossenfelder}.
\newblock \emph{\aap}, 664:\penalty0 A40, Aug. 2022.
\newblock \doi{10.1051/0004-6361/202243216}.

\bibitem[{Muni} et~al.(2025){Muni}, {Pontzen}, {Read}, {Agertz}, {Rey},
  {Taylor}, {Kim}, and {Gray}]{Muni2025}
C.~{Muni} et al.
\newblock \emph{\mnras}, 536\penalty0 (1):\penalty0 314--323, Jan. 2025.
\newblock \doi{10.1093/mnras/stae2748}.

\bibitem[{Murugeshan} et~al.(2024){Murugeshan}, {Deg}, {Westmeier}, {Shen},
  {For}, {Spekkens}, {Wong}, {Staveley-Smith}, {Catinella}, {Lee-Waddell},
  {D{\'e}nes}, {Rhee}, {Cortese}, {Goliath}, {Halloran}, {van der Hulst},
  {Kamphuis}, {Koribalski}, {Kraan-Korteweg}, {Lelli}, {Venkataraman},
  {Verdes-Montenegro}, and {Yu}]{Murugeshan2024}
C.~{Murugeshan} et al.
\newblock \emph{\pasa}, 41:\penalty0 e088, Nov. 2024.
\newblock \doi{10.1017/pasa.2024.91}.

\bibitem[{Naik} et~al.(2019){Naik}, {Puchwein}, {Davis}, {Sijacki}, and
  {Desmond}]{Naik2019}
A.~P. {Naik} et al.
\newblock \emph{\mnras}, 489\penalty0 (1):\penalty0 771--787, Oct. 2019.
\newblock \doi{10.1093/mnras/stz2131}.

\bibitem[{Navarro} et~al.(1996){Navarro}, {Frenk}, and {White}]{Navarro1996}
J.~F. {Navarro}, C.~S. {Frenk}, and S.~D.~M. {White}.
\newblock \emph{Astrophys. J.}, 462:\penalty0 563, May 1996.
\newblock \doi{10.1086/177173}.

\bibitem[{Nestor Shachar} et~al.(2023){Nestor Shachar}, {Price}, {F{\"o}rster
  Schreiber}, {Genzel}, {Shimizu}, {Tacconi}, {{\"U}bler}, {Burkert}, {Davies},
  {Dekel}, {Herrera-Camus}, {Lee}, {Liu}, {Lutz}, {Naab}, {Neri}, {Renzini},
  {Saglia}, {Schuster}, {Sternberg}, {Wisnioski}, and {Wuyts}]{Nestor2023}
A.~{Nestor Shachar} et al.
\newblock \emph{\apj}, 944\penalty0 (1):\penalty0 78, Feb. 2023.
\newblock \doi{10.3847/1538-4357/aca9cf}.

\bibitem[{Noordermeer} et~al.(2005){Noordermeer}, {van der Hulst}, {Sancisi},
  {Swaters}, and {van Albada}]{Noordermeer2005}
E.~{Noordermeer} et al.
\newblock \emph{Astron. Astrophys.}, 442:\penalty0 137--157, Oct. 2005.
\newblock \doi{10.1051/0004-6361:20053172}.

\bibitem[{Ntampaka} et~al.(2015){Ntampaka}, {Trac}, {Sutherland}, {Battaglia},
  {P{\'o}czos}, and {Schneider}]{Ntampaka2015}
M.~{Ntampaka} et al.
\newblock \emph{\apj}, 803\penalty0 (2):\penalty0 50, Apr. 2015.
\newblock \doi{10.1088/0004-637X/803/2/50}.

\bibitem[{Obreschkow} et~al.(2009){Obreschkow}, {Kl{\"o}ckner}, {Heywood},
  {Levrier}, and {Rawlings}]{Obreschkow2009}
D.~{Obreschkow} et al.
\newblock \emph{\apj}, 703\penalty0 (2):\penalty0 1890--1903, Oct. 2009.
\newblock \doi{10.1088/0004-637X/703/2/1890}.

\bibitem[{O'Brien} et~al.(2010){O'Brien}, {Freeman}, and {van der
  Kruit}]{OBrien2010}
J.~C. {O'Brien}, K.~C. {Freeman}, and P.~C. {van der Kruit}.
\newblock \emph{Astron. Astrophys.}, 515:\penalty0 A62, June 2010.
\newblock \doi{10.1051/0004-6361/200912567}.

\bibitem[{Oh} et~al.(2015){Oh}, {Hunter}, {Brinks}, {Elmegreen}, {Schruba},
  {Walter}, {Rupen}, {Young}, {Simpson}, {Johnson}, {Herrmann}, {Ficut-Vicas},
  {Cigan}, {Heesen}, {Ashley}, and {Zhang}]{Oh2015}
S.-H. {Oh} et al.
\newblock \emph{Astron. J.}, 149\penalty0 (6):\penalty0 180, June 2015.
\newblock \doi{10.1088/0004-6256/149/6/180}.

\bibitem[{Olling}(1995)]{Olling1995}
R.~P. {Olling}.
\newblock \emph{\aj}, 110:\penalty0 591, Aug. 1995.
\newblock \doi{10.1086/117545}.

\bibitem[{Oman} et~al.(2015){Oman}, {Navarro}, {Fattahi}, {Frenk}, {Sawala},
  {White}, {Bower}, {Crain}, {Furlong}, {Schaller}, {Schaye}, and
  {Theuns}]{Oman2015}
K.~A. {Oman} et al.
\newblock \emph{Mon. Not. R. Astron. Soc.}, 452\penalty0 (4):\penalty0
  3650--3665, Oct. 2015.
\newblock \doi{10.1093/Mon. Not. R. Astron. Soc./stv1504}.

\bibitem[{Oosterloo} et~al.(2007){Oosterloo}, {Fraternali}, and
  {Sancisi}]{Oosterloo07}
T.~{Oosterloo}, F.~{Fraternali}, and R.~{Sancisi}.
\newblock \emph{\aj}, 134\penalty0 (3):\penalty0 1019, Sept. 2007.
\newblock \doi{10.1086/520332}.

\bibitem[{Ott} et~al.(2012){Ott}, {Stilp}, {Warren}, {Skillman}, {Dalcanton},
  {Walter}, {de Blok}, {Koribalski}, and {West}]{Ott2012}
J.~{Ott} et al.
\newblock \emph{Astron. J.}, 144:\penalty0 123, Oct. 2012.
\newblock \doi{10.1088/0004-6256/144/4/123}.

\bibitem[{Panpanich} and {Burikham}(2018)]{Panpanich2018}
S.~{Panpanich} and P.~{Burikham}.
\newblock \emph{\prd}, 98\penalty0 (6):\penalty0 064008, Sept. 2018.
\newblock \doi{10.1103/PhysRevD.98.064008}.

\bibitem[{Paranjape} and {Sheth}(2022)]{Paranjape2022}
A.~{Paranjape} and R.~K. {Sheth}.
\newblock \emph{\mnras}, 517\penalty0 (1):\penalty0 130--139, Nov. 2022.
\newblock \doi{10.1093/mnras/stac2689}.

\bibitem[{Persic} et~al.(1996){Persic}, {Salucci}, and {Stel}]{Persic1996}
M.~{Persic}, P.~{Salucci}, and F.~{Stel}.
\newblock \emph{\mnras}, 281\penalty0 (1):\penalty0 27--47, July 1996.
\newblock \doi{10.1093/mnras/278.1.27}.

\bibitem[{Petersen} and {Lelli}(2020)]{Petersen2020}
J.~{Petersen} and F.~{Lelli}.
\newblock \emph{\aap}, 636:\penalty0 A56, Apr. 2020.
\newblock \doi{10.1051/0004-6361/201936964}.

\bibitem[{Pickering} et~al.(1997){Pickering}, {Impey}, {van Gorkom}, and
  {Bothun}]{Pickering1997}
T.~E. {Pickering}, C.~D. {Impey}, J.~H. {van Gorkom}, and G.~D. {Bothun}.
\newblock \emph{\aj}, 114:\penalty0 1858, Nov. 1997.
\newblock \doi{10.1086/118611}.

\bibitem[{Pillepich} et~al.(2019){Pillepich}, {Nelson}, {Springel}, {Pakmor},
  {Torrey}, {Weinberger}, {Vogelsberger}, {Marinacci}, {Genel}, {van der Wel},
  and {Hernquist}]{Pillepich2019}
A.~{Pillepich} et al.
\newblock \emph{\mnras}, 490\penalty0 (3):\penalty0 3196--3233, Dec. 2019.
\newblock \doi{10.1093/mnras/stz2338}.

\bibitem[{Ponomareva} et~al.(2018){Ponomareva}, {Verheijen}, {Papastergis},
  {Bosma}, and {Peletier}]{Ponomareva2018}
A.~A. {Ponomareva} et al.
\newblock \emph{\mnras}, 474\penalty0 (4):\penalty0 4366--4384, Mar. 2018.
\newblock \doi{10.1093/mnras/stx3066}.

\bibitem[{Ponomareva} et~al.(2021){Ponomareva}, {Mulaudzi}, {Maddox}, {Frank},
  {Jarvis}, {Di Teodoro}, {Glowacki}, {Kraan-Korteweg}, {Oosterloo}, {Adams},
  {Pan}, {Prandoni}, {Rajohnson}, {Sinigaglia}, {Adams}, {Heywood}, {Bowler},
  {Hatfield}, {Collier}, and {Sekhar}]{Ponomareva2021}
A.~A. {Ponomareva} et al.
\newblock \emph{\mnras}, 508\penalty0 (1):\penalty0 1195--1205, Nov. 2021.
\newblock \doi{10.1093/mnras/stab2654}.

\bibitem[{Pontzen} and {Governato}(2012)]{Pontzen2012}
A.~{Pontzen} and F.~{Governato}.
\newblock \emph{\mnras}, 421\penalty0 (4):\penalty0 3464--3471, Apr. 2012.
\newblock \doi{10.1111/j.1365-2966.2012.20571.x}.

\bibitem[{Posti}(2022)]{Posti22}
L.~{Posti}.
\newblock \emph{Research Notes of the American Astronomical Society},
  6\penalty0 (11):\penalty0 233, Nov. 2022.
\newblock \doi{10.3847/2515-5172/aca0df}.

\bibitem[{Powell} et~al.(2021){Powell}, {Vegetti}, {McKean}, {Spingola},
  {Rizzo}, and {Stacey}]{Powell21}
D.~{Powell} et al.
\newblock \emph{\mnras}, 501\penalty0 (1):\penalty0 515--530, Feb. 2021.
\newblock \doi{10.1093/mnras/staa2740}.

\bibitem[{Price} et~al.(2021){Price}, {Shimizu}, {Genzel}, {{\"U}bler},
  {F{\"o}rster Schreiber}, {Tacconi}, {Davies}, {Coogan}, {Lutz}, {Wuyts},
  {Wisnioski}, {Nestor}, {Sternberg}, {Burkert}, {Bender}, {Contursi},
  {Davies}, {Herrera-Camus}, {Lee}, {Naab}, {Neri}, {Renzini}, {Saglia},
  {Schruba}, and {Schuster}]{Price2021}
S.~H. {Price} et al.
\newblock \emph{\apj}, 922\penalty0 (2):\penalty0 143, Dec. 2021.
\newblock \doi{10.3847/1538-4357/ac22ad}.

\bibitem[{Puglisi} et~al.(2023){Puglisi}, {Dudzevi{\v{c}}i{\={u}}t{\.{e}}},
  {Swinbank}, {Gillman}, {Tiley}, {Bower}, {Cirasuolo}, {Cortese},
  {Glazebrook}, {Harrison}, {Ibar}, {Molina}, {Obreschkow}, {Oman}, {Schaller},
  {Shankar}, and {Sharples}]{Puglisi2023}
A.~{Puglisi} et al.
\newblock \emph{\mnras}, 524\penalty0 (2):\penalty0 2814--2835, Sept. 2023.
\newblock \doi{10.1093/mnras/stad1966}.

\bibitem[{Read} et~al.(2016){Read}, {Iorio}, {Agertz}, and
  {Fraternali}]{Read2016b}
J.~I. {Read}, G.~{Iorio}, O.~{Agertz}, and F.~{Fraternali}.
\newblock \emph{Mon. Not. R. Astron. Soc.}, 462\penalty0 (4):\penalty0
  3628--3645, Nov. 2016.
\newblock \doi{10.1093/Mon. Not. R. Astron. Soc./stw1876}.

\bibitem[{Ren} et~al.(2019){Ren}, {Kwa}, {Kaplinghat}, and {Yu}]{Ren2019}
T.~{Ren}, A.~{Kwa}, M.~{Kaplinghat}, and H.-B. {Yu}.
\newblock \emph{Physical Review X}, 9\penalty0 (3):\penalty0 031020, July 2019.
\newblock \doi{10.1103/PhysRevX.9.031020}.

\bibitem[{Rhee} et~al.(2004){Rhee}, {Valenzuela}, {Klypin}, {Holtzman}, and
  {Moorthy}]{Rhee2004}
G.~{Rhee} et al.
\newblock \emph{\apj}, 617\penalty0 (2):\penalty0 1059--1076, Dec. 2004.
\newblock \doi{10.1086/425565}.

\bibitem[{Rhee} et~al.(2023){Rhee}, {Meyer}, {Popping}, {Bellstedt}, {Driver},
  {Robotham}, {Whiting}, {Baldry}, {Brough}, {Brown}, {Bunton}, {Dodson},
  {Holwerda}, {Hopkins}, {Koribalski}, {Lee-Waddell}, {L{\'o}pez-S{\'a}nchez},
  {Loveday}, {Mahony}, {Roychowdhury}, {Rozgonyi}, and
  {Staveley-Smith}]{Rhee2023}
J.~{Rhee} et al.
\newblock \emph{\mnras}, 518\penalty0 (3):\penalty0 4646--4671, Jan. 2023.
\newblock \doi{10.1093/mnras/stac3065}.

\bibitem[{Rizzo} et~al.(2020){Rizzo}, {Vegetti}, {Powell}, {Fraternali},
  {McKean}, {Stacey}, and {White}]{Rizzo20}
F.~{Rizzo} et al.
\newblock \emph{\nat}, 584\penalty0 (7820):\penalty0 201--204, Aug. 2020.
\newblock \doi{10.1038/s41586-020-2572-6}.

\bibitem[{Rizzo} et~al.(2023){Rizzo}, {Roman-Oliveira}, {Fraternali},
  {Frickmann}, {Valentino}, {Brammer}, {Zanella}, {Kokorev}, {Popping},
  {Whitaker}, {Kohandel}, {Magdis}, {Di Mascolo}, {Ikeda}, {Jin}, and
  {Toft}]{Rizzo2023}
F.~{Rizzo} et al.
\newblock \emph{\aap}, 679:\penalty0 A129, Nov. 2023.
\newblock \doi{10.1051/0004-6361/202346444}.

\bibitem[{Robles} et~al.(2017){Robles}, {Bullock}, {Elbert}, {Fitts},
  {Gonz{\'a}lez-Samaniego}, {Boylan-Kolchin}, {Hopkins}, {Faucher-Gigu{\`e}re},
  {Kere{\v{s}}}, and {Hayward}]{Robles2017}
V.~H. {Robles} et al.
\newblock \emph{\mnras}, 472\penalty0 (3):\penalty0 2945--2954, Dec. 2017.
\newblock \doi{10.1093/mnras/stx2253}.

\bibitem[{Rogstad} et~al.(1974){Rogstad}, {Lockhart}, and
  {Wright}]{Rogstad1974}
D.~H. {Rogstad}, I.~A. {Lockhart}, and M.~C.~H. {Wright}.
\newblock \emph{Astrophys. J.}, 193:\penalty0 309--319, Oct. 1974.
\newblock \doi{10.1086/153164}.

\bibitem[{Romeo}(1992)]{Romeo1992}
A.~B. {Romeo}.
\newblock \emph{\mnras}, 256\penalty0 (2):\penalty0 307--320, May 1992.
\newblock \doi{10.1093/mnras/256.2.307}.

\bibitem[{Rubin} et~al.(1978){Rubin}, {Ford}, and {Thonnard}]{Rubin1978}
V.~C. {Rubin}, W.~K. {Ford}, Jr., and N.~{Thonnard}.
\newblock \emph{Astrophys. J.}, 225:\penalty0 L107--L111, Nov. 1978.
\newblock \doi{10.1086/182804}.

\bibitem[{Sales} et~al.(2022){Sales}, {Wetzel}, and {Fattahi}]{Sales2022}
L.~V. {Sales}, A.~{Wetzel}, and A.~{Fattahi}.
\newblock \emph{Nature Astronomy}, 6:\penalty0 897--910, June 2022.
\newblock \doi{10.1038/s41550-022-01689-w}.

\bibitem[{Sancisi}(2004)]{Sancisi2004}
R.~{Sancisi}.
\newblock In \emph{Dark Matter in Galaxies}, volume 220 of \emph{{IAU
  Symposia}}, page 233, July 2004.

\bibitem[{Sanders}(2010)]{Sanders2010}
R.~H. {Sanders}.
\newblock {The Dark Matter Problem: A Historical Perspective}, 2010.

\bibitem[{Schoenmakers} et~al.(1997){Schoenmakers}, {Franx}, and {de
  Zeeuw}]{Schoen1997}
R.~H.~M. {Schoenmakers}, M.~{Franx}, and P.~T. {de Zeeuw}.
\newblock \emph{Mon. Not. R. Astron. Soc.}, 292\penalty0 (2):\penalty0
  349--364, Dec. 1997.
\newblock \doi{10.1093/Mon. Not. R. Astron. Soc./292.2.349}.

\bibitem[{Schombert} and {McGaugh}(2014)]{Schombert2014}
J.~{Schombert} and S.~{McGaugh}.
\newblock \emph{Pub. Astron. Soc. Austr.}, 31:\penalty0 e036, Sept. 2014.
\newblock \doi{10.1017/Pub. Astron. Soc. Austr..2014.32}.

\bibitem[{Schombert} et~al.(2019){Schombert}, {McGaugh}, and
  {Lelli}]{Schombert2019}
J.~{Schombert}, S.~{McGaugh}, and F.~{Lelli}.
\newblock \emph{Mon. Not. R. Astron. Soc.}, 483\penalty0 (2):\penalty0
  1496--1512, Feb. 2019.
\newblock \doi{10.1093/Mon. Not. R. Astron. Soc./sty3223}.

\bibitem[{Shelest} and {Lelli}(2020)]{Shelest2020}
A.~{Shelest} and F.~{Lelli}.
\newblock \emph{\aap}, 641:\penalty0 A31, Sept. 2020.
\newblock \doi{10.1051/0004-6361/202038184}.

\bibitem[{Sicking}(1997)]{Sicking1997}
F.~J. {Sicking}.
\newblock \emph{{The thickness of the HI gas layer in spiral galaxies}}.
\newblock PhD thesis, -, Dec. 1997.

\bibitem[{Simon} et~al.(2003){Simon}, {Bolatto}, {Leroy}, and
  {Blitz}]{Simon2003}
J.~D. {Simon}, A.~D. {Bolatto}, A.~{Leroy}, and L.~{Blitz}.
\newblock \emph{Astrophys. J.}, 596\penalty0 (2):\penalty0 957--981, Oct. 2003.
\newblock \doi{10.1086/378200}.

\bibitem[{Simon} et~al.(2005){Simon}, {Bolatto}, {Leroy}, {Blitz}, and
  {Gates}]{Simon2005}
J.~D. {Simon} et al.
\newblock \emph{Astrophys. J.}, 621\penalty0 (2):\penalty0 757--776, Mar. 2005.
\newblock \doi{10.1086/427684}.

\bibitem[Skordis and Z\l{}o\ifmmode~\acute{s}\else
  \'{s}\fi{}nik(2021)]{Skordis2021}
C.~Skordis and T.~Z\l{}o\ifmmode~\acute{s}\else \'{s}\fi{}nik.
\newblock \emph{Phys. Rev. Lett.}, 127:\penalty0 161302, Oct 2021.
\newblock \doi{10.1103/PhysRevLett.127.161302}.

\bibitem[{Smith} et~al.(2019){Smith}, {Bureau}, {Davis}, {Cappellari}, {Liu},
  {North}, {Onishi}, {Iguchi}, and {Sarzi}]{Smith2019}
M.~D. {Smith} et al.
\newblock \emph{\mnras}, 485\penalty0 (3):\penalty0 4359--4374, May 2019.
\newblock \doi{10.1093/mnras/stz625}.

\bibitem[{Spekkens} and {Sellwood}(2007)]{Spekkens2007}
K.~{Spekkens} and J.~A. {Sellwood}.
\newblock \emph{Astrophys. J.}, 664\penalty0 (1):\penalty0 204--214, July 2007.
\newblock \doi{10.1086/518471}.

\bibitem[{Spergel} and {Steinhardt}(2000)]{Spergel2000}
D.~N. {Spergel} and P.~J. {Steinhardt}.
\newblock \emph{\prl}, 84\penalty0 (17):\penalty0 3760--3763, Apr. 2000.
\newblock \doi{10.1103/PhysRevLett.84.3760}.

\bibitem[{Spitzer}(1942)]{Spitzer1942}
J.~{Spitzer}, Lyman.
\newblock \emph{Astrophys. J.}, 95:\penalty0 329, May 1942.
\newblock \doi{10.1086/144407}.

\bibitem[{Stott} et~al.(2016){Stott}, {Swinbank}, {Johnson}, {Tiley}, {Magdis},
  {Bower}, {Bunker}, {Bureau}, {Harrison}, {Jarvis}, {Sharples}, {Smail},
  {Sobral}, {Best}, and {Cirasuolo}]{Stott2016}
J.~P. {Stott} et al.
\newblock \emph{\mnras}, 457\penalty0 (2):\penalty0 1888--1904, Apr. 2016.
\newblock \doi{10.1093/mnras/stw129}.

\bibitem[{Swaters} et~al.(2002){Swaters}, {van Albada}, {van der Hulst}, and
  {Sancisi}]{Swaters2002a}
R.~A. {Swaters}, T.~S. {van Albada}, J.~M. {van der Hulst}, and R.~{Sancisi}.
\newblock \emph{Astron. Astrophys.}, 390:\penalty0 829--861, Aug. 2002.
\newblock \doi{10.1051/0004-6361:20011755}.

\bibitem[{Swaters} et~al.(2009){Swaters}, {Sancisi}, {van Albada}, and {van der
  Hulst}]{Swaters2009}
R.~A. {Swaters}, R.~{Sancisi}, T.~S. {van Albada}, and J.~M. {van der Hulst}.
\newblock \emph{Astron. Astrophys.}, 493:\penalty0 871--892, Jan. 2009.
\newblock \doi{10.1051/0004-6361:200810516}.

\bibitem[{Swaters} et~al.(2012){Swaters}, {Sancisi}, {van der Hulst}, and {van
  Albada}]{Swaters2012}
R.~A. {Swaters}, R.~{Sancisi}, J.~M. {van der Hulst}, and T.~S. {van Albada}.
\newblock \emph{\mnras}, 425\penalty0 (3):\penalty0 2299--2308, Sept. 2012.
\newblock \doi{10.1111/j.1365-2966.2012.21599.x}.

\bibitem[{Swaters} et~al.(2014){Swaters}, {Bershady}, {Martinsson}, {Westfall},
  {Andersen}, and {Verheijen}]{Swaters2014}
R.~A. {Swaters} et al.
\newblock \emph{Astrophys. J.}, 797\penalty0 (2):\penalty0 L28, Dec. 2014.
\newblock \doi{10.1088/2041-8205/797/2/L28}.

\bibitem[{Tiley} et~al.(2019){Tiley}, {Swinbank}, {Harrison}, {Smail},
  {Turner}, {Schaller}, {Stott}, {Sobral}, {Theuns}, {Sharples}, {Gillman},
  {Bower}, {Bunker}, {Best}, {Richard}, {Bacon}, {Bureau}, {Cirasuolo}, and
  {Magdis}]{Tiley2019}
A.~L. {Tiley} et al.
\newblock \emph{\mnras}, 485\penalty0 (1):\penalty0 934--960, May 2019.
\newblock \doi{10.1093/mnras/stz428}.

\bibitem[{Trachternach} et~al.(2008){Trachternach}, {de Blok}, {Walter},
  {Brinks}, and {Kennicutt}]{Trachternach2008}
C.~{Trachternach} et al.
\newblock \emph{Astron. J.}, 136\penalty0 (6):\penalty0 2720--2760, Dec. 2008.
\newblock \doi{10.1088/0004-6256/136/6/2720}.

\bibitem[{van Albada} and {Sancisi}(1986)]{vanAlbada1986}
T.~S. {van Albada} and R.~{Sancisi}.
\newblock \emph{Royal Society of London Philosophical Transactions Series A},
  320:\penalty0 447--464, Dec. 1986.

\bibitem[{van Albada} et~al.(1985){van Albada}, {Bahcall}, {Begeman}, and
  {Sancisi}]{vanAlbada1985}
T.~S. {van Albada}, J.~N. {Bahcall}, K.~{Begeman}, and R.~{Sancisi}.
\newblock \emph{Astrophys. J.}, 295:\penalty0 305--313, Aug. 1985.
\newblock \doi{10.1086/163375}.

\bibitem[{Verheijen}(2001)]{Verheijen2001b}
M.~A.~W. {Verheijen}.
\newblock \emph{Astrophys. J.}, 563:\penalty0 694--715, Dec. 2001.
\newblock \doi{10.1086/323887}.

\bibitem[{Verheijen} and {Sancisi}(2001)]{Verheijen2001}
M.~A.~W. {Verheijen} and R.~{Sancisi}.
\newblock \emph{Astron. Astrophys.}, 370:\penalty0 765--867, May 2001.
\newblock \doi{10.1051/0004-6361:20010090}.

\bibitem[{Villanueva-Domingo} et~al.(2022){Villanueva-Domingo},
  {Villaescusa-Navarro}, {Angl{\'e}s-Alc{\'a}zar}, {Genel}, {Marinacci},
  {Spergel}, {Hernquist}, {Vogelsberger}, {Dave}, and
  {Narayanan}]{Villanueva-Domingo2022}
P.~{Villanueva-Domingo} et al.
\newblock \emph{\apj}, 935\penalty0 (1):\penalty0 30, Aug. 2022.
\newblock \doi{10.3847/1538-4357/ac7aa3}.

\bibitem[{V{\u{a}}r{\u{a}}{\c{s}}teanu}
  et~al.(2025){V{\u{a}}r{\u{a}}{\c{s}}teanu}, {Jarvis}, {Ponomareva},
  {Desmond}, {Heywood}, {Yasin}, {Maddox}, {Glowacki}, {Maksymowicz-Maciata},
  {Pi{\~n}a}, and {Pan}]{Varasteanu2025}
A.~A. {V{\u{a}}r{\u{a}}{\c{s}}teanu} et al.
\newblock \emph{\mnras}, 541\penalty0 (3):\penalty0 2366--2392, Aug. 2025.
\newblock \doi{10.1093/mnras/staf1079}.

\bibitem[{Walter} et~al.(2008){Walter}, {Brinks}, {de Blok}, {Bigiel},
  {Kennicutt}, {Thornley}, and {Leroy}]{Walter2008}
F.~{Walter} et al.
\newblock \emph{Astron. J.}, 136:\penalty0 2563, Dec. 2008.
\newblock \doi{10.1088/0004-6256/136/6/2563}.

\bibitem[{Wang} et~al.(2016){Wang}, {Koribalski}, {Serra}, {van der Hulst},
  {Roychowdhury}, {Kamphuis}, and {Chengalur}]{Wang2016}
J.~{Wang} et al.
\newblock \emph{Mon. Not. R. Astron. Soc.}, 460\penalty0 (2):\penalty0
  2143--2151, Aug. 2016.
\newblock \doi{10.1093/Mon. Not. R. Astron. Soc./stw1099}.

\bibitem[{Warren} et~al.(1992){Warren}, {Quinn}, {Salmon}, and
  {Zurek}]{Warren1992}
M.~S. {Warren}, P.~J. {Quinn}, J.~K. {Salmon}, and W.~H. {Zurek}.
\newblock \emph{\apj}, 399:\penalty0 405, Nov. 1992.
\newblock \doi{10.1086/171937}.

\bibitem[{Westmeier} et~al.(2022){Westmeier}, {Deg}, {Spekkens}, {Reynolds},
  {Shen}, {Gaudet}, {Goliath}, {Huynh}, {Venkataraman}, {Lin}, {O'Beirne},
  {Catinella}, {Cortese}, {D{\'e}nes}, {Elagali}, {For}, {J{\'o}zsa},
  {Howlett}, {van der Hulst}, {Jurek}, {Kamphuis}, {Kilborn}, {Kleiner},
  {Koribalski}, {Lee-Waddell}, {Murugeshan}, {Rhee}, {Serra}, {Shao},
  {Staveley-Smith}, {Wang}, {Wong}, {Zwaan}, {Allison}, {Anderson}, {Ball},
  {Bock}, {Brodrick}, {Bunton}, {Cooray}, {Gupta}, {Hayman}, {Mahony}, {Moss},
  {Ng}, {Pearce}, {Raja}, {Roxby}, {Voronkov}, {Warhurst}, {Courtois}, and
  {Said}]{Westmeier2022}
T.~{Westmeier} et al.
\newblock \emph{\pasa}, 39:\penalty0 e058, Nov. 2022.
\newblock \doi{10.1017/pasa.2022.50}.

\bibitem[{Wisnioski} et~al.(2019){Wisnioski}, {F{\"o}rster Schreiber},
  {Fossati}, {Mendel}, {Wilman}, {Genzel}, {Bender}, {Wuyts}, {Davies},
  {{\"U}bler}, {Bandara}, {Beifiori}, {Belli}, {Brammer}, {Chan}, {Davies},
  {Fabricius}, {Galametz}, {Lang}, {Lutz}, {Nelson}, {Momcheva}, {Price},
  {Rosario}, {Saglia}, {Seitz}, {Shimizu}, {Tacconi}, {Tadaki}, {van~Dokkum},
  and {Wuyts}]{Wisnioski2019}
E.~{Wisnioski} et al.
\newblock \emph{\apj}, 886\penalty0 (2):\penalty0 124, Dec. 2019.
\newblock \doi{10.3847/1538-4357/ab4db8}.

\bibitem[{Workman} et~al.(2022){Workman}, {Burkert}, {Crede}, {Klempt},
  {Thoma}, {Tiator}, {Agashe}, {Aielli}, {Allanach}, {Amsler}, {Antonelli},
  {Aschenauer}, {Asner}, {Baer}, {Banerjee}, {Barnett}, {Baudis}, {Bauer},
  {Beatty}, {Belousov}, {Beringer}, {Bettini}, {Biebel}, {Black}, {Blucher},
  {Bonventre}, {Bryzgalov}, {Buchmuller}, {Bychkov}, {Cahn}, {Carena},
  {Ceccucci}, {Cerri}, {Chivukula}, {Cowan}, {Cranmer}, {Cremonesi},
  {D'Ambrosio}, {Damour}, {de Florian}, {de Gouv{\^e}a}, {DeGrand}, {de Jong},
  {Demers}, {Dobrescu}, {D'Onofrio}, {Doser}, {Dreiner}, {Eerola}, {Egede},
  {Eidelman}, {El-Khadra}, {Ellis}, {Eno}, {Erler}, {Ezhela}, {Fetscher},
  {Fields}, {Freitas}, {Gallagher}, {Gershtein}, {Gherghetta},
  {Gonzalez-Garcia}, {Goodman}, {Grab}, {Gritsan}, {Grojean}, {Groom},
  {Gr{\"u}newald}, {Gurtu}, {Gutsche}, {Haber}, {Hamel}, {Hanhart},
  {Hashimoto}, {Hayato}, {Hebecker}, {Heinemeyer}, {Hern{\'a}ndez-Rey},
  {Hikasa}, {Hisano}, {H{\"o}cker}, {Holder}, {Hsu}, {Huston}, {Hyodo},
  {Ianni}, {Kado}, {Karliner}, {Katz}, {Kenzie}, {Khoze}, {Klein}, {Krauss},
  {Kreps}, {Kri{\v{z}}an}, {Krusche}, {Kwon}, {Lahav}, {Laiho}, {Lellouch},
  {Lesgourgues}, {Liddle}, {Ligeti}, {Lin}, {Lippmann}, {Liss}, {Littenberg},
  {Louren{\c{c}}o}, {Lugovsky}, {Lugovsky}, {Lusiani}, {Makida}, {Maltoni},
  {Mannel}, {Manohar}, {Marciano}, {Masoni}, {Matthews}, {Mei{\ss}ner},
  {Melzer-Pellmann}, {Mikhasenko}, {Miller}, {Milstead}, {Mitchell},
  {M{\"o}nig}, {Molaro}, {Moortgat}, {Moskovic}, {Nakamura}, {Narain}, {Nason},
  {Navas}, {Nelles}, {Neubert}, {Nevski}, {Nir}, {Olive}, {Patrignani},
  {Peacock}, {Petrov}, {Pianori}, {Pich}, {Piepke}, {Pietropaolo}, {Pomarol},
  {Pordes}, {Profumo}, {Quadt}, {Rabbertz}, {Rademacker}, {Raffelt},
  {Ramsey-Musolf}, {Ratcliff}, {Richardson}, {Ringwald}, {Robinson}, {Roesler},
  {Rolli}, {Romaniouk}, {Rosenberg}, {Rosner}, {Rybka}, {Ryskin}, {Ryutin},
  {Sakai}, {Sarkar}, {Sauli}, {Schneider}, {Sch{\"o}nert}, {Scholberg},
  {Schwartz}, {Schwiening}, {Scott}, {Sefkow}, {Seljak}, {Sharma}, {Sharpe},
  {Shiltsev}, {Signorelli}, {Silari}, {Simon}, {Sj{\"o}strand}, {Skands},
  {Skwarnicki}, {Smoot}, {Soffer}, {Sozzi}, {Spanier}, {Spiering}, {Stahl},
  {Stone}, {Sumino}, {Syphers}, {Takahashi}, {Tanabashi}, {Tanaka},
  {Ta{\v{s}}evsk{\'y}}, {Terao}, and {Terashi}]{2022PTEP.2022h3C01W}
R.~L. {Workman} et al.
\newblock \emph{Progress of Theoretical and Experimental Physics},
  2022\penalty0 (8):\penalty0 083C01, Aug. 2022.
\newblock \doi{10.1093/ptep/ptac097}.

\bibitem[{Xi} et~al.(2024){Xi}, {Peng}, {Staveley-Smith}, {For}, {Liu}, {Chen},
  {Yu}, {Ding}, {Guo}, {Zou}, {Xue}, {Wang}, {Brink}, {Zheng}, {Filippenko},
  {Yang}, {Wei}, {Dai}, {Li}, {He}, {Jiang}, {Moiseev}, and {Kotov}]{Xi2024}
H.~{Xi} et al.
\newblock \emph{\apjl}, 966\penalty0 (2):\penalty0 L36, May 2024.
\newblock \doi{10.3847/2041-8213/ad4357}.

\end{thebibliography}
